\documentclass[11pt]{article}

\usepackage{epsf}
\usepackage{amsmath}
\usepackage{amssymb}
\usepackage{graphicx}
\usepackage{dcolumn}
\usepackage{bm}
\usepackage{lscape}
\usepackage{tabularx}
\usepackage[titles]{tocloft}
\usepackage{cite}

\newcommand{\append}[1]{\protect\refstepcounter{section}
                \section*{Appendix \thesection \, #1}
                \addcontentsline{toc}{section}{Appendix \thesection: #1}}
\def\thefootnote{\fnsymbol{footnote}}

\begin{document}

\begin{center}
\Large\bf\boldmath
\vspace*{2.0cm}SuperIso v3.0: A program for calculating flavor physics observables in 2HDM and supersymmetry
\unboldmath
\end{center}

\vspace{0.8cm}
\begin{center}
F. Mahmoudi\footnote{Electronic address: \tt mahmoudi@in2p3.fr\\
\hspace*{0.48cm} URL: \tt http://superiso.in2p3.fr}\\[0.4cm]
{\sl Laboratoire de Physique Corpusculaire de Clermont-Ferrand (LPC),\\ Universit\'e Blaise Pascal, CNRS/IN2P3, 63177 Aubi\`ere Cedex, France}
\end{center}
\vspace{0.9cm}
\begin{abstract}
\noindent We describe \verb?SuperIso v3.0? which is a public program for evaluation of flavor physics observables in the Standard Model (SM), general two-Higgs-doublet model (2HDM), minimal supersymmetric Standard Model (MSSM) and next to minimal supersymmetric Standard Model (NMSSM). \verb?SuperIso v3.0?, in addition to the branching ratio of $B \to X_s \gamma$ and the isospin asymmetry of $B \to K^* \gamma$, incorporates other flavor observables such as the branching ratio of $B_s \to \mu^+ \mu^-$, the branching ratio of $B \to \tau \nu_\tau$, the branching ratio of $B \to D \tau \nu_\tau$, the branching ratio of $K \to \mu \nu_\mu$ and the branching ratios of $D_s \to \tau \nu_\tau$ and $D_s \to \mu \nu_\mu$. The calculation of the branching ratio of $B \to X_s \gamma$ is also improved, as it includes NNLO Standard Model contributions. The program also computes the muon anomalous magnetic moment $(g-2)$. Nine sample models are included in the package, namely SM, 2HDM, and mSUGRA, NUHM, AMSB and GMSB for the MSSM, and CNMSSM, NGMSB and NNUHM for the NMSSM. \verb?SuperIso? uses a SUSY Les Houches Accord file (SLHA1 or SLHA2) as input, which can be either generated automatically by the program via a call to external spectrum calculators, or provided by the user. The calculation of the observables is detailed in the Appendices, where a suggestion for the allowed intervals for each observable is also provided.
\end{abstract}
\newpage
\tableofcontents
\newpage
\listoftables
\newpage
\def\thefootnote{\arabic{footnote}}
\setcounter{footnote}{0}
\section{Introduction} 
Along with the direct searches for new physics and particles, indirect searches appear as very important complementary tools to explore physics beyond the Standard Model (SM). The presence of new particles as virtual states in processes involving only ordinary particles as external states provides us with the opportunities to study the indirect effects of new physics. This is the case for example of rare $B$ decays or the anomalous magnetic moment of the charged leptons. We investigate here the extension of the Higgs sector in the two-Higgs-doublet model (2HDM) and supersymmetry (SUSY), which are the most popular extensions of the Standard Model.\\
\\
Phenomenological interests of indirect constraints are numerous. This includes, in first place, tests of the SM predictions. By studying the indirect constraints it is also possible to reveal indirect effects of new physics. This aspect can be complementary to direct searches or even used as guideline. Finally, the results of indirect searches can be employed in order to check consistencies with the direct search results.\\
\\
The main purpose of the \verb?SuperIso? program is to offer the possibility to evaluate the most important indirect observables and constraints.\\
\\
\verb?SuperIso? \cite{superiso} was in its first versions devoted to the calculation of the isospin symmetry breaking in $B \to K^* \gamma$ decays in the minimal supersymmetric extension of the Standard Model (MSSM) with minimal flavor violation. This observable imposes stringent constraints on supersymmetric models \cite{ahmady}, which justifies a dedicated program. The calculation of the $b \to s \gamma$ branching ratio was also included in the first version and has been improved by adding the Standard Model NNLO contributions since version 2. Also, a broader set of flavor physics observables has been implemented. This includes the branching ratio of $B_s \to \mu^+ \mu^-$, the branching ratio of $B_u \to \tau \nu_\tau$, the branching ratios of $B \to D^0 \tau \nu_\tau$ and $B \to D^0 e \nu_e$, the branching ratio of $K \to \mu \nu_\mu$ and the branching ratios of $D_s \to \tau \nu_\tau$ and $D_s \to \mu \nu_\mu$. The calculation of the anomalous magnetic moment of the muon is also implemented in the program.\\
\\
\verb?SuperIso? has been extended to the general two-Higgs-doublet model since version 2.6 and to the next to minimal supersymmetric extension of the Standard Model (NMSSM) since version 3.0 \cite{superiso3}.\\
\\
\verb?SuperIso? uses a SUSY Les Houches Accord file \cite{slha,slha2} as input, which can be either generated automatically by the program via a call to SOFTSUSY \cite{softsusy}, ISAJET \cite{isajet}, NMSSMTools \cite{nmssmtools}, or provided by the user. \verb?SuperIso? can also use the LHA inspired format for the 2HDM, generated by 2HDMC \cite{2hdmc}, which will be described in Appendix \ref{lha_2hdm}. The program is able to perform the calculations automatically for different types of 2HDM (I--IV), for different supersymmetry breaking scenarios, such as the minimal Supergravity Grand Unification (mSUGRA) also called Constrained MSSM (CMSSM), the Non-Universal Higgs Mass model (NUHM), the Anomaly Mediated Supersymmetry Breaking scenario (AMSB) and the Gauge Mediated Supersymmetry Breaking scenario (GMSB), and for the NMSSM scenarios namely CNMSSM, NGMSB and NNUHM.\\
\\
In the following, we first discuss the content of the \verb?SuperIso? package and give the list of the main routines. Then we describe the procedure to use \verb?SuperIso?, and introduce the inputs and outputs of the program. Finally, we present some examples of results obtained with \verb?SuperIso?. In the Appendices, a complete description of the formulas used to calculate the different observables in \verb?SuperIso? is given for reference, as well as suggestions for the allowed intervals.\\
%
\section{Content of the SuperIso v3.0 package}%
\label{content}
\verb?SuperIso? is a C program respecting the C99 standard, devoted to the calculation of the most constraining flavor physics observables. Ten main programs are provided in the package, but the users are also invited to write their own main programs. \verb?slha.c? can scan files written following the SUSY Les Houches Accord formats, and calculates the corresponding observables. \verb?sm.c? provides the values of the observables in the Standard Model while \verb?thdm.c? computes the observables in 2HDM types I--IV and requires \verb?2HDMC? \cite{2hdmc} for the generation of the input file containing the Higgs masses and couplings. The main programs \verb?msugra.c?, \verb?amsb.c?, \verb?gmsb.c?, and \verb?nuhm.c? have to be linked to the \verb?SOFTSUSY? \cite{softsusy} and/or the \verb?ISASUGRA/ISAJET? \cite{isajet} packages, in order to compute supersymmetric mass spectra and couplings within respectively the mSUGRA, AMSB, GMSB or NUHM scenarios. The programs \verb?cnmssm.c?, \verb?ngmsb.c?, and \verb?nnuhm.c? have to be linked to the \verb?NMSSMTools? \cite{nmssmtools} program to compute supersymmetric mass spectra and couplings within respectively the CNMSSM, NGMSB or NNUHM scenarios.\\
\\
The computation of the different observables in \verb?SuperIso? proceeds following three main steps:
\begin{itemize}
\item Generation of a SLHA file with \verb?ISAJET?, \verb?SOFTSUSY?, \verb?NMSSMTools? or \verb?2HDMC? (or supply of a SLHA file by the user),\vspace*{-0.2cm}
\item Scan of the SLHA file,\vspace*{-0.2cm}
\item Calculation of the observables.
\end{itemize}
The last point hides a complex procedure: to compute the inclusive branching ratio of $b \rightarrow s \gamma$ for example, \verb?SuperIso? needs first to compute the Wilson coefficients at matching scale, and then to evolve them using the Renormalization Group Equations (RGE) to a lower scale, before using them to compute the branching ratio.\\
\subsection{Parameter structure}
The package \verb?SuperIso v3.0? has been deeply revised and expanded as compared to the first release, and many of the original routines have been modified. The code still relies on the definition of a structure in \verb?src/include.h?, which has been also expanded, in particular in order to support the new parameters of the SLHA2 format. This structure is defined as follows:\\
\begin{verbatim}
typedef struct parameters
/* structure containing all the scanned parameters from the SLHA file */
{
	int SM;
	int model; /* mSUGRA = 1, GMSB = 2, AMSB = 3 */
	int generator; /* ISAJET = 1, SOFTSUSY = 2 */
	double Q; /* Qmax ; default = M_EWSB = sqrt(m_stop1*mstop2) */
	double m0,m12,tan_beta,sign_mu,A0; /* mSUGRA parameters */
	double Lambda,Mmess,N5,cgrav,m32; /* AMSB, GMSB parameters */
	double mass_Z,mass_W,mass_b,mass_top_pole,mass_tau_pole; /* SM parameters */
	double inv_alpha_em,alphas_MZ,alpha,Gfermi,GAUGE_Q; /* SM parameters */
	double charg_Umix[3][3],charg_Vmix[3][3],stop_mix[3][3],sbot_mix[3][3],
	stau_mix[3][3],neut_mix[6][6],mass_neut[6]; /* mass mixing matrices */
	double Min,M1_Min,M2_Min,M3_Min,At_Min,Ab_Min,Atau_Min,M2H1_Min,M2H2_Min,
	mu_Min,M2A_Min,tb_Min,mA_Min; /* optional input parameters at scale Min */
	double MeL_Min,MmuL_Min,MtauL_Min,MeR_Min,MmuR_Min,
	MtauR_Min; /* optional input parameters at scale Min */
	double MqL1_Min,MqL2_Min,MqL3_Min,MuR_Min,McR_Min,MtR_Min,
	MdR_Min,MsR_Min,MbR_Min; /* optional input parameters at scale Min */
	double N51,N52,N53,M2H1_Q,M2H2_Q; /* optional input parameters (N51...3: GMSB) */
	double mass_d,mass_u,mass_s,mass_c,mass_t,mass_e,mass_nue,mass_mu,
	mass_num,mass_tau,mass_nut; /* SM masses */
	double mass_gluon,mass_photon,mass_Z0; /* SM masses */
	double mass_h0,mass_H0,mass_A0,mass_H,mass_dnl,mass_upl,mass_stl,mass_chl,
	mass_b1,mass_t1; /* Higgs & superparticle masses */
	double mass_el,mass_nuel,mass_mul,mass_numl,mass_tau1,mass_nutl,mass_gluino,
	mass_cha1,mass_cha2; /* superparticle masses */
	double mass_dnr,mass_upr,mass_str,mass_chr,mass_b2,mass_t2,mass_er,mass_mur,
	mass_tau2; /* superparticle masses */
	double mass_nuer,mass_numr,mass_nutr,mass_graviton,
	mass_gravitino; /* superparticle masses */
	double gp,g2,g3,YU_Q,yut[4],YD_Q,yub[4],YE_Q,yutau[4]; /* gauge & Yukawa couplings */
	double HMIX_Q,mu_Q,tanb_GUT,Higgs_VEV,mA2_Q,MSOFT_Q,M1_Q,M2_Q,
	M3_Q; /* parameters at scale Q */
	double MeL_Q,MmuL_Q,MtauL_Q,MeR_Q,MmuR_Q,MtauR_Q,MqL1_Q,MqL2_Q,MqL3_Q,MuR_Q,McR_Q,
	MtR_Q,MdR_Q,MsR_Q,MbR_Q; /* masses at scale Q */
	double AU_Q,A_u,A_c,A_t,AD_Q,A_d,A_s,A_b,AE_Q,A_e,A_mu,A_tau; /*trilinear couplings*/
	
	/* SLHA2 */
	int NMSSM,Rparity,CPviolation,Flavor;
	double mass_nutau2,mass_e2,mass_nue2,mass_mu2,mass_numu2,mass_d2,mass_u2,mass_s2,
	mass_c2;
	double CKM_lambda,CKM_A,CKM_rho,CKM_eta;
	double PMNS_theta12,PMNS_theta23,PMNS_theta13,PMNS_delta13,PMNS_alpha1,PMNS_alpha2;
	double lambdaNMSSM_Min,kappaNMSSM_Min,AlambdaNMSSM_Min,AkappaNMSSM_Min,
	lambdaSNMSSM_Min,xiFNMSSM_Min,xiSNMSSM_Min,mupNMSSM_Min,mSp2NMSSM_Min,mS2NMSSM_Min,
	mass_H03,mass_A02,NMSSMRUN_Q,lambdaNMSSM,kappaNMSSM,AlambdaNMSSM,AkappaNMSSM,
	lambdaSNMSSM,xiFNMSSM,xiSNMSSM,mupNMSSM,mSp2NMSSM,mS2NMSSM; /* NMSSM parameters */
	double PMNSU_Q,CKM_Q,MSE2_Q,MSU2_Q,MSD2_Q,MSL2_Q,MSQ2_Q,TU_Q,TD_Q,TE_Q;
	int NMSSMcoll,NMSSMtheory,NMSSMups1S,NMSSMetab1S; /* NMSSM exclusions */
	
	double CKM[4][4]; /* CKM matrix */
	double H0_mix[4][4],A0_mix[4][4]; /* Higgs mixing matrices */
	double sU_mix[7][7],sD_mix[7][7],sE_mix[7][7], sNU_mix[4][4]; /* mixing matrices */
	double sCKM_msq2[4][4],sCKM_msl2[4][4],sCKM_msd2[4][4],sCKM_msu2[4][4],
	sCKM_mse2[4][4]; /* super CKM matrices */
	double PMNS_U[4][4]; /* PMNS mixing matrices */
	double TU[4][4],TD[4][4],TE[4][4]; /* trilinear couplings */
	
	/* non-SLHA */
	double mass_b_1S,mass_b_pole,mtmt;
	double Lambda5; /* Lambda QCD */
	
	/* Flavor constants */
	double f_B,f_Bs,f_Ds,m_B,m_Bs,m_Ds,m_K,m_Kstar,m_D,life_B,life_Bs,life_Ds;
	
	/* Higgs widths */
	double width_h0,width_H0,width_A0,width_H,width_H03,width_A02;

	/* 2HDM */
	int THDM_model;
	double lambda_u[4][4],lambda_d[4][4],lambda_l[4][4];
}
parameters;

\end{verbatim}
This structure contains all the important parameters and is called by most of the main functions in the program. 
\subsection{Main routines}
The main routines of \verb?SuperIso? are explained in the following. The complete list of implemented routines can be found in \verb?src/include.h?.
\begin{itemize}
\item \verb?void Init_param(struct parameters* param)? \\
\\
This function initializes the \verb?param? structure, setting all the parameters to 0, apart from the SM masses and couplings, which receive the values given in the PDG \cite{PDG}.\\

\item \verb?int Les_Houches_Reader(char name[], struct parameters* param)? \\
\\
This routine reads the SLHA file whose name is contained in \verb?name?, and put all the read parameters in the structure \verb?param?. This function has been updated to the SLHA2 format. This routine can also read the LHA inspired format for the 2HDM described in Appendix \ref{lha_2hdm}. A negative value for \verb?param->model? indicates a problem in reading the SLHA file, or a model not yet included in SuperIso (such as $R$-parity breaking models). In this case, \verb?Les_Houches_Reader? returns 0, otherwise 1.\\

\item \verb?int test_slha(char name[])?\\
\\
This routine checks if the SLHA file whose name is contained in \verb?name? is valid, and if so returns 1. If not, -1 means that in the SLHA generator the computation did not succeed ({\it e.g.} because of tachyonic particles), -2 means that the considered model is not currently implemented in \verb?SuperIso?, and -3 that the provided file is either not in the SLHA format, or some important elements are missing.\\

\item \verb?int softsusy_sugra(double m0, double m12, double tanb, double A0,?\\
\verb?double sgnmu, double mtop, double mbot, double alphas_mz, char name[])? 
\item \verb?int isajet_sugra(double m0, double m12, double tanb, double A0, ?\\
\verb?double sgnmu, double mtop, char name[])? 
\item \verb?int softsusy_gmsb(double Lambda, double Mmess, double tanb, int N5,?\\
\verb? double cGrav, double sgnmu, double mtop, double mbot, double alphas_mz,?\\
\verb? char name[])? 
\item \verb?int softsusy_amsb(double m0, double m32, double tanb, double sgnmu,?\\
\verb?double mtop, double mbot, double alphas_mz, char name[])? 
\item \verb?int softsusy_nuhm(double m0, double m12, double tanb, double A0, double mu,?\\ 
\verb?double mA, double mtop, double mbot, double alphas_mz, char name[])?\\
\\
The above routines call \verb?SOFTSUSY? or \verb?ISAJET? to compute the mass spectrum corresponding to the input parameters (more details are given in the next sections), and return a SLHA file whose name has to be specified in the string \verb?name?.\\

\item \verb?int thdmc_types(double l1, double l2, double l3, double l4, double l5,?\\
\verb?double l6, double l7, double m12_2, double tanb, int type, char name[])?\\
\\
This routine calls \verb?2HDMC? to compute the masses and couplings corresponding to the 2HDM input parameters, and returns a LHA inspired file whose name has to be specified in the string \verb?name?.\\

\item \verb?int nmssmtools_cnmssm(double m0, double m12, double tanb, double A0,?\\
\verb?double lambda, double AK, double sgnmu, double mtop, double mbot,?\\
\verb?double alphas_mz, char name[])?
\item \verb?int nmssmtools_nnuhm(double m0, double m12, double tanb, double A0, ?\\
\verb?double MHDGUT, double MHUGUT, double lambda, double AK, double sgnmu,?\\
\verb?double mtop, double mbot, double alphas_mz, char name[])?
\item \verb?int nmssmtools_ngmsb(double Lambda, double Mmess, double tanb, int N5,?\\
\verb?double lambda, double AK, double Del_h, double sgnmu, double mtop,?\\
\verb?double mbot, double alphas_mz, char name[])?\\
\\
The above routines call \verb?NMSSMTools? to compute the mass spectrum corresponding to the input parameters (more details are given in the next sections) and return the SLHA2 file \verb?name?.\\

\item \verb?double alphas_running(double Q, double mtop, double mbot, ?\\
\verb?struct parameters* param)? \\
\\
This function computes the strong coupling constant at the energy scale \verb?Q? using the parameters in \verb?param?, provided the top quark mass \verb?mtop? and bottom quark mass \verb?mbot? used for the matching between the scales corresponding to different flavor numbers are specified. The main formula for calculating $\alpha_s$ is given in Appendix~\ref{alphas}. The prototype of this function has changed since version 2.\\
\item \verb?double running_mass(double quark_mass, double Qinit, double Qfin, double mtop,?\\
\verb?double mbot, struct parameters* param)? \\
\\
This function calculates the running quark mass at the scale \verb?Qfin?, for a quark of mass \verb?quark_mass? at the scale \verb?Qinit? using the structure \verb?param?, knowing the matching scales \verb?mtop? and \verb?mbot?. A description of the quark mass calculations can be found in Appendix~\ref{quarkmasses}.\\
\item \verb?void CW_calculator(double C0w[], double C1w[], double C2w[], double mu_W,?\\
\verb?struct parameters* param)?
\item \verb?void C_calculator_base1(double C0w[], double C1w[], double C2w[], double mu_W,?\\
\verb?double C0b[], double C1b[], double C2b[], double mu, struct parameters* param)?
\item \verb?void C_calculator_base2(double C0w[], double C1w[], double mu_W, double C0b[],?\\
\verb?double C1b[], double mu, struct parameters* param)?\\
\\
These three routines replace \verb?C_calculator? in the first version.\\
The procedure \verb?CW_calculator? computes the LO contributions to the Wilson coefficients \verb?C0w[]?, the NLO contributions \verb?C1w[]? and the NNLO contributions \verb?C2w[]? at the matching scale \verb?mu_W?, using the parameters of \verb?param?, as described in Appendix~\ref{wilsonMW}.\\
\\
\verb?C_calculator_base1? evolves the LO, NLO and NNLO Wilson coefficients \verb?C0w[]?, \verb?C1w[]?, \verb?C2w[]? initially at scale \verb?mu_W? towards \verb?C0b[]?, \verb?C1b[]?, \verb?C2b[]? at scale \verb?mu?, in the standard operator basis described in Appendix~\ref{wilsonbasis1}. \\
\verb?C_calculator_base2? evolves the LO and NLO Wilson coefficients \verb?C0w[]?, \verb?C1w[]? initially at scale \verb?mu_W? towards \verb?C0b[]?, \verb?C1b[]? at scale \verb?mu?, in the traditional operator basis described in Appendix~\ref{wilsonbasis2}. 

\item \verb?double bsgamma(double C0[], double C1[], double C2[], double mu, double mu_W,?\\
\verb?struct parameters* param)? \\
\\
This function has replaced the previous calculation of $b \rightarrow s \gamma$, which was performed at NLO accuracy. Here, knowing the LO, NLO and NNLO Wilson coefficients \verb?C0[]?, \verb?C1[]?, \verb?C2[]? at scale \verb?mu?, and given the matching scale \verb?mu_W? this procedure computes the inclusive branching ratio of $b \rightarrow s \gamma$ at NNLO, as described in Appendix~\ref{bsgamma}.\\
The container routine \verb?double bsgamma_calculator(char name[])?, in which \verb?name? contains the name of a SLHA file, automatizes the whole calculation, as it first calls \verb?Init_param? and \verb?Les_Houches_Reader?, then \verb?CW_calculator? and \verb?C_calculator_base1?, and finally \verb?bsgamma?.

\item \verb?double delta0(double C0[],double C0_spec[],double C1[],double C1_spec[],?\\
\verb?struct parameters* param,double mub,double muspec, double lambda_h)? \\
\\
This function computes the isospin asymmetry as described in Appendix~\ref{isospin}, using both the LO and NLO parts of the Wilson coefficients at scale \verb?mub? (\verb?C0[]? and \verb?C1[]?), and at the spectator scale \verb?muspec? (\verb?C0_spec[]? and \verb?C1_spec[]?), with the additional input $\Lambda_h$ in GeV. Compared to the first version, the calculation has been improved, and all the involved integrals have been coded in separate routines. Again, an automatic container routine which only needs the name of a SLHA file is provided:
\verb?double delta0_calculator(char name[])?.

\item \verb?int excluded_mass_calculator(char name[])? \\%
\\
This routine, with the name of the SLHA file in the argument, checks whether the parameter space point is excluded by the collider constraints on the particle masses and if so returns 1. The implemented mass limits are given in Appendix~\ref{constraints}, and can be updated by the users in \verb?src/excluded_masses.c? . These limits are valid only in the MSSM.

\item \verb?int NMSSM_collider_excluded(char name[])?
\item \verb?int NMSSM_theory_excluded(char name[])? \\%
\\
These two routines only apply to the SLHA file \verb?name? generated by NMSSMTools, as they need NMSSMTools specific outputs. They respectively check if a parameter space point is excluded by collider constraints \cite{ellwanger2005} or by theoretical constraints (such as unphysical global minimum). The output $1$ means that the point is excluded.

\item \verb?int charged_LSP_calculator(char name[])? \\%
\\
This routine, with the name of the SLHA file in the argument, checks whether the LSP is charged, and if so returns 1.
\end{itemize}

\noindent The following routines encode newly implemented observables:
\begin{itemize}
\item \verb?double Btaunu_calculator(char name[])?\\
\verb?double RBtaunu_calculator(char name[])?\\
\\
These two routines compute respectively the branching ratio of the leptonic decay $B_u \to \tau \nu_\tau$ and the ratio ${\rm BR}(B_u \to \tau \nu_\tau)/{\rm BR}(B_u \to \tau \nu_\tau)^{\rm{SM}}$ as described in Appendix~\ref{Btaunu}. These leptonic decays occur at tree level, and we consider also higher order SUSY corrections to the Yukawa coupling. These routines only need the name of a SLHA file as input.

\item \verb?double BDtaunu_calculator(char name[])?\\
\verb?double BDtaunu_BDenu_calculator(char name[])?\\
\\
These routines compute respectively the branching ratio of the semileptonic decay $B \to D^0 \tau \nu_\tau$ and the ratio $\rm{BR}(B \to D^0 \tau \nu_\tau)/\rm{BR}(B \to D^0 e \nu_e)$ as described in Appendix~\ref{BDtaunu}. These semileptonic decays occur at tree level, and we consider also higher order SUSY corrections to the Yukawa coupling.

\item \verb?double Bsmumu_calculator(char name[])?\\
\\
This function computes the branching ratio of the rare decay $B_s \to \mu^+ \mu^-$ considering full one loop corrections as described in Appendix~\ref{Bsmumu}.

\item \verb?double Kmunu_pimunu_calculator(char name[])?\\
\verb?double Rl23_calculator(char name[])?\\
\\
These functions compute respectively the ratio $\rm{BR}(K \to \mu \nu_\mu)/\rm{BR}(\pi \to \mu \nu_\mu)$ and the observable $R_{\ell23}$ as described in Appendix~\ref{Kmunu}. These leptonic decays occur at tree level, and we consider also higher order SUSY corrections to the Yukawa coupling.

\item \verb?double Dstaunu_calculator(char name[])?\\
\verb?double Dsmunu_calculator(char name[])?\\
\\
These routines compute respectively the branching ratios of the leptonic decays $D_s \to \tau \nu_\tau$ and $D_s \to \mu \nu_\mu$ as described in Appendix~\ref{Dslnu}. These leptonic decays occur at tree level, and we consider also higher order SUSY corrections to the Yukawa coupling.

\item \verb?double muon_gm2_calculator(char name[])?\\
\\
This routine computes the muon anomalous magnetic moment ($\delta a_\mu$), considering one loop contributions as well as the main two loop contributions, as described in Appendix~\ref{muon}. 
\end{itemize}
%
\newpage
\section{Compilation and installation instructions}%
\label{compilation}
The main structure of the \verb?SuperIso? package is unchanged since the first version, and the spirit of the program relies on the idea of simplicity of use.\\
\\
The \verb?SuperIso? package\footnote{An alternative package including the calculation of the relic density, {\tt SuperIso Relic} \cite{superiso_relic}, is also available at: {\tt http://superiso.in2p3.fr/relic} .} can be downloaded from:\\
\\
{\tt http://superiso.in2p3.fr}\\
\\
The following main directory is created after unpacking:\\
\\
\verb?superiso_vX.X?\\
\\
It contains the \verb?src/? directory, in which all the source files can be found. The main directory contains also a \verb?Makefile?, a \verb?README?, ten sample main programs (\verb?sm.c?, \verb?thdm.c?, \verb?msugra.c?, \verb?amsb.c?, \verb?gmsb.c?, \verb?nuhm.c?, \verb?cnmssm.c?, \verb?ngmsb.c?, \verb?nnuhm.c? and \verb?slha.c?) and one example of input file in the SUSY Les Houches Accord format (\verb?example.lha?). The compilation options should be defined in the \verb?Makefile?, as well as the path to the ISAJET \verb?isasugra.x? and SOFTSUSY \verb?softpoint.x? executable files, the path to NMSSMTools main directory, and the path to 2HDMC directory, when needed.\\
\verb?SuperIso? is written for a C compiler respecting the C99 standard. In particular, it has been tested successfully with the GNU C Compiler and the Intel C Compiler on Linux and Mac 32-bits or 64-bits machines, and with the latest versions of \verb?SOFTSUSY?, \verb?ISAJET?, \verb?NMSSMTools? and \verb?2HDMC?. Additional information can be found in the \verb?README? file.\\
\\
To compile the library, type\\
\\
\verb?make?\\
\\
\noindent This creates the \verb?libisospin.a? in \verb?src/?. Then, to compile one of the ten programs provided in the main directory, type\\
\\
\verb?make name? \qquad or \qquad \verb?make name.c?\\
\\
\noindent where \verb?name? can be \verb?sm?, \verb?thdm?, \verb?msugra?, \verb?amsb?, \verb?gmsb?, \verb?nuhm?, \verb?cnmssm?, \verb?ngmsb?, \verb?nnuhm? or \verb?slha?. This generates an executable program with the \verb?.x? extension. Note that \verb?sm? and \verb?slha? do not need any additional program, but \verb?msugra?, \verb?amsb?, \verb?gmsb? and \verb?nuhm? need either \verb?ISAJET? or \verb?SOFTSUSY?, \verb?cnmssm?, \verb?ngmsb? and \verb?nnuhm? require \verb?NMSSMTools? and \verb?2HDMC? is necessary for \verb?thdm?.\\
\\
\verb?sm.x? calculates the different observables described in the Appendices in the Standard Model, using the parameters of Table~\ref{tabparam}.\\
\\
\verb?slha.x? calculates the different observables using the parameters contained in the SLHA file whose name has to be passed as input parameter.\\
\\
\verb?thdm.x? calculates the observables in 2HDM (general model or types I--IV), starting first by calculating the mass spectrum and couplings thanks to \verb?2HDMC?.\\
\\
\verb?amsb.x?, \verb?gmsb.x?, \verb?msugra.x? and \verb?nuhm.x? compute the observables, first by calculating the mass spectrum and couplings thanks to \verb?ISAJET? (for \verb?msugra.x? only) and/or \verb?SOFTSUSY? within respectively the AMSB, GMSB, mSUGRA or NUHM parameter spaces. \\
\\
\verb?cnmssm.x?, \verb?ngmsb.x? or \verb?nnuhm.x? compute the observables, after calculating the mass spectrum and couplings thanks to \verb?NMSSMTools? within respectively the CNMSSM, NGMSB or NNUHM parameter spaces.\\
\\
For all these programs, arguments referring to the usual input parameters have to be passed to the program. If not, a message will describe which parameters have to be specified.

\section{Input and output description}%
\label{sample}
The inputs of \verb?slha?, \verb?msugra?, \verb?amsb?, \verb?gmsb? and \verb?nuhm? are the same as in the previous version, but the outputs are more numerous. 
\subsection{SLHA input file}
The program \verb?slha.x? reads the needed parameters in the given input SLHA file and calculates the observables. For example, the command
\begin{verbatim}
./slha.x example.lha
\end{verbatim}
returns
\begin{verbatim}
delta0=8.259e-02
BR_bsgamma=2.974e-04
BR_Btaunu=1.097e-04
Rtaunu=9.966e-01
BR_Kmunu/BR_pimunu=6.454e-01
Rl23=1.000e+00
BR_BDtaunu=6.966e-03
BR_BDtaunu/BR_BDenu=2.973e-01
BR_Bsmumu=3.470e-09
BR_Dstaunu=4.813e-02
BR_Dsmunu=4.969e-03
a_muon=2.045e-10
excluded_mass=0
\end{verbatim}
where \verb?delta0? refers to the isospin symmetry breaking in $B \to K^* \gamma$ decays, \verb?BR_bsgamma? the branching ratio of $B \to X_s \gamma$, \verb?BR_Btaunu? the branching ratio of $B_u \to \tau \nu_\tau$, \verb?Rtaunu? the normalized ratio to the SM value,  \verb?BR_Kmunu/BR_pimunu? the ratio ${\rm BR}(K \to \mu \nu_\mu)/{\rm BR}(\pi \to \mu \nu_\mu)$, \verb?Rl23? the ratio $R_{\ell 23}$, \verb?BR_BDtaunu? the branching ratio of $B \to D^0 \tau \nu_\tau$, \verb?BR_BDtaunu/BR_BDenu? the ratio ${\rm BR}(B \to D^0 \tau \nu_\tau)/{\rm BR}(B \to D^0 e \nu_e)$, \verb?BR_Bsmumu? the branching ratio of $B_s \to \mu^+ \mu^-$, \verb?BR_Dstaunu? and \verb?BR_Dsmunu? the branching ratios of $D_s \to \tau \nu_\tau$ and $D_s \to \mu \nu_\mu$ respectively, \verb?a_muon? the deviation in the anomalous magnetic moment of the muon, and \verb?excluded_mass? indicates that the point is already excluded by the direct searches if it is equal to 1. More details on the definitions and calculations of these observables are given in the appendices. If the SLHA file provided to \verb?slha.x? is inconsistent, a message will be displayed:
\begin{itemize}
\item \verb?Invalid point? means that the SLHA generator had not succeeded in generating the mass spectrum ({\it e.g.} due to the presence of tachyonic particles).\vspace*{-0.2cm}
\item \verb?Model not yet implemented? means that the SLHA file is intended for a model not implemented in SuperIso, such as $R$-parity violating models.\vspace*{-0.2cm}
\item \verb?Invalid SLHA file? means that the SLHA file is invalid and misses important parameters.
\end{itemize}
\subsection{SM main program}
The program \verb?sm.x? is a standalone program which computes the different observables in the Standard Model. No argument is necessary for this program. The command
\begin{verbatim}
./sm.x
\end{verbatim}
returns
\begin{verbatim}
delta0=8.203e-02
BR_bsgamma=3.061e-04
BR_Btaunu=1.102e-04
Rtaunu=1.000e+00
BR_Kmunu/BR_pimunu=6.454e-01
Rl23=1.000e+00
BR_BDtaunu=6.980e-03
BR_BDtaunu/BR_BDenu=2.975e-01
BR_Bsmumu=3.395e-09
BR_Dstaunu=4.822e-02
BR_Dsmunu=4.975e-03
\end{verbatim}

\subsection{mSUGRA inputs}
The program \verb?msugra.x? computes the observables in the mSUGRA parameter space, using \verb?ISAJET? and/or \verb?SOFTSUSY? to generate the mass spectra. If only one of these generators is available the corresponding \verb?#define? in \verb?msugra.c? has to be commented. The necessary arguments to this program are:
\begin{itemize}
 \item $m_0$: universal scalar mass at GUT scale,\vspace*{-0.2cm}
 \item $m_{1/2}$: universal gaugino mass at GUT scale,\vspace*{-0.2cm}
 \item $A_0$: trilinear soft breaking parameter at GUT scale,\vspace*{-0.2cm}
 \item $\tan\beta$: ratio of the two Higgs vacuum expectation values.\\
\end{itemize}
Optional arguments can also be given:
\begin{itemize}
 \item $sign(\mu)$: sign of Higgsino mass term, positive by default,\vspace*{-0.2cm}
 \item $m_t^{pole}$: top quark pole mass, by default 172.4 GeV,\vspace*{-0.2cm}
 \item $\overline{m}_b(\overline{m}_b)$: scale independent b-quark mass, by default 4.2 GeV (option only available for \verb?SOFTSUSY?),\vspace*{-0.2cm}
 \item $\alpha_s(M_Z)$: strong coupling constant at the $Z$-boson mass, by default 0.1176 (option only available for \verb?SOFTSUSY?).
\end{itemize}
If the arguments are not specified, a message will describe the needed parameters in the correct order.\\
\\
With \verb?SOFTSUSY? 3.0.11 and \verb?ISAJET? 7.79, running the program with:
\begin{verbatim}
./msugra.x 500 500 0 50
\end{verbatim}
returns
\begin{verbatim}
SLHA file generated by SOFTSUSY                                               
delta0=9.967e-02                                                              
BR_bsgamma=2.223e-04                                                          
BR_Btaunu=6.507e-05                                                           
Rtaunu=5.905e-01                                                              
BR_Kmunu/BR_pimunu=6.429e-01                                                  
Rl23=9.980e-01                                                                
BR_BDtaunu=6.289e-03                                                          
BR_BDtaunu/BR_BDenu=2.681e-01
BR_Bsmumu=4.007e-08
BR_Dstaunu=4.794e-02
BR_Dsmunu=4.951e-03
a_muon=2.049e-09
charged_LSP=0
excluded_mass=0

SLHA file generated by ISAJET
delta0=1.009e-01
BR_bsgamma=2.169e-04
BR_Btaunu=7.245e-05
Rtaunu=6.582e-01
BR_Kmunu/BR_pimunu=6.434e-01
Rl23=9.984e-01
BR_BDtaunu=6.404e-03
BR_BDtaunu/BR_BDenu=2.733e-01
BR_Bsmumu=3.335e-08
BR_Dstaunu=4.793e-02
BR_Dsmunu=4.950e-03
a_muon=2.084e-09
charged_LSP=0
excluded_mass=0
\end{verbatim}
corresponding to the observables described in the previous section.

\subsection{AMSB inputs}
The program \verb?amsb.x? computes the observables using the corresponding parameters generated by \verb?SOFTSUSY? in the AMSB scenario. The necessary arguments to this program are:
\begin{itemize}
 \item $m_0$: universal scalar mass at GUT scale,\vspace*{-0.2cm}
 \item $m_{3/2}$: gravitino mass at GUT scale,\vspace*{-0.2cm}
 \item $\tan\beta$: ratio of the two Higgs vacuum expectation values.
\end{itemize}
Optional arguments are the same as for mSUGRA. If the input parameters are absent, a message will ask for them.\\
\\
With \verb?SOFTSUSY? 3.0.11, running the program with:
\begin{verbatim}
./amsb.x 500 5000 5 -1
\end{verbatim}
returns
\begin{verbatim}
delta0=7.671e-02
BR_bsgamma=3.295e-04
BR_Btaunu=1.097e-04
Rtaunu=9.953e-01
BR_Kmunu/BR_pimunu=6.454e-01
Rl23=1.000e+00
BR_BDtaunu=6.971e-03
BR_BDtaunu/BR_BDenu=2.972e-01
BR_Bsmumu=3.349e-09
BR_Dstaunu=4.818e-02
BR_Dsmunu=4.975e-03
a_muon=-6.598e-10
excluded_mass=1
\end{verbatim}

\subsection{GMSB inputs}
The program \verb?gmsb.x? computes the observables using the GMSB parameters generated by \verb?SOFTSUSY?. The necessary arguments to this program are:
\begin{itemize}
 \item $\Lambda$: scale of the SUSY breaking in GeV (usually 10000-100000 GeV),\vspace*{-0.2cm}
 \item $M_{mess}$: messenger mass scale ($> \Lambda$),\vspace*{-0.2cm}
 \item $N_5$: equivalent number of $5+\bar{5}$ messenger fields,\vspace*{-0.2cm}
 \item $\tan\beta$: ratio of the two Higgs vacuum expectation values.
\end{itemize}
Optional arguments are the same as for mSUGRA, with an additional one:
\begin{itemize}
 \item $c_{Grav}$ ($\ge 1$): ratio of the gravitino mass to its value for a breaking scale $\Lambda$, 1 by default.
\end{itemize}
Again, in the case of lack of arguments, a message will be displayed.\\
\\
With \verb?SOFTSUSY? 3.0.11, running the program with:
\begin{verbatim}
./gmsb.x 2e4 5e6 1 10
\end{verbatim}
returns
\begin{verbatim}
delta0=6.339e-02
BR_bsgamma=4.260e-04
BR_Btaunu=8.077e-05
Rtaunu=7.330e-01
BR_Kmunu/BR_pimunu=6.439e-01
Rl23=9.988e-01
BR_BDtaunu=6.542e-03
BR_BDtaunu/BR_BDenu=2.789e-01
BR_Bsmumu=3.805e-09
BR_Dstaunu=4.806e-02
BR_Dsmunu=4.962e-03
a_muon=3.774e-08
excluded_mass=1
\end{verbatim}

\subsection{NUHM inputs}
The program \verb?nuhm.x? computes the observables using the NUHM parameters generated by \verb?SOFTSUSY?. The necessary arguments to this program are the same as for mSUGRA, with two additional ones, the values of $\mu$ and $m_A$:
\begin{itemize}
 \item $m_0$: universal scalar mass at GUT scale,\vspace*{-0.2cm}
 \item $m_{1/2}$: universal gaugino mass at GUT scale,\vspace*{-0.2cm}
 \item $A_0$: trilinear soft breaking parameter at GUT scale,\vspace*{-0.2cm}
 \item $\tan\beta$: ratio of the two Higgs vacuum expectation values,\vspace*{-0.2cm}
 \item $\mu$: $\mu$ parameter,\vspace*{-0.2cm}
 \item $m_A$: CP-odd Higgs mass.
\end{itemize}
Optional arguments can also be given:
\begin{itemize}
 \item $m_t^{pole}$: top quark pole mass, by default 172.4 GeV,\vspace*{-0.2cm}
 \item $\overline{m}_b(\overline{m}_b)$: scale independent b-quark mass, by default 4.2 GeV,\vspace*{-0.2cm}
 \item $\alpha_s(M_Z)$: strong coupling constant at the $Z$-boson mass, by default 0.1176.
\end{itemize}
In the absence of arguments, a message will be shown.\\
\\
With \verb?SOFTSUSY? 3.0.11, running the program with:
\begin{verbatim}
./nuhm.x 500 500 0 50 500 500
\end{verbatim}
returns
\begin{verbatim}
delta0=1.066e-01
BR_bsgamma=2.014e-04
BR_Btaunu=6.576e-05
Rtaunu=5.967e-01
BR_Kmunu/BR_pimunu=6.429e-01
Rl23=9.981e-01
BR_BDtaunu=6.300e-03
BR_BDtaunu/BR_BDenu=2.686e-01
BR_Bsmumu=3.650e-08
BR_Dstaunu=4.795e-02
BR_Dsmunu=4.951e-03
a_muon=2.244e-09
excluded_mass=0
\end{verbatim}

\subsection{CNMSSM inputs}
The program \verb?cnmssm.x? computes the observables using the CNMSSM parameters generated by \verb?NMSSMTools?\footnote{As the soft singlet mass $m^2_S$ and the singlet self coupling
$\kappa$ are both determined in terms of the other parameters through the minimization equations of the Higgs potential in {\tt NMSSMTools}, what we call CNMSSM here is a partially constrained NMSSM.}.
The necessary arguments to this program are:
\begin{itemize}
 \item $m_0$: universal scalar mass at GUT scale,\vspace*{-0.2cm}
 \item $m_{1/2}$: universal gaugino mass at GUT scale,\vspace*{-0.2cm}
 \item $A_0$: trilinear soft breaking parameter at GUT scale,\vspace*{-0.2cm}
 \item $\tan\beta$: ratio of the two Higgs vacuum expectation values,\vspace*{-0.2cm}
 \item $\lambda$: cubic Higgs coupling.
\end{itemize}
Optional arguments can also be given:
\begin{itemize}
 \item $sign(\mu)$: sign of Higgsino mass term, positive by default,\vspace*{-0.2cm}
 \item $A_\kappa$: trilinear soft breaking parameter at GUT scale, by default $A_\kappa=A_0$,\vspace*{-0.2cm}
 \item $m_t^{pole}$: top quark pole mass, by default 172.4 GeV,\vspace*{-0.2cm}
 \item $\overline{m}_b(\overline{m}_b)$: scale independent b-quark mass, by default 4.2 GeV,\vspace*{-0.2cm}
 \item $\alpha_s(M_Z)$: strong coupling constant at the $Z$-boson mass, by default 0.1176.
\end{itemize}
In the absence of arguments, a message will be shown.\\
\\
With \verb?NMSSMTools? 2.2.0, running the program with:
\begin{verbatim}
./cnmssm.x 500 500 0 50 0.01
\end{verbatim}
returns
\begin{verbatim}
SLHA file generated by NMSSMtools
delta0=9.945e-02
BR_bsgamma=2.232e-04
BR_Btaunu=6.478e-05
Rtaunu=5.878e-01
BR_Kmunu/BR_pimunu=6.429e-01
Rl23=9.980e-01
BR_BDtaunu=6.284e-03
BR_BDtaunu/BR_BDenu=2.679e-01
BR_Bsmumu=3.909e-08
BR_Dstaunu=4.794e-02
BR_Dsmunu=4.951e-03
a_muon=2.120e-09
charged_LSP=0
collider_excluded=0
theory_excluded=0
\end{verbatim}

\subsection{NGMSB inputs}
The program \verb?ngmsb.x? computes the observables using the NGMSB parameters generated by \verb?NMSSMTools?. The necessary arguments to this program are:
\begin{itemize}
 \item $\Lambda$: scale of the SUSY breaking in GeV (usually 10000-100000 GeV),\vspace*{-0.2cm}
 \item $M_{mess}$: messenger mass scale ($> \Lambda$),\vspace*{-0.2cm}
 \item $N_5$: equivalent number of $5+\bar{5}$ messenger fields,\vspace*{-0.2cm}
 \item $\tan\beta$: ratio of the two Higgs vacuum expectation values,\vspace*{-0.2cm}
 \item $\lambda$: cubic Higgs coupling.
\end{itemize}
Optional arguments are the same as for CNMSSM, with an additional one:
\begin{itemize}
 \item $\Delta_H$: 0 by default.
\end{itemize}
\verb?NMSSMTools? allows also for other optional parameters such as $\mu'$, $B'$, $\xi_S$ and $\xi_F$.\\
\\
With \verb?NMSSMTools? 2.2.0, running the program with:
\begin{verbatim}
./ngmsb.x 1e5 2e10 5 20 0.1 1 -1000
\end{verbatim}
returns
\begin{verbatim}
delta0=8.104e-02
BR_bsgamma=3.089e-04
BR_Btaunu=1.094e-04
Rtaunu=9.928e-01
BR_Kmunu/BR_pimunu=6.454e-01
Rl23=1.000e+00
BR_BDtaunu=6.968e-03
BR_BDtaunu/BR_BDenu=2.970e-01
BR_Bsmumu=1.096e-09
BR_Dstaunu=4.818e-02
BR_Dsmunu=4.975e-03
a_muon=8.081e-10
collider_excluded=1
theory_excluded=1
\end{verbatim}

\subsection{NNUHM inputs}
The program \verb?nnuhm.x? computes the observables using the NNUHM parameters generated by \verb?NMSSMTools?. The necessary arguments to this program are the same as for CNMSSM, with two additional ones:
\begin{itemize}
 \item $m_0$: universal scalar mass at GUT scale,\vspace*{-0.2cm}
 \item $m_{1/2}$: universal gaugino mass at GUT scale,\vspace*{-0.2cm}
 \item $A_0$: trilinear soft breaking parameter at GUT scale,\vspace*{-0.2cm}
 \item $\tan\beta$: ratio of the two Higgs vacuum expectation values,\vspace*{-0.2cm}
 \item $\lambda$: cubic Higgs coupling,\vspace*{-0.2cm}
 \item $M_{H_D}$: down Higgs mass parameter at GUT scale,\vspace*{-0.2cm}
 \item $M_{H_U}$: up Higgs mass parameter at GUT scale.
\end{itemize}
Optional arguments can also be given:
\begin{itemize}
 \item $sign(\mu)$: sign of Higgsino mass term, positive by default,\vspace*{-0.2cm}
 \item $A_\kappa$: trilinear soft breaking parameter at GUT scale, by default $A_\kappa=A_0$,\vspace*{-0.2cm}
 \item $m_t^{pole}$: top quark pole mass, by default 172.4 GeV,\vspace*{-0.2cm}
 \item $\overline{m}_b(\overline{m}_b)$: scale independent b-quark mass, by default 4.2 GeV,\vspace*{-0.2cm}
 \item $\alpha_s(M_Z)$: strong coupling constant at the $Z$-boson mass, by default 0.1176.
\end{itemize}
In the absence of arguments, a message will be shown.\\
\\
With \verb?NMSSMTools? 2.2.0, running the program with:
\begin{verbatim}
./nnuhm.x 500 500 0 50 0.1 500 500
\end{verbatim}
returns
\begin{verbatim}
SLHA file generated by NMSSMtools
delta0=9.954e-02
BR_bsgamma=2.232e-04
BR_Btaunu=6.487e-05
Rtaunu=5.887e-01
BR_Kmunu/BR_pimunu=6.429e-01
Rl23=9.980e-01
BR_BDtaunu=6.286e-03
BR_BDtaunu/BR_BDenu=2.679e-01
BR_Bsmumu=2.108e-07
BR_Dstaunu=4.794e-02
BR_Dsmunu=4.951e-03
a_muon=2.121e-09
charged_LSP=0
collider_excluded=1
theory_excluded=0
\end{verbatim}
\subsection{2HDM inputs}
The program \verb?thdm.x? computes the observables using the 2HDM parameters generated by \verb?2HDMC?. The necessary arguments to this program are:
\begin{itemize}
 \item $type$: Yukawa type (1-4),\vspace*{-0.2cm}
 \item $\tan\beta$: ratio of the two Higgs vacuum expectation values,\vspace*{-0.2cm}
 \item $m_A$: CP-odd Higgs mass.
\end{itemize}
Optional arguments can also be given:
\begin{itemize}
 \item $\lambda_1,\cdots,\lambda_7$: Higgs potential parameters,\vspace*{-0.2cm}
 \item $m_{12}^2$: Higgs potential parameter alternative to $m_A$.
\end{itemize}
By specifying the Higgs potential parameters, it is possible to do the calculations in general 2HDM. If not specified, the optional arguments are set to the default tree level MSSM-like values:
\begin{eqnarray*}
&&\lambda_1=\lambda_2=\frac{g^2+g'^2}{4} \;,\qquad \lambda_3=\frac{g^2-g'^2}{4} \;,\qquad \lambda_4=-\frac{g^2}{2} \;, \qquad \lambda_5=\lambda_6=\lambda_7=0 \;,\\
&& m_{12}^2=m_A^2\cos\beta\sin\beta \;.
\end{eqnarray*}
In the absence of the necessary arguments, a message will be shown.\\
\\
With \verb?2HDMC? 1.0.5, running the program with:
\begin{verbatim}
./thdm.x 4 10 300
\end{verbatim}
returns
\begin{verbatim}
delta0=7.840e-02
BR_bsgamma=2.469e-04
BR_Btaunu=1.103e-04
Rtaunu=1.001e+00
BR_Kmunu/BR_pimunu=6.347e-01
Rl23=1.000e+00
BR_BDtaunu=6.979e-03
BR_BDtaunu/BR_BDenu=2.975e-01
BR_Bsmumu=3.273e-09
BR_Dstaunu=4.818e-02
BR_Dsmunu=4.943e-03
a_muon=-1.514e-12
\end{verbatim}

\noindent Using the aforementioned main programs as examples, the user is encouraged to write his/her own programs in order to, for example, perform scans in a given supersymmetric scenario.
\section{Results}
\label{result}
\begin{figure}[p]
\begin{center}
\includegraphics[width=9.9cm]{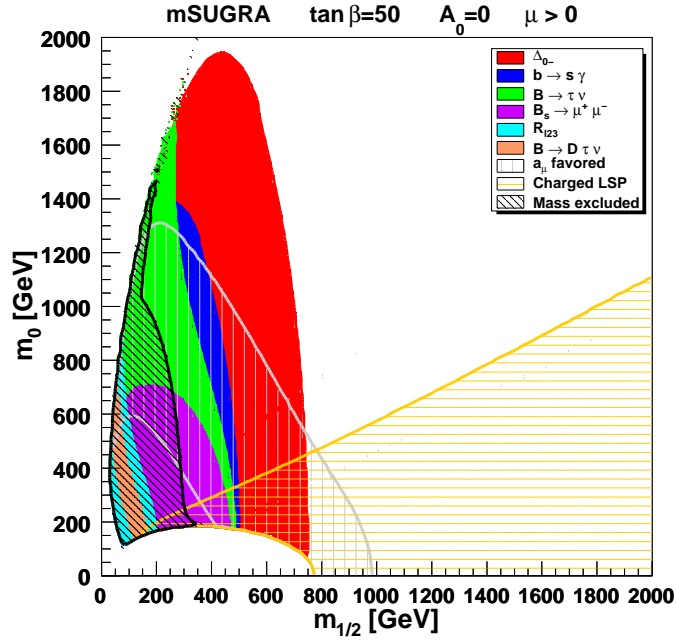}
\caption{Constraints in mSUGRA $(m_{1/2} - m_0)$ parameter plane. For the description of the various colored zones see the text. The contours are superimposed in the order given in the legend.}
\label{fig1}
\end{center}
\end{figure}

\begin{figure}
\begin{center}
\includegraphics[width=9.9cm]{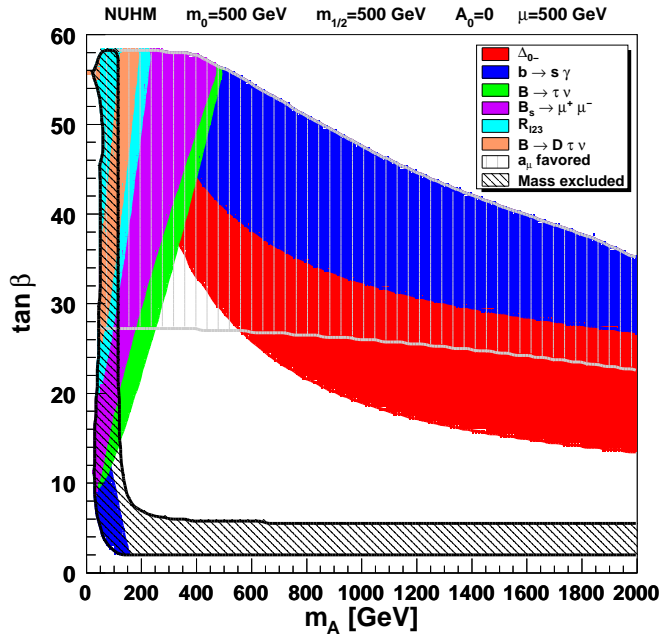} 
\caption{Constraints in NUHM $(m_A - \tan\beta)$ parameter plane. For the description of the various colored zones see the text. The contours are superimposed in the order given in the legend.}
\label{fig2}
\end{center}
\end{figure}

We illustrate in this section the constraints on the SUSY parameter space that can be obtained using observables calculated with \verb?SuperIso v3.0?. For more extended discussions about the constraints obtained using \verb?SuperIso?, see for example \cite{mahmoudi,allanach,mahmoudi2,eriksson,abdussalam,mahmoudi3,horton}. In Figures~\ref{fig1} and \ref{fig2}, two examples of the obtained constraints in the mSUGRA and NUHM scenarios are displayed. The different areas in the figures correspond to the following observables:
\begin{itemize}
 \item red region: excluded by the isospin asymmetry, \vspace*{-0.3cm}
 \item blue region: excluded by the inclusive branching ratio of $b \to s \gamma$, \vspace*{-0.3cm}
 \item black hatched region: excluded by the collider mass limits, \vspace*{-0.3cm}
 \item violet region: excluded by the branching ratio of $B_s \to \mu^+ \mu^-$, \vspace*{-0.3cm}
 \item gray hatched region: \textbf{favored} by the anomalous magnetic moment of the muon, \vspace*{-0.3cm}
 \item yellow hatched region: charged LSP, disfavored by cosmology, \vspace*{-0.3cm}
 \item green region: excluded by the branching ratio of $B_u \to \tau \nu_\tau$, \vspace*{-0.3cm}
 \item orange region: excluded by the branching ratio of $B \to D^0 \tau \nu_\tau$, \vspace*{-0.3cm}
 \item cyan region: excluded by the branching ratio of $K \to \mu \nu_\mu$.
\end{itemize}
The allowed interval for each observable is given in Appendix~\ref{constraints}.\\
\\
In Figure~\ref{fig1}, the exclusion regions in the mSUGRA parameter plane $(m_{1/2} - m_0)$ for $\tan\beta=50$, $A_0=0$ and $\mu>0$ are displayed. One can notice that small values of $m_{0}$ and $m_{1/2}$ are disfavored by the observables. The unfilled region in the bottom left corner corresponds to points with tachyonic particles.\\
In Figure~\ref{fig2}, the exclusion zones are displayed in the NUHM parameter plane $(m_A - \tan\beta)$ for $m_{0}=500$ GeV, $m_{1/2}=500$ GeV, $A_0=0$ and $\mu=500$ GeV. Most observables tend to disfavor the high $\tan\beta$ region in this plane. The white top right triangle corresponds to a region where tachyonic particles are encountered.
%
\section{Conclusion}
%
%
\verb?SuperIso v3.0? features many new additions and improvements as compared to the first version of the program. It is now able to compute numerous flavor physics observables -- as well as the muon anomalous magnetic moment -- which have already proven to be very useful in the exploration of the MSSM and NMSSM parameter spaces. Investigating the indirect constraints has many interesting phenomenological impacts, and can provide us with important information before the LHC data. They can also be used as guidelines for the LHC direct searches, and will be very valuable for the consistency checks.\\
\\
In spite of the numerous changes, the spirit of the program is still based on the simplicity of use. The code will continue to incorporate other $B$ physics observables. Also, the extension of the program to beyond minimal flavor violation is under development.
\newpage
\appendix
%
\append{QCD coupling}%
\label{alphas}
The $\alpha_s$ evolution is expressed as \cite{PDG}:
\begin{eqnarray}
\alpha_s(\mu)&=&\frac{4\pi}{\beta_0\ln(\mu^2/\Lambda_{n_f}^2)}\left\{1-\frac{\beta_1}{\beta_0^2}\frac{\ln\bigl(\ln(\mu^2/\Lambda_{n_f}^2)\bigr)}{\ln(\mu^2/\Lambda_{n_f}^2)}+\frac{ \beta_1^2}{\beta_0^4\ln^2(\mu^2/\Lambda_{n_f}^2)}\right.\nonumber \\
&& \qquad \times \left.\left[\left(\ln\bigl(\ln(\mu^2/\Lambda_{n_f}^2)\bigr)-\frac{1}{2}\right)^2+\frac{\beta_2\beta_0}{2 \beta_1^2}-\frac{5}{4}\right]\right\} \;,
\end{eqnarray}
with
\begin{equation}
\beta_0=11-\frac{2}{3}n_f~,~~~~\beta_1=102-\frac{38}{3}n_f~,~~~~\beta_2=2857-\frac{5033}{9}n_f+\frac{325}{27}n_f^2\;. \label{betas}
\end{equation}
$n_f$ denotes the number of active flavors (\textit{i.e.} 4 for energies between the charm and the bottom masses, 5 between the bottom and the top masses, and 6 beyond the top mass). We compute the associated $\Lambda_{n_f}$ by requiring continuity of $\alpha_s$. In particular in \verb?SuperIso?, $\Lambda_{5}$ is calculated so that $\alpha_s(M_Z)$ matches the value given in the input SLHA file, and then $\Lambda_{4}$ and $\Lambda_{6}$ are calculated if needed by imposing the continuity at the bottom and top mass scales respectively, which are also input parameters.
%
\append{Evolution of quark masses}%
\label{quarkmasses}
We use the following three loop formula to compute the pole mass of quarks \cite{PDG}:
\begin{eqnarray}
m_q^{\rm{pole}}&=&\overline{m}_q(\overline{m}_q)\Bigg\{1+\frac{4 \alpha_s(\overline{m}_q)}{3\pi}\nonumber\\
&&+\left[-1.0414 \sum_{k=1}^{n_{f_l}} \left(1-\frac{4}{3}\frac{\overline{m}_{q_k}}{\overline{m}_q}\right)+13.4434\right]\left[\frac{ \alpha_s(\overline{m}_q)}{\pi} \right]^2\\
&& +\left[0.6527 n_{f_l}^2 - 26.655 n_{f_l} +190.595\right] \left[\frac{ \alpha_s(\overline{m}_q)}{\pi} \right]^3 \Bigg\} \;, \nonumber
\end{eqnarray}
where $n_{f_l}$ is the number of flavors $q_k$ lighter than $q$.\\
\\
For the $\overline{MS}$ top mass, we use \cite{PDG}
\begin{equation}
\overline{m}_t(\overline{m}_t) = m_t^{\rm{pole}} \left(1-\frac43 \frac{\alpha_s(m_t^{\rm{pole}})}{\pi}\right) \;.
\end{equation}
\\
The running mass of the quarks is given by \cite{running}
\begin{equation}
\overline{m}_q(\mu_1) = \frac{R\bigl(\alpha_s(\mu_1)\bigr)}{R\bigl(\alpha_s(\mu_2)\bigr)} \overline{m}_q(\mu_2) \;,
\end{equation}
where
\begin{eqnarray}
R(\alpha_s) &=& \left(\frac{\beta_0}{2}\frac{\alpha_s}{\pi} \right)^
{2\gamma_0/\beta_0} \Bigg\{1 + \left(2\frac{\gamma_1}{\beta_0} - 
\frac{\beta_1\gamma_0}{\beta_0^2} \right)\frac{\alpha_s}{\pi} \\
&& + \frac{1}{2}\left[\left(2\frac{\gamma_1}{\beta_0} - \frac{\beta_1 \gamma_0}
{\beta_0^2} \right)^2 + 2\frac{\gamma_2}{\beta_0} - 
\frac{\beta_1\gamma_1}{\beta_0^2} - \frac{\beta_2\gamma_0}{16\beta_0^2} + 
\frac{\beta_1^2\gamma_0}{2\beta_0^3} \right] 
\left(\frac{\alpha_s}{\pi} \right)^2 + O(\alpha_s^3)\Bigg\}\;,\nonumber
\end{eqnarray}
with
\begin{eqnarray}
\gamma_0 &=& 2\;,\\
\gamma_1 &=& \frac{101}{12} - \frac{5}{18} n_f\;,\\
\gamma_2 &=& \frac{1}{32}\left[1249 - \left(\frac{2216}{27} + \frac{160}{3}\zeta(3) 
\right)n_f - \frac{140}{81}n_f^2\right]\;,
\end{eqnarray}
with $n_f$ the number of active flavors, and $\beta$'s given in Eq. (\ref{betas}).\\
\\
The $1S$ bottom quark mass is given by \cite{hoang}:
\begin{equation}
m_b^{1S} = m_b^{\rm pole}\,\bigg[1-\Delta^{\mbox{\tiny LO}}-\Delta^{\mbox{\tiny NLO}}-\Delta^{\mbox{\tiny NNLO}}\bigg]\;,
\end{equation}
where
\begin{eqnarray}
\Delta^{\mbox{\tiny LO}} & = & \frac{C_F^2\,\alpha_s^2(\mu_b)}{8}\;,\\[2mm]  
\Delta^{\mbox{\tiny NLO}} & = & \frac{C_F^2\,\alpha_s^2(\mu_b)}{8}\, \Big(\frac{\alpha_s(\mu_b)}{\pi}\Big)\,\bigg[\,\beta_0^\prime\,(L + 1) + \frac{a_1}{2} \,\bigg]\;,\\[2mm] 
\Delta^{\mbox{\tiny NNLO}} & = & \frac{C_F^2\,\alpha_s^2(\mu_b)}{8}\, \Big(\frac{\alpha_s(\mu_b)}{\pi}\Big)^2\,
\bigg[\,\beta_0^{\prime 2}\,\bigg(\, \frac{3}{4} L^2 +  L + \frac{\zeta(3)}{2} + \frac{\pi^2}{24} +\frac{1}{4} \,\bigg) + \beta_0^\prime\,\frac{a_1}{2}\,\bigg(\, \frac{3}{2}\,L + 1\,\bigg)\nonumber\\
& & + \frac{\beta_1^\prime}{4}\,(L + 1) +\frac{a_1^2}{16} + \frac{a_2}{8} + \bigg(\, C_A - \frac{C_F}{48} \,\bigg)\, C_F \pi^2 \,\bigg]\;,
\end{eqnarray}
with
\begin{equation}
L \equiv \ln\left(\frac{\mu_b}{C_F\,\alpha_s(\mu_b)\,m_b^{\rm pole}}\right)\;,
\end{equation}\\
\begin{equation}
\zeta(3) \approx 1.2020569\;, \label{zeta3}
\end{equation}
and $\mu_b = O(m_b)$. \\
\\
Also,
\begin{eqnarray}
\beta_0^\prime & = & \frac{11}{3}\,C_A - \frac{4}{3}\,T\,n_{f_l}\nonumber\;,\\[2mm]
\beta_1^\prime & = & \frac{34}{3}\,C_A^2 -\frac{20}{3}C_A\,T\,n_{f_l}- 4\,C_F\,T\,n_{f_l}\nonumber\;,\\[2mm]
a_1 & = &  \frac{31}{9}\,C_A - \frac{20}{9}\,T\,n_{f_l}\;,\\[2mm]
a_2 & = & \bigg(\,\frac{4343}{162}+4\,\pi^2-\frac{\pi^4}{4}
 +\frac{22}{3}\,\zeta(3)\,\bigg)\,C_A^2 
-\bigg(\,\frac{1798}{81}+\frac{56}{3}\,\zeta(3)\,\bigg)\,C_A\,T\,n_{f_l}
\nonumber\\[2mm]
&& -\bigg(\,\frac{55}{3}-16\,\zeta(3)\,\bigg)\,C_F\,T\,n_{f_l} 
+\bigg(\,\frac{20}{9}\,T\,n_{f_l}\,\bigg)^2\;.\nonumber
\end{eqnarray}
In the above equations $C_A=3$, $T=1/2$, $n_{f_l}=4$ and $C_F=4/3$.
%
\append{Wilson coefficients at matching scale}%
\label{wilsonMW}
The effective Hamiltonian describing the $b\to s\gamma$ transitions has the following generic structure:
\begin{equation}
{\cal H}_{eff}=\frac{G_F}{\sqrt{2}}\sum_{p=u,c} V^*_{ps}V_{pb} \sum_{i=1}^8 C_i(\mu) \, O_i\;,
\end{equation}
where $G_F$ is the Fermi coupling constant, $V_{ij}$ are elements of the CKM matrix, $O_i(\mu)$ are the relevant operators and $C_i(\mu)$ are the corresponding Wilson coefficients evaluated at the scale $\mu$.\\
The Wilson coefficients are given in the standard operator basis \cite{chetyrkin}:
\begin{equation}
\label{standard_basis}
\begin{array}{rl}
O_1 ~= & (\bar{s} \gamma_{\mu} T^a P_L c)(\bar{c} \gamma^{\mu} T^a P_L b)\;,\\[2mm]
O_2 ~= & (\bar{s} \gamma_{\mu} P_L c)(\bar{c} \gamma^{\mu} P_L b)\;,\\[2mm]
O_3 ~= & (\bar{s} \gamma_{\mu} P_L b) {\displaystyle\sum_q} (\bar{q} \gamma^{\mu} q)\;,\\[2mm]     
O_4 ~= & (\bar{s} \gamma_{\mu} T^a P_L b) {\displaystyle\sum_q} (\bar{q} \gamma^{\mu} T^a q)\;,\\[2mm]    
O_5 ~= & (\bar{s} \gamma_{\mu_1}\gamma_{\mu_2}\gamma_{\mu_3} P_L b) {\displaystyle\sum_q} (\bar{q} \gamma^{\mu_1}\gamma^{\mu_2}\gamma^{\mu_3} q)\;,\\[2mm]
O_6 ~= & (\bar{s} \gamma_{\mu_1}\gamma_{\mu_2}\gamma_{\mu_3} T^a P_L b) {\displaystyle\sum_q} (\bar{q} \gamma^{\mu_1}\gamma^{\mu_2}\gamma^{\mu_3} T^a q)\;,\\[2mm]
O_7 ~= & \dfrac{e}{16\pi^2} \left[ \bar{s} \sigma^{\mu \nu} (m_s P_L + m_b P_R) b \right] F_{\mu \nu}\;,\\[3mm]
O_8 ~= & \dfrac{g}{16\pi^2} \left[ \bar{s} \sigma^{\mu \nu} (m_s P_L + m_b P_R) T^a b \right] G_{\mu \nu}^a\;,
\end{array}
\end{equation}
where $P_{R,L}=(1\pm \gamma_5)/2$. In this basis, the effective Wilson coefficients are defined as \cite{buras}:
\begin{equation}
C_i^{\rm eff}(\mu) = \left\{ \begin{array}{ll}
C_i(\mu), & \mbox{ for $i = 1, ..., 6$ ,} \\[1mm] 
C_7(\mu) + {\displaystyle\sum_{j=1}^6} y_j C_j(\mu), & \mbox{ for $i = 7$ ,} \\[1mm]
C_8(\mu) + {\displaystyle\sum_{j=1}^6 z_j} C_j(\mu), & \mbox{ for $i = 8$ .}
\end{array} \right.
\end{equation}\\
where $\vec{y} = (0, 0, -\frac{1}{3}, -\frac{4}{9}, -\frac{20}{3}, -\frac{80}{9})$ and $\vec{z} = (0, 0, 1, -\frac{1}{6}, 20,
-\frac{10}{3})$.\\

\subsection{Standard Model contributions}
We express here the SM contributions to the Wilson coefficients following \cite{bobeth,misiak}.\\
\\
The LO coefficients read:
\begin{eqnarray}
\begin{array}{lclclcl}
C^{c(0)}_2(\mu_W) &=& -1\;, \\[3mm] 
C^{c(0)}_7(\mu_W) &=& \dfrac{23}{36}\;,\hspace*{3.7cm} && C^{t(0)}_7(\mu_W) &=& -\dfrac{1}{2} A^t_0(x_{tW})\;,  \\[3mm] 
C^{c(0)}_8(\mu_W) &=& \dfrac{1}{3}\;, &&  C^{t(0)}_8(\mu_W) &=& -\dfrac{1}{2} F^t_0(x_{tW})\;,
\end{array}\label{wilsonLO}
\end{eqnarray}
\\
the NLO coefficients are:\\
\begin{eqnarray}
\begin{array}{lclclcl}
C^{c(1)}_1(\mu_W) &=& -15 - 6 L\;, \hspace*{3cm}& \hspace{3mm} & &&  \\[3mm] 
C^{c(1)}_2(\mu_W) &=& 0\;, &&&&  \\[3mm] 
C^{c(1)}_3(\mu_W) &=& 0\;, && 
C^{t(1)}_3(\mu_W) &=& 0\;,  \\[3mm] 
C^{c(1)}_4(\mu_W) &=& \dfrac{7}{9} - \dfrac{2}{3} L\;, &&
C^{t(1)}_4(\mu_W) &=& E^t_0(x_{tW})\;,  \\[3mm] 
C^{c(1)}_5(\mu_W) &=& 0\;, && 
C^{t(1)}_5(\mu_W) &=& 0\;,  \\[3mm] 
C^{c(1)}_6(\mu_W) &=& 0\;, && 
C^{t(1)}_6(\mu_W) &=& 0\;,  \\[3mm] 
C^{c(1)}_7(\mu_W) &=& -\dfrac{713}{243} - \dfrac{4}{81} L\;, &&
C^{t(1)}_7(\mu_W) &=& -\dfrac{1}{2} A^t_1(x_{tW})\;,  \\[3mm] 
C^{c(1)}_8(\mu_W) &=& -\dfrac{91}{324}  + \dfrac{4}{27} L\;, &&
C^{t(1)}_8(\mu_W) &=& -\dfrac{1}{2} F^t_1(x_{tW})\;,  \\[3mm] 
\end{array}\label{wilsonNLO}
\end{eqnarray}
\\
\\
\\
and the NNLO coefficients are:\\
\begin{eqnarray}
\begin{array}{lcl}
C^{c(2)}_1(\mu_W) &=& T(x_{tW})-\dfrac{7987}{72} -\dfrac{17}{3} \pi^2 -\dfrac{475}{6} L - 17 L^2\;,\\[5mm]
C^{c(2)}_2(\mu_W) &=& -\dfrac{127}{18} -\dfrac{4}{3} \pi^2 -\dfrac{46}{3} L - 4 L^2\;,\\[5mm]
C^{c(2)}_3(\mu_W) &=& \dfrac{680}{243} + \dfrac{20}{81} \pi^2 +\dfrac{68}{81} L  + \dfrac{20}{27} L^2\;,\\[5mm]
C^{c(2)}_4(\mu_W) &=& -\dfrac{950}{243} -\dfrac{10}{81} \pi^2 -\dfrac{124}{27} L -\dfrac{10}{27} L^2\;,\\[5mm]
C^{c(2)}_5(\mu_W) &=& -\dfrac{68}{243} - \dfrac{2}{81} \pi^2 -\dfrac{14}{81} L - \dfrac{2}{27} L^2\;,\\[5mm]
C^{c(2)}_6(\mu_W) &=& -\dfrac{85}{162} -\dfrac{5}{108} \pi^2 -\dfrac{35}{108} L - \dfrac{5}{36} L^2\;,\\[8mm]
C^{t(2)}_3(\mu_W) &=& G^t_1(x_{tW})\;,\\[5mm]
C^{t(2)}_4(\mu_W) &=& E^t_1(x_{tW})\;,\\[5mm]
C^{t(2)}_5(\mu_W) &=& -\dfrac{1}{10} G^t_1(x_{tW}) + \dfrac{2}{15} E^t_0(x_{tW})\;,\\[5mm]
C^{t(2)}_6(\mu_W) &=& -\dfrac{3}{16} G^t_1(x_{tW}) + \dfrac{1}{4} E^t_0(x_{tW})\;,\\[5mm]
\end{array}\label{wilsonNNLO}
\end{eqnarray}\\
\\
where 
\begin{eqnarray}
x_{tW} &=& \left( \frac{\overline{m}_t(\mu_W)}{M_W} \right)^2\;,\\
L &=& \ln \left(\frac{\mu_W^2}{M_W^2}\right)\;,
\end{eqnarray}
and $\mu_W = O(M_W)$.
\vspace{6cm}\\
The necessary functions in Eqs. (\ref{wilsonLO}-\ref{wilsonNNLO}) are:\\
\begin{eqnarray}
A^t_0(x) &=& \dfrac{-3x^3+2x^2}{2(1-x)^4} \ln x + \dfrac{22x^3-153x^2+159x-46}{36(1-x)^3}\;,\\[7mm]
E^t_0(x) &=& \dfrac{-9x^2+16x-4}{6(1-x)^4} \ln x + \dfrac{-7x^3-21x^2+42x+4}{36(1-x)^3}\;,\\[7mm]
F^t_0(x) &=& \dfrac{3x^2}{2(1-x)^4} \ln x + \dfrac{5x^3-9x^2+30x-8}{12(1-x)^3}\;,\\[7mm]
A^t_1(x) &=& \dfrac{32x^4+244x^3-160x^2+16x}{9(1-x)^4} \mbox{Li}_2\left(1-\dfrac{1}{x}\right)\\
&&+ \dfrac{-774x^4-2826x^3+1994x^2-130x+8}{81(1-x)^5} \ln x \nonumber\\
&&+ \dfrac{-94x^4-18665x^3+20682x^2-9113x+2006}{243(1-x)^4}\nonumber\\
&&+ \left[   \dfrac{-12x^4-92x^3+56x^2}{3(1-x)^5} \ln x+ \dfrac{-68x^4-202x^3-804x^2+794x-152}{27(1-x)^4} \right] \ln \dfrac{\mu_W^2}{m_t^2}\;,\nonumber\\[7mm]
E^t_1(x) &=& \dfrac{515x^4-614x^3-81x^2-190x+40}{54(1-x)^4}\mbox{Li}_2\left(1-\dfrac{1}{x}\right)\\
&&+ \dfrac{-1030x^4+435x^3+1373x^2+1950x-424}{108(1-x)^5} \ln x \nonumber\\ 
&&+ \dfrac{-29467x^4+45604x^3-30237x^2+66532x-10960}{1944(1-x)^4}\nonumber\\ 
&& + \left[ \dfrac{133x^4-2758x^3-2061x^2+11522x-1652}{324(1-x)^4} \right. \nonumber\\
&& \left. + \dfrac{-1125x^3+1685x^2+380x-76}{54(1-x)^5} \ln x\right] \ln \dfrac{\mu_W^2}{m_t^2}\;, 
\nonumber
\end{eqnarray}
as well as:\\
\begin{eqnarray}
F^t_1(x) &=& \dfrac{4x^4-40x^3-41x^2-x}{3(1-x)^4} \mbox{Li}_2\left(1-\dfrac{1}{x}\right) \\
&& +\dfrac{-144x^4+3177x^3+3661x^2+250x-32}{108(1-x)^5} \ln x \nonumber\\ 
&& +\dfrac{-247x^4+11890x^3+31779x^2-2966x+1016}{648(1-x)^4} \nonumber\\ 
&& + \left[ \dfrac{17x^3+31x^2}{(1-x)^5} \ln x+ \dfrac{-35x^4+170x^3+447x^2+338x-56}{18(1-x)^4} \right] \ln \dfrac{\mu_W^2}{m_t^2}\;, \nonumber\\ [8mm]
G^t_1(x) &=& \dfrac{10x^4-100x^3+30x^2+160x-40}{27(1-x)^4} \mbox{Li}_2\left(1-\dfrac{1}{x}\right) \\
&& + \dfrac{30x^3-42x^2-332x+68}{81(1-x)^4} \ln x +\dfrac{-6x^3-293x^2+161x+42}{81(1-x)^3} \nonumber\\
&& + \left[ \dfrac{90x^2-160x+40}{27(1-x)^4} \ln x 
       + \dfrac{35x^3+105x^2-210x-20}{81(1-x)^3} \right] \ln \dfrac{\mu_W^2}{m_t^2}\;,\nonumber\\[8mm] 
T(x) &=& -(16x+8)\sqrt{4x-1} \; \mbox{Cl}_2\left(2 \arcsin \dfrac{1}{2\sqrt{x}}\right) +\left(16x+\dfrac{20}{3}\right) \ln x \\
&& + 32x + \dfrac{112}{9}\;.\nonumber
\end{eqnarray}
\\
The integral representations for the functions $\mbox{Li}_2$ and $\mbox{Cl}_2$ are as follows:\\
\begin{eqnarray}
\mbox{Li}_2(z) &=& -\int_0^z dt \frac{\ln (1-t)}{t}\;, \label{Li2}\\
\mbox{Cl}_2(x) &=& {\rm Im}\left[ \mbox{Li}_2(e^{ix}) \right] = -\int_0^x d\theta \ln |2 \sin(\theta/2)|\;.
\end{eqnarray}
The remaining NNLO coefficients take the form:\\
\begin{eqnarray}
C_7^{c(2)}(\mu_W) &=& C_7^{c(3)}(\mu_W=M_W) 
    + \frac{13763}{2187} \ln\frac{\mu_W^2}{M_W^2} 
    + \frac{814}{729}  \ln^2\frac{\mu_W^2}{M_W^2}\;,\\[5mm]
C_8^{c(2)}(\mu_W) &=& C_8^{c(3)}(\mu_W=M_W) 
    + \frac{16607}{5832} \ln\frac{\mu_W^2}{M_W^2} 
    + \frac{397}{486} \ln^2\frac{\mu_W^2}{M_W^2}\;,\\[5mm]
C_7^{t(3)}(\mu_W) &=& C_7^{t(3)}(\mu_W=m_t) \\[2mm]
&& + \ln\frac{\mu_W^2}{m_t^2}  
  \left[ \frac{-592 x^5 - 22 x^4 + 12814 x^3 - 6376 x^2 + 512 x}{27 (x - 1)^5} \,
  \mbox{Li}_2\left(1-\frac{1}{x}\right) \right. \nonumber\\ 
&& \left.+\frac{-26838 x^5 + 25938 x^4 + 627367 x^3 - 331956 x^2 + 16989 x - 460}{729 (x - 1)^6} \ln x
\right. \nonumber\\ 
&& \left.+ \frac{34400 x^5 + 276644 x^4 - 2668324 x^3 + 1694437 x^2 - 323354 x + 53077}{2187 (x - 1)^5} \right] \nonumber\\ 
&& + \ln^2\frac{\mu_W^2}{m_t^2} \left[ \frac{-63 x^5 + 532 x^4 + 2089 x^3 - 1118 x^2}{9 (x - 1)^6} \ln x
\right. \nonumber\\
&& \left.+ \frac{1186 x^5 - 2705 x^4 - 24791 x^3 - 16099 x^2 + 19229 x - 2740}{162 (x - 1)^5} \right] \;,\nonumber\\[5mm]
C_8^{t(3)}(\mu_W) &=& C_8^{t(3)}(\mu_W=m_t)\\[2mm]
&& + \ln\frac{\mu_W^2}{m_t^2} \left[  \frac{-148 x^5 + 1052 x^4 - 4811 x^3 - 3520 x^2 - 61 x}{18 (x - 1)^5} \,   \mbox{Li}_2\left(1-\frac{1}{x}\right) \right. \nonumber\\ 
&& \left.+ \frac{-15984 x^5 + 152379 x^4 - 1358060 x^3 - 1201653 x^2 - 74190 x +  9188}{1944 (x - 1)^6} \ln x
\right. \nonumber\\ 
&& \left.+ \frac{109669 x^5 - 1112675 x^4 + 6239377 x^3 + 8967623 x^2 + 768722 x - 42796}{11664 (x - 1)^5} \right] \nonumber\\ 
&& + \ln^2\frac{\mu_W^2}{m_t^2} \left[   \frac{-139 x^4 - 2938 x^3 - 2683 x^2}{12 (x - 1)^6} \ln x 
\right. \nonumber\\ 
&& \left.+ \frac{1295 x^5 - 7009 x^4 + 29495 x^3 + 64513 x^2 + 17458 x - 2072}{216 (x - 1)^5} \right]\;.\nonumber
\end{eqnarray}\\
As regard to the three-loop quantities $C_7^{c(3)}(\mu_W=M_W)$, $C_8^{c(3)}(\mu_W=M_W)$, $C_7^{t(3)}(\mu_W=m_t)$ and $C_8^{t(3)}(\mu_W=m_t)$, we have access to their
expansions at $x\to1$ and $x\to\infty$.\\
Denoting $z = 1/x$ and $w=1-z$, the coefficients become:
\begin{eqnarray}
C_7^{c(3)}(\mu_W=M_W) &\simeq&  1.525 -0.1165 z +0.01975 z \ln z +0.06283 z^2 +0.005349 z^2 \ln z \\
&& +0.01005 z^2 \ln^2z -0.04202 z^3 +0.01535 z^3 \ln z -0.00329 z^3 \ln^2 z \nonumber\\
&& +0.002372 z^4 -0.0007910 z^4 \ln z +{\cal O}(z^5)\;,\nonumber\\[5mm]
C_7^{c(3)}(\mu_W=M_W) &\simeq& 1.432 +0.06709 w +0.01257 w^2 +0.004710 w^3 +0.002373w^4 \\
&& +0.001406w^5 +\!0.0009216 w^6 +\!0.00064730 w^7 +\!0.0004779 w^8 +\!{\cal O}(w^9)\;,\nonumber\\ [5mm]
C_8^{c(3)}(\mu_W=M_W) &\simeq& -1.870 +0.1010 z -0.1218 z \ln z +0.1045 z^2 -0.03748 z^2 \ln z\\
&& +0.01151 z^2 \ln^2 z -0.01023 z^3 +0.004342 z^3 \ln z +0.0003031 z^3 \ln^2 z \nonumber\\
&& -0.001537 z^4 +0.0007532 z^4 \ln z +{\cal O}(z^5)\;, \nonumber\\ [5mm]
C_8^{c(3)}(\mu_W=M_W) &\simeq& -1.676 -0.1179 w -0.02926 w^2 -0.01297 w^3 -0.007296 w^4\\
&& -0.004672 w^5 -0.003248 w^6 -0.002389 w^7 -0.001831 w^8 +{\cal O}(w^9)\;,\nonumber\\[5mm]
C_7^{t(3)}(\mu_W=m_t) &\simeq& 12.06 +12.93 z +3.013 z \ln z +96.71 z^2 +52.73 z^2 \ln z +147.9 z^3\\
&& +187.7 z^3 \ln z -144.9 z^4 +236.1 z^4 \ln z +{\cal O}(z^5)\;,\nonumber\\ [5mm]
C_7^{t(3)}(\mu_W=m_t) &\simeq& 11.74 +0.3642 w +0.1155 w^2 -0.003145 w^3 -0.03263 w^4  \\
&& -0.03528 w^5-0.03076 w^6 -0.02504 w^7 -0.01985 w^8 +{\cal O}(w^9)\;,\nonumber\\ [5mm]
C_8^{t(3)}(\mu_W=m_t) &\simeq& -0.8954 -7.043 z -98.34 z^2 -46.21 z^2 \ln z -127.1 z^3 \\
&& -181.6 z^3 \ln z +535.8 z^4 -76.76 z^4 \ln z +{\cal O}(z^5)\;,\nonumber\\ [5mm]
C_8^{t(3)}(\mu_W=m_t) &\simeq& -0.6141 -0.8975 w -0.03492 w^2 +0.06791 w^3 +0.07966 w^4 \\
&& +0.07226 w^5 +0.06132 w^6 +0.05096 w^7 +0.04216 w^8 +{\cal O}(w^9)\;.\nonumber
\end{eqnarray}
\\
For $n=0,1,2$ and $i=1,\cdots,8$, the Wilson coefficients are obtained using:
\begin{equation}
C^{(n)}_i = C_i^{t(n)} - C_i^{c(n)}\;.
\end{equation}

\subsection{Charged Higgs contributions}
\noindent At the Leading Order, the relevant charged Higgs contributions to the Wilson coefficients are given by \cite{ciuchini}:
\begin{equation}
\delta C_{7,8}^{H(0)}(\mu_W) =\frac{\lambda_{tt}^2}{3} F_{7,8}^{(1)}(x_{tH^\pm})-\lambda_{tt}\lambda_{bb} F_{7,8}^{(2)}(x_{tH^{\pm}}) \;,
\end{equation}
where
\begin{equation}
x_{tH^{\pm}}=\frac{\overline{m}_t^2(\mu_W )}{M_{H^\pm}^2}\;,
\end{equation}
and
\begin{eqnarray}
F_7^{(1)}(x)&=&\frac{x(7-5x-8x^2)}{24(x-1)^3}+\frac{x^2(3x-2)}{4(x-1)^4}\ln x \;,\nonumber\\ F_8^{(1)}(x)&=&\frac{x(2+5x-x^2)}{8(x-1)^3}-\frac{3x^2}{4(x-1)^4}\ln x\;,\label{F7812}\\
F_7^{(2)}(x)&=&\frac{x(3-5x)}{12(x-1)^2}+\frac{x(3x-2)}{6(x-1)^3}\ln x \;,\nonumber\\ F_8^{(2)}(x)&=&\frac{x(3-x)}{4(x-1)^2}-\frac{x}{2(x-1)^3}\ln x \;.\nonumber
\end{eqnarray}
$\lambda_{tt},\lambda_{bb}$ are the Yukawa couplings. In supersymmetry, they read:
\begin{equation}
\lambda_{tt}=-\frac{1}{\lambda_{bb}}=\frac{1}{\tan \beta}\;.
\end{equation}
For the different types of 2HDM, the Yukawa couplings are summarized in Table~\ref{2hdmyuk}.\\
\begin{table}
\centering
\begin{tabular*}{0.7\columnwidth}{@{\extracolsep{\fill}}ccccccc}
\hline
Type & $\lambda_{UU}$ & $\lambda_{DD}$ & $\lambda_{LL}$ \\
\hline
I & $\cot\beta$ & $\cot\beta$ & $\cot\beta$ \\
II & $\cot\beta$ & $-\tan\beta$ & $-\tan\beta$ \\
III & $\cot\beta$ & $-\tan\beta$ & $\cot\beta$ \\
IV & $\cot\beta$ & $\cot\beta$ & $-\tan\beta$ \\
\hline
\end{tabular*}
\caption[Yukawa couplings for the four types of 2HDM]{Yukawa couplings for the four types of 2HDM. $U$, $D$ and $L$ stand respectively for the up-type quarks, the down-type quarks and the leptons.}
\label{2hdmyuk}
\end{table}
\\
At the NLO, the charged Higgs contributions can be written in the form \cite{ciuchini}:
\begin{eqnarray}
\delta C_7^{(1)}(\mu_W) &=& G_7^H(x_{tH^{\pm}}) + \Delta_7^H(x_{tH^{\pm}}) \ln\frac{\mu_W^2}{M_{H^\pm}^2}-\frac49 E^H(x_{tH^{\pm}})\;,\\
\delta C_8^{(1)}(\mu_W) &=& G_8^H(x_{tH^{\pm}}) + \Delta_8^H(x_{tH^{\pm}}) \ln\frac{\mu_W^2}{M_{H^\pm}^2} -\frac16 E^H(x_{tH^{\pm}}) \;,
\end{eqnarray}
with\\
\begin{eqnarray}
G_7^H(x) &= &  \lambda_{tt}\lambda_{bb}\frac{4}{3}x\left[ \frac{4(-3+7x-2x^2)}{3(x-1)^3}{\rm Li}_2 \left( 1 - \frac{1}{x} \right)+\frac{8-14x-3x^2}{3(x-1)^4}\ln^2x  \right.\\
&& \left. +\frac{7-13x+2x^2}{(x-1)^3} +\frac{2(-3-x+12x^2-2x^3)}{3(x-1)^4}\ln x\right] \nonumber\\ 
&& +\lambda_{tt}^2\frac{2}{9}x\left[ \frac{x(18-37x+8x^2)}{(x-1)^4}{\rm Li}_2 \left( 1 - \frac{1}{x} \right) \right. +\frac{x(-14+23x+3x^2)}{(x-1)^5}\ln^2x\nonumber \\
&& \left. +\frac{-50+251x-174x^2-192x^3+21x^4}{9(x-1)^5}\ln x +\frac{797-5436x+7569x^2-1202x^3}{108(x-1)^4}\right] \;,\nonumber
\end{eqnarray} 
\begin{eqnarray}
\Delta_7^H(x) &= & \lambda_{tt}\lambda_{bb}\frac{2}{9}x\left[ \frac{21-47x+8x^2}{(x-1)^3}+\frac{2(-8+14x+3x^2)}{(x-1)^4}\ln x \right]\\ 
&& +\lambda_{tt}^2\frac{2}{9}x\left[ \frac{-31-18x+135x^2-14x^3}{6(x-1)^4}+\frac{x(14-23x-3x^2)}{(x-1)^5}\ln x \right]\;,\nonumber \\[2mm]
G_8^H(x) &= & \lambda_{tt}\lambda_{bb}\frac{1}{3}x\left[ \frac{-36+25x-17x^2}{2(x-1)^3}{\rm Li}_2  \left( 1 - \frac{1}{x} \right)+\frac{19+17x}{(x-1)^4}\ln^2x  \right. \\ 
&& \left. +\frac{-3-187x+12x^2-14x^3}{4(x-1)^4}\ln x+\frac{3(143-44x+29x^2)}{8(x-1)^3}\right] \nonumber\\ 
&& +\lambda_{tt}^2\frac{1}{6}x\left[ \frac{x(30-17x+13x^2)}{(x-1)^4}{\rm Li}_2 \left( 1 - \frac{1}{x} \right)-\frac{x(31+17x)}{(x-1)^5}\ln^2x \right.\nonumber \\ 
&&\left.  +\frac{-226+817x+1353x^2+318x^3+42x^4}{36(x-1)^5}\ln x  +\frac{1130-18153x+7650x^2-4451x^3}{216(x-1)^4}\right]\;,\nonumber\\[2mm]
\Delta_8^H(x) &=& \lambda_{tt}\lambda_{bb}\frac{1}{3}x\left[ \frac{81-16x+7x^2}{2(x-1)^3}-\frac{19+17x}{(x-1)^4}\ln x \right]\\
&& +\lambda_{tt}^2\frac{1}{6}x\left[ \frac{-38-261x+18x^2-7x^3}{6(x-1)^4}+\frac{x(31+17x)}{(x-1)^5}\ln x \right] \;,\nonumber \\[2mm]
E^H(x)&=&\lambda_{tt}^2\left[ \frac{x(16-29x+7x^2)}{36(x-1)^3}+\frac{x(3x-2)}{6(x-1)^4}\ln x \right] \;.
\end{eqnarray}
\subsection{Supersymmetric contributions}
The relevant chargino contributions are given by \cite{ciuchini2,gomez}:
\begin{eqnarray}
\delta C_{7,8}^{\chi}(\mu) &=& - \sum_{k=1}^2 \sum_{i=1}^2 \left\{ \frac{2}{3} |\Gamma_{ki}|^2 \frac{M_W^2}{m_{\tilde t_k}^2} F_{7,8}^{(1)}(x_{\tilde t_k\chi_i^{\pm}}) + \Gamma_{ki}^*\Gamma_{ki}' \frac{M_W}{m_{\chi_i^{\pm}}} F_{7,8}^{(3)}(x_{\tilde t_k\chi_i^{\pm}})\right\} \\
&& + \sum_{i=1}^2 \left\{ \frac{2}{3} |\tilde \Gamma_{1i}|^2\frac{M_W^2}{m_{\tilde{q}_{12}}^2} F_{7,8}^{(1)}(x_{\tilde{q}_{12}\chi_i^{\pm}})+ \tilde \Gamma_{1i}^*\tilde \Gamma_{1i}' \frac{M_W}{m_{\chi_i^{\pm}}} F_{7,8}^{(3)}(x_{\tilde{q}_{12}\chi_i^{\pm}})\right\}\;,\nonumber
\end{eqnarray}\\
where $x_{ij}=m_i^2/m_j^2$, and $m_{\tilde{q}_{12}}$ is the common mass of the up and charm squarks, that we consider identical ($m_{\tilde{q}_{12}} \approx m_{\tilde{u}} \approx m_{\tilde{c}}$). The functions $F_{7,8}^{(3)}(x)$ take the form
\begin{eqnarray}
F_7^{(3)}(x) &=& \frac{(5-7x)}{6(x-1)^2} +\frac{x (3x-2)}{3(x-1)^3}\ln x \;,\nonumber\\
F_8^{(3)}(x) &=& \frac{(1+x)}{2(x-1)^2} -\frac{x}{(x-1)^3}\ln x \;,
\end{eqnarray}
and $F_{7,8}^{(1)}(x)$ are given in Eq. (\ref{F7812}).
Moreover
\begin{eqnarray}
\Gamma_{ij} &=& (D^{\tilde{t}}_{1i})^{*}V^{*}_{j1} -\frac{\overline{m}_t(\mu)}{\sqrt 2 M_W\sin\beta} (D^{\tilde{t}}_{2i})^{*} V^*_{j2} \;,\nonumber\\
\Gamma_{ij}' &=& \frac{ (D^{\tilde{t}}_{1i})^{*} U_{j2}} {\sqrt 2 \cos\beta (1+\epsilon_{b}^*\tan\beta)} \;,
\end{eqnarray}\\
where $U$ and $V$ are the chargino mixing matrices, following the diagonalizing convention:
\begin{equation}
U \begin{pmatrix} M_2 & M_{W} \sqrt{2} \sin \beta \cr M_{W} \sqrt{2} \cos \beta & \mu \end{pmatrix} V^{-1}\;,
\end{equation}
and $D^{\tilde{q}}$ is the squark $\tilde{q}$ mixing matrix such as:
\begin{equation} 
D^{\tilde{q}} = \begin{pmatrix} \cos\theta_{\tilde{q}} & -\sin\theta_{\tilde{q}}  \cr \sin\theta_{\tilde{q}} & \cos\theta_{\tilde{q}} \end{pmatrix} \;.
\end{equation}
$\epsilon_{b}$, which will be given below, is a two loop SUSY correction, with effects enhanced by factors of $\tan\beta$.\\
\\
$\tilde \Gamma_{ij}$ and $\tilde \Gamma_{ij}'$  are obtained  from $\Gamma_{ij}$ and $\Gamma_{ij}'$ by replacing the matrix $D^{\tilde{t}}$ by the unity matrix. 
In the following, we adopt the notations:
\begin{equation}
\cos \theta_{\tilde{q}} = D^{\tilde{q}}_{11} \equiv  c_{\tilde{q}}\;,\;\;
\sin \theta_{\tilde{q}} = D^{\tilde{q}}_{21} \equiv s_{\tilde{q}} \; .
\end{equation}
The leading $\tan\beta$ corrections are contained in the following expressions for $\epsilon_b$, $\epsilon_b^\prime$ and $\epsilon_0^\prime$ \cite{micromegas}, which are evaluated at a typical SUSY scale, $\mu_s$. $\epsilon_b$ can be split into two parts:
\begin{equation}
\epsilon_b=\epsilon_0+\epsilon_2 \;,\label{epsb}
\end{equation}
with
\begin{eqnarray}
\epsilon_0  &=&\frac{2\,\alpha_s(\mu_s)}{3\,\pi} \frac{A_b/\tan\beta-\mu}{m_{\tilde{g}}} H_2(x_{\tilde{b}_1 \tilde{g}},x_{\tilde{b}_2 \tilde{g}}) \label{eps0}\\
&&+ \frac{\alpha(M_Z)\mu M_2}{4 \sin^2\theta_W \pi} \left[\frac{c_{\tilde{b}}^2}{2m_{\tilde{b}_1}^2} H_2\left(\displaystyle\frac
{M_2^2}{m_{\tilde{b}_1}^2}, \displaystyle\frac
{\mu^2}{m_{\tilde{b}_1}^2}\right)+\frac{s_{\tilde{b}}^2}{2 m_{\tilde{b}_2}^2}H_2\left(\displaystyle\frac
{M_2^2}{m_{\tilde{b}_2}^2}, \displaystyle\frac
{\mu^2}{m_{\tilde{b}_2}^2}\right)\right] \nonumber\;,
\end{eqnarray}\\
and
\begin{eqnarray}
\epsilon_2  &=&\frac{ \tilde{y}_t^2(\mu_s)}{16\, \pi^2} \,\sum_{i=1}^2 U_{i2}\frac{\mu/\tan\beta-A_t}{m_{\chi^\pm_i}}\,H_2(x_{\tilde{t}_1 \chi^\pm_i},x_{\tilde{t}_2 \chi^\pm_i}) \,V_{i2} \label{eps2}\\
&&+ \frac{\alpha(M_Z)\mu M_2}{4 \sin^2\theta_W \pi} \left[\frac{c_{\tilde{t}}^2}{m_{\tilde{t}_1}^2}H_2\left(\displaystyle\frac
{M_2^2}{m_{\tilde{t}_1}^2}, \displaystyle\frac
{\mu^2}{m_{\tilde{t}_1}^2}\right)+\frac{s_{\tilde{t}}^2}{m_{\tilde{t}_2}^2}H_2\left(\displaystyle\frac
{M_2^2}{m_{\tilde{t}_2}^2}, \displaystyle\frac
{\mu^2}{m_{\tilde{t}_2}^2}\right)\right] \nonumber\;,
\end{eqnarray}\\
where $A_q$ is the trilinear coupling of the quark $q$. $y_q$ and $\tilde{y}_q$ are the ordinary and supersymmetric Yukawa couplings of the quark $q$ respectively. The function $H_2$ is defined as:
\begin{equation}
H_2(x,y) = \frac{x\ln\, x}{(1-x)(x-y)} + \frac{y\ln\, y}{(1-y)(y-x)}\;. \label{Hxy}
\end{equation}
We neglect the neutralino mixing matrices and we assume that the chargino masses are given by $\mu$ and $M_2$.\\
\begin{eqnarray}
\epsilon_b^\prime (t)&=& \frac{2\,\alpha_s(\mu_s)}{3\,\pi}\frac{A_b/\tan\beta-\mu}{m_{\tilde{g}}} \left[ c_{\tilde{t}}^2 c^2_{\tilde{b}}\,H_2(x_{\tilde{t}_1 \tilde{g}},x_{\tilde{b}_2 \tilde{g}})  \right.\\
&& \left. + c_{\tilde{t}}^2 s^2_{\tilde{b}}
\,H_2(x_{\tilde{t}_1 \tilde{g}},x_{\tilde{b}_1 \tilde{g}}) + s_{\tilde{t}}^2 c^2_{\tilde{b}} \,H_2(x_{\tilde{t}_2 \tilde{g}},x_{\tilde{b}_2 \tilde{g}}) + s_{\tilde{t}}^2 s^2_{\tilde{b}} \,H_2(x_{\tilde{t}_2 \tilde{g}},x_{\tilde{b}_1 \tilde{g}})\right]  \nonumber \\
&& +\, \frac{ y_t^2(\mu_s)}{16\, \pi^2} \sum_{i=1}^{n_{\chi^0}}\,N_{i4}^*\frac{A_t-\mu/\tan\beta}{m_{\chi^0_i}} \left[ c_{\tilde{t}}^2 c^2_{\tilde{b}}\,H_2(x_{\tilde{t}_2 \chi^0_i},x_{\tilde{b}_1 \chi^0_i}) \right. \nonumber\\
&&\left. + c_{\tilde{t}}^2 s^2_{\tilde{b}}\,    H_2(x_{\tilde{t}_2 \chi^0_i},x_{\tilde{b}_2 \chi^0_i})\label{epsbt} +\,s_{\tilde{t}}^2 c^2_{\tilde{b}}\, H_2(x_{\tilde{t}_1 \chi^0_i},x_{\tilde{b}_1 \chi^0_i}) + s_{\tilde{t}}^2 s^2_{\tilde{b}}\, H_2(x_{\tilde{t}_1 \chi^0_i},x_{\tilde{b}_2 \chi^0_i}) \right] N_{i3} \nonumber \\
&&+ \frac{\alpha(M_Z)\mu M_2}{4 \sin^2\theta_W \pi} \left[\frac{c_{\tilde{b}}^2}{m_{\tilde{b}_1}^2}H_2\left(\displaystyle\frac
{M_2^2}{m_{\tilde{b}_1}^2}, \displaystyle\frac
{\mu^2}{m_{\tilde{b}_1}^2}\right) +\frac{s_{\tilde{b}}^2}{m_{\tilde{b}_2}^2}
H_2\left(\displaystyle\frac
{M_2^2}{m_{\tilde{b}_2}^2}, \displaystyle\frac
{\mu^2}{m_{\tilde{b}_2}^2}\right) \right. \nonumber\\
&& + \left.\frac{c_{\tilde{t}}^2}{2 m_{\tilde{t}_1}^2}H_2\left(\displaystyle\frac
{M_2^2}{m_{\tilde{t}_1}^2}, \displaystyle\frac
{\mu^2}{m_{\tilde{t}_1}^2}\right)+\frac{s_{\tilde{t}}^2}{2m_{\tilde{t}_2}^2} H_2\left(\displaystyle\frac
{M_2^2}{m_{\tilde{t}_2}^2}, \displaystyle\frac
{\mu^2}{m_{\tilde{t}_2}^2}\right) \right] \nonumber\;.
\end{eqnarray}\\
In the above equation, $N$ represents the neutralino mixing matrix and $n_{\chi^0}$ the number of neutralinos, {\it i.e.} four in the MSSM and five in the NMSSM.\\
\\
The last $\epsilon$ correction reads:\\
\begin{eqnarray}
\epsilon_0^\prime&=& -\frac{2\,\alpha_s(\mu_s)}{3\,\pi} \frac{\mu+A_t/\tan\beta}{m_{\tilde{g}}} \left[ c_{\tilde{t}}^2
\,H_2(x_{\tilde{t}_2 \tilde{g}},x_{\tilde{s} \tilde{g}}) \;+\; s_{\tilde{t}}^2 H_2(x_{\tilde{t}_1 \tilde{g}},x_{\tilde{s} \tilde{g}}) \right]\\
&& +\, \frac{ y_b^2(\mu_s)}{16\, \pi^2} \sum_{i=1}^{n_{\chi^0}}\,N_{i4}^*\frac{\mu/\tan\beta}{m_{\chi^0_i}}
\, \left[ c_{\tilde{t}}^2 c^2_{\tilde{b}}\,H_2(x_{\tilde{t}_1 \chi^0_i},x_{\tilde{b}_2 \chi^0_i}) + c_{\tilde{t}}^2 s^2_{\tilde{b}}\,H_2(x_{\tilde{t}_1 \chi^0_i},x_{\tilde{b}_1 \chi^0_i}) \right.\nonumber \\
&&\left.+\,s_{\tilde{t}}^2 c^2_{\tilde{b}}\, H_2(x_{\tilde{t}_2 \chi^0_i},x_{\tilde{b}_2 \chi^0_i}) + s_{\tilde{t}}^2 s^2_{\tilde{b}}\, H_2(x_{\tilde{t}_2 \chi^0_i},x_{\tilde{b}_1 \chi^0_i}) \right] N_{i3}\nonumber\;.
\end{eqnarray}\\
The SM and charged Higgs contributions at the $\mu_W$ scale are affected by $\epsilon_b$, $\epsilon_b^\prime$ and $\epsilon_0^\prime$ as the following:
\begin{eqnarray}
\delta C_{7,8}^{(SM,\tan\beta)}(\mu_W) &=& \frac{\bigl( \epsilon_b -\epsilon^\prime_b(t)\bigr) \tan\beta}{1+\epsilon_b\tan\beta} \, F_{7,8}^{(2)}(x_{tW})\;,\\
\delta C_{7,8}^{(H,\tan\beta)}(\mu_W) &=& -\frac{\bigl( \epsilon^\prime_0 +\epsilon_b\bigr) \tan\beta}{1+\epsilon_b\tan\beta} \, F_{7,8}^{(2)}(x_{tH^\pm})\;.
\end{eqnarray}\\
Higher order charged Higgs contributions are expressed as \cite{dambrosio}:\\
\begin{equation}
\delta C_{7,8}^{(H,\tan^2\beta)}(\mu_W)=-\frac{ \epsilon_2 \epsilon_1^\prime \tan^2\beta}{(1+\epsilon_0\tan\beta)(1+\epsilon_b\tan\beta)} \, F_{7,8}^{(2)}(x_{tH^\pm})\;,
\end{equation}\\
where $F_{7,8}^{(2)}(x)$ are given in Eq. (\ref{F7812}) and $\epsilon_1^\prime$ reads \cite{freitas}:
\begin{eqnarray}
\epsilon_1^\prime &=& \frac{1}{16 \pi^2} \left[ \frac{A_b y_b^2}{\mu }  
 H_2 \left( \frac{m_{{\tilde t}_L}^2}{\mu^2}, \frac{m_{{\tilde b}_R}^2}{\mu^2} \right)
-g^2 \frac{M_2}{\mu} H_2 \left( \frac{m_{{\tilde t}_L}^2}{\mu^2}, \frac{M_2^2}{\mu^2}\right)\right]\;.
\end{eqnarray}
Finally, we add the neutral Higgs contributions \cite{buras3,hofer}. In the MSSM, they are expressed as
\begin{eqnarray}
\delta^{H^0}C_{7(8)}^{(0)}(\mu_W) &=& \frac{a_{7(8)}}{72}\frac{\epsilon_2}{\cos^2\beta(1+\epsilon_0\tan\beta)
(1+\epsilon_b\tan\beta)^2} \\
&& \times \sum_{S=h^0,H^0,A^0}
\frac{\overline m_b^2}{M^2_S} (x^{S*}_u - x^{S*}_d \tan\beta) (x^{S}_d + x^{S}_u \epsilon_b) \;,\nonumber
\end{eqnarray}
where $a_7=1$, $a_8=-3$ and, for $S=(h^0,H^0,A^0)$,
\begin{equation}
x^S_d=(-\sin\alpha,\cos\alpha,i \sin\beta)\;,\qquad
x^S_u=(\cos\alpha,\sin\alpha,-i \cos\beta) \;.
\end{equation}
In the NMSSM, they are generalized to
\begin{eqnarray}
\delta^{H^0}C_{7(8)}^{(0)}(\mu_W) &=& \frac{a_{7(8)}}{72}\frac{\epsilon_2}{\cos^2\beta(1+\epsilon_0\tan\beta)
(1+\epsilon_b\tan\beta)^2} \\
&& \times \sum_{i=1}^3
\frac{\overline m_b^2}{m^2_i} (U^{H*}_{i1} + U^{H*}_{i2} \tan\beta) (U^H_{i2} + U^H_{i1} \epsilon_b) \;,\nonumber\\
\delta^{A^0}C_{7(8)}^{(0)}(\mu_W) &=& \frac{a_{7(8)}}{72}\frac{\epsilon_2}{\cos^2\beta(1+\epsilon_0\tan\beta)
(1+\epsilon_b\tan\beta)^2} \\
&& \times \sum_{j=1}^2
\frac{\overline m_b^2}{m^2_j} (U^{A*}_{j1} + U^{A*}_{j2} \tan\beta) (U^A_{j2} + U^A_{j1} \epsilon_b) \;,\nonumber
\end{eqnarray}
where $i=h^0,H^0,H_3^0$ and $j=a_1^0,A_2^0$.\\
\\
The CP-even and CP-odd Higgs mixing matrices are respectively \cite{heng2008}\\
\begin{equation}
 U^H=\left(\begin{array}{ccc}
(\cos\theta_H -\sin\theta_H v\delta_+/x)/\tan\beta & \cos\theta_H & -\sin\theta_H\\
(\sin\theta_H +\cos\theta_H v\delta_+/x) & \sin\theta_H & \cos\theta_H\\
1 & -1/\tan\beta & -v\delta_+/x\tan\beta
\end{array}\right) \;. \label{Hmix}
\end{equation}
and
\begin{equation}
 U^A=\left(\begin{array}{ccc}
\cos\theta_A \sin\beta & \cos\theta_A \cos\beta & \sin\theta_A\\
-\sin\theta_A \sin\beta & -\sin\theta_A \cos\beta & \cos\theta_A\\
-\cos\beta & \sin\beta & 0
\end{array}\right) \;. \label{Amix}
\end{equation}\\
where $\theta_H$ and $\theta_A$ are the mixing angles, $x$ is the scalar VEV, $v\approx 246$ GeV and
\begin{equation}
 \delta_+ = \frac{\sqrt2 A_\lambda + 2 \kappa x}{\sqrt2 A_\lambda + \kappa x} \;\;.
\end{equation}
The complete Wilson coefficients $C_i$ are obtained by adding the different contributions.
\append{Renormalization group equations}%
\subsection{RGE in the standard operator basis}
\label{wilsonbasis1}%
The transformations of the Wilson coefficients from the matching scale $\mu_W$ to the scale $\mu_b$ in the standard operator basis (Eq. (\ref{standard_basis})) are given by \cite{czakon}:
\begin{eqnarray}
\vec{C}^{(0)\rm eff}(\mu_b) &=& U^{(0)} \vec{C}^{(0)\rm eff}(\mu_W)\;,\\[2mm]
\vec{C}^{(1)\rm eff}(\mu_b) &=& \eta \left[U^{(0)} \vec{C}^{(1)\rm eff}(\mu_W)+U^{(1)} \vec{C}^{(0)\rm eff}(\mu_W)\right]\;,\\[2mm]
\vec{C}^{(2)\rm eff}(\mu_b) &=& \eta^2 \left[ U^{(0)} \vec{C}^{(2)\rm eff}(\mu_W)+U^{(1)} \vec{C}^{(1)\rm eff}(\mu_W)+U^{(2)} \vec{C}^{(0)\rm eff}(\mu_W)\right] \;,
\end{eqnarray}
where $\eta=\alpha_s(\mu_W)/\alpha_s(\mu_b)$ and $\vec{C}=\{C_1, \cdots, C_8\}$. The $U^{(n)}$ matrix elements read
\begin{equation}
U^{(n)}_{kl} = \sum_{j=0}^n \sum_{i=1}^8 m^{(nj)}_{kli} \eta^{a_i-j}\;.
\end{equation}
The powers $a_i$ are given in Table~\ref{tabai}. The $m^{(nj)}_{kli}$ relevant in our calculations for the $U^{(n)}_{kl}$ are given in Tables~\ref{mnjLO}--\ref{mnjNNLO}.

\begin{table}[!t]
\begin{tabular}{|c|c|c|c|c|c|c|c|c|}
\hline
$~~~~i~~~~$&$~~~~1~~~~$&$~~~~2~~~~$&$~~~~3~~~~$&$~~~~4~~~~$&$~~~~5~~~~$&$~~~~6~~~~$&$~~~~7~~~~$&$~~~~8~~~~$\\[1mm]
\hline
$a_i$&$14/23$&$16/23$&$6/23$&$-12/23$&$  0.4086 $&$ -0.4230 $&$ -0.8994 $&$  0.1456 $\\
\hline
\end{tabular}
\caption[$a_i$ numbers]{ The numbers $a_i$.\label{tabai}}
\end{table}

\begin{table}[p]
\begin{center}
\begin{tabular}{|c|c|c|c|c|c|c|c|c|}\hline
 $i$&1&2&3&4&5&6&7&8\\\hline
 $m^{(00)}_{11i}$&0&0&0.3333&0.6667&0&0&0&0\\[1mm]
 $m^{(00)}_{12i}$&0&0&1&$-$1&0&0&0&0\\[1mm]
 $m^{(00)}_{21i}$&0&0&0.2222&$-$0.2222&0&0&0&0\\[1mm]
 $m^{(00)}_{22i}$&0&0&0.6667&0.3333&0&0&0&0\\[1mm]
 $m^{(00)}_{31i}$&0&0&0.0106&0.0247&$-$0.0129&$-$0.0497&0.0092&0.0182\\[1mm]
 $m^{(00)}_{32i}$&0&0&0.0317&$-$0.0370&$-$0.0659&0.0595&$-$0.0218&0.0335\\[1mm]
 $m^{(00)}_{34i}$&0&0&0&0&$-$0.1933&0.1579&0.1428&$-$0.1074\\[1mm]
 $m^{(00)}_{41i}$&0&0&0.0159&$-$0.0741&0.0046&0.0144&0.0562&$-$0.0171\\[1mm]
 $m^{(00)}_{42i}$&0&0&0.0476&0.1111&0.0237&$-$0.0173&$-$0.1336&$-$0.0316\\[1mm]
 $m^{(00)}_{44i}$&0&0&0&0&0.0695&$-$0.0459&0.8752&0.1012\\[1mm]
 $m^{(00)}_{51i}$&0&0&$-$0.0026&$-$0.0062&0.0018&0.0083&$-$0.0004&$-$0.0009\\[1mm]
 $m^{(00)}_{52i}$&0&0&$-$0.0079&0.0093&0.0094&$-$0.0100&0.0010&$-$0.0017\\[1mm]
 $m^{(00)}_{54i}$&0&0&0&0&0.0274&$-$0.0264&$-$0.0064&0.0055\\[1mm]
 $m^{(00)}_{61i}$&0&0&$-$0.0040&0.0185&0.0021&$-$0.0136&$-$0.0043&0.0012\\[1mm]
 $m^{(00)}_{62i}$&0&0&$-$0.0119&$-$0.0278&0.0108&0.0163&0.0103&0.0023\\[1mm]
 $m^{(00)}_{64i}$&0&0&0&0&0.0317&0.0432&$-$0.0675&$-$0.0074\\[1mm]
 $m^{(00)}_{71i}$&0.5784&$-$0.3921&$-$0.1429&0.0476&$-$0.1275&0.0317&0.0078&$-$0.0031\\[1mm]
 $m^{(00)}_{72i}$&2.2996&$-$1.0880&$-$0.4286&$-$0.0714&$-$0.6494&$-$0.0380&$-$0.0185&$-$0.0057\\[1mm]
 $m^{(00)}_{73i}$&8.0780&$-$5.2777&0&0&$-$2.8536&0.1281&0.1495&$-$0.2244\\[1mm]
 $m^{(00)}_{74i}$&5.7064&$-$3.8412&0&0&$-$1.9043&$-$0.1008&0.1216&0.0183\\[1mm]
 $m^{(00)}_{75i}$&202.9010&$-$149.4668&0&0&$-$55.2813&2.6494&0.7191&$-$1.5213\\[1mm]
 $m^{(00)}_{76i}$&86.4618&$-$59.6604&0&0&$-$25.4430&$-$1.2894&0.0228&$-$0.0917\\[1mm]
 $m^{(00)}_{77i}$&0&1&0&0&0&0&0&0\\[1mm]
 $m^{(00)}_{78i}$&2.6667&$-$2.6667&0&0&0&0&0&0\\[1mm]
 $m^{(00)}_{81i}$&0.2169&0&0&0&$-$0.1793&$-$0.0730&0.0240&0.0113\\[1mm]
 $m^{(00)}_{82i}$&0.8623&0&0&0&$-$0.9135&0.0873&$-$0.0571&0.0209\\[1mm]
 $m^{(00)}_{84i}$&2.1399&0&0&0&$-$2.6788&0.2318&0.3741&$-$0.0670\\[1mm]
 $m^{(00)}_{88i}$&1&0&0&0&0&0&0&0\\[1mm]
\hline
\end{tabular}
\caption[$m^{(00)}_{kli}$ values]{Values of the $m^{(00)}_{kli}$ relevant for $U^{(0)}_{kl}$ \cite{czakon}.\label{mnjLO}}
\end{center}
\end{table}

\begin{table}[p]
\begin{center}
\begin{tabular}{|c|c|c|c|c|c|c|c|c|}\hline
 $i$&1&2&3&4&5&6&7&8\\\hline
 $m^{(10)}_{12i}$&0&0&$-$2.9606&$-$4.0951&0&0&0&0\\[1mm] 
 $m^{(11)}_{12i}$&0&0&5.9606&1.0951&0&0&0&0\\[1mm]
 $m^{(10)}_{22i}$&0&0&$-$1.9737&1.3650&0&0&0&0\\[1mm] 
 $m^{(11)}_{22i}$&0&0&1.9737&$-$1.3650&0&0&0&0\\[1mm]
 $m^{(10)}_{32i}$&0&0&$-$0.0940&$-$0.1517&$-$0.2327&0.2288&0.1455&$-$0.4760\\[1mm] 
 $m^{(11)}_{32i}$&0&0&$-$0.5409&1.6332&1.6406&$-$1.6702&$-$0.2576&$-$0.2250\\[1mm]
 $m^{(10)}_{42i}$&0&0&$-$0.1410&0.4550&0.0836&$-$0.0664&0.8919&0.4485\\[1mm] 
 $m^{(11)}_{42i}$&0&0&2.2203&2.0265&$-$4.1830&$-$0.7135&$-$1.8215&0.7996\\[1mm]
 $m^{(10)}_{52i}$&0&0&0.0235&0.0379&0.0330&$-$0.0383&$-$0.0066&0.0242\\[1mm] 
 $m^{(11)}_{52i}$&0&0&0.0400&$-$0.1861&$-$0.1669&0.1887&0.0201&0.0304\\[1mm]
 $m^{(10)}_{62i}$&0&0&0.0352&$-$0.1138&0.0382&0.0625&$-$0.0688&$-$0.0327\\[1mm] 
 $m^{(11)}_{62i}$&0&0&$-$0.2614&$-$0.1918&0.4197&0.0295&0.1474&$-$0.0640\\[1mm]
 $m^{(10)}_{71i}$&0.0021&$-$1.4498&0.8515&0.0521&0.6707&0.1220&$-$0.0578&0.0355\\[1mm] 
 $m^{(11)}_{71i}$&$-$4.3519&3.0646&1.5169&$-$0.5013&0.3934&$-$0.6245&0.2268&0.0496\\[1mm]
 $m^{(10)}_{72i}$&9.9372&$-$7.4878&1.2688&$-$0.2925&$-$2.2923&$-$0.1461&0.1239&0.0812\\[1mm] 
 $m^{(11)}_{72i}$&$-$17.3023&8.5027&4.5508&0.7519&2.0040&0.7476&$-$0.5385&0.0914\\[1mm]
 $m^{(10)}_{74i}$&$-$8.6840&8.5586&0&0&0.7579&0.4446&0.3093&0.4318\\[1mm] 
 $m^{(11)}_{74i}$&$-$42.9356&30.0198&0&0&5.8768&1.9845&3.5291&$-$0.2929\\[1mm]
 $m^{(10)}_{77i}$&0&7.8152&0&0&0&0&0&0\\[1mm] 
 $m^{(11)}_{77i}$&0&$-$7.8152&0&0&0&0&0&0\\[1mm]
 $m^{(10)}_{78i}$&17.9842&$-$18.7604&0&0&0&0&0&0\\[1mm] 
 $m^{(11)}_{78i}$&$-$20.0642&20.8404&0&0&0&0&0&0\\[1mm]
 $m^{(10)}_{82i}$&3.7264&0&0&0&$-$3.2247&0.3359&0.3812&$-$0.2968\\[1mm] 
 $m^{(11)}_{82i}$&$-$5.8157&0&1.4062&$-$3.9895&3.2850&3.6851&$-$0.1424&0.6492\\[1mm]
 $m^{(10)}_{88i}$&6.7441&0&0&0&0&0&0&0\\[1mm] 
 $m^{(11)}_{88i}$&$-$6.7441&0&0&0&0&0&0&0\\[1mm]
\hline
\end{tabular}
\caption[$m^{(1j)}_{kli}$ values]{Values of the $m^{(1j)}_{kli}$ relevant for $U^{(1)}_{kl}$ \cite{czakon}.\label{mnjNLO}}
\end{center}
\end{table}
\begin{table}[!ht]
\begin{center}
\begin{tabular}{|c|c|c|c|c|c|c|c|c|}\hline
 $i$&1&2&3&4&5&6&7&8\\\hline
 $m^{(20)}_{72i}$&$-$212.4136&167.6577&5.7465&$-$3.7262&28.8574&$-$2.1262&2.2903&0.1462\\[1mm] 
 $m^{(21)}_{72i}$&$-$74.7681&58.5182&$-$13.4731&3.0791&7.0744&2.8757&3.5962&$-$1.2982\\[1mm] 
 $m^{(22)}_{72i}$&31.4443&$-$18.1165&23.2117&13.2771&$-$19.8699&4.0279&$-$8.6259&2.6149\\[1mm]
 $m^{(20)}_{77i}$&0&44.4252&0&0&0&0&0&0\\[1mm] 
 $m^{(21)}_{77i}$&0&$-$61.0768&0&0&0&0&0&0\\[1mm] 
 $m^{(22)}_{77i}$&0&16.6516&0&0&0&0&0&0\\[1mm]
 $m^{(20)}_{78i}$&15.4051&$-$18.7662&0&0&0&0&0&0\\[1mm] 
 $m^{(21)}_{78i}$&$-$135.3141&146.6159&0&0&0&0&0&0\\[1mm] 
 $m^{(22)}_{78i}$&36.4636&$-$44.4043&0&0&0&0&0&0\\[1mm]
\hline
\end{tabular}
\caption[$m^{(2j)}_{kli}$ values]{Values of the $m^{(2j)}_{kli}$ relevant for $U^{(2)}_{kl}$ \cite{czakon}.\label{mnjNNLO}}
\end{center}
\end{table}
\subsection{RGE in the traditional operator basis}
\label{wilsonbasis2}%
The operators in the traditional basis can be expressed as \cite{buras}
\begin{eqnarray}
P_1 & =& (\bar{s}^\alpha_L \gamma_\mu c^\beta_L)(\bar{c}^\beta_L \gamma^\mu b^\alpha_L)\;, \nonumber\\  
P_2 & =& (\bar{s}^\alpha_L \gamma_\mu c^\alpha_L) (\bar{c}^\beta_L \gamma^\mu b^\beta_L) \;, \nonumber\\      
P_3 & =& (\bar{s}^\alpha_L \gamma_\mu b^\alpha_L) \sum_q (\bar{q}^\beta_L \gamma^\mu q^\beta_L) \;, \\
P_4 & =& (\bar{s}^\alpha_L \gamma_\mu b^\beta_L) \sum_q (\bar{q}^\beta_L \gamma^\mu q^\alpha_L) \; , \nonumber\\
P_5 & =& (\bar{s}^\alpha_L \gamma_\mu b^\alpha_L)\sum_q (\bar{q}^\beta_R \gamma^\mu q^\beta_R) \; , \nonumber\\    
P_6 & =& (\bar{s}^\alpha_L \gamma_\mu b^\beta_L) \sum_q (\bar{q}^\beta_R \gamma^\mu 
q^\alpha_R) \; .\nonumber
\end{eqnarray}
The operators $P_7$ and $P_8$ are identical to $O_7$ and $O_8$, and the $C_7$ and $C_8$ coefficients are therefore indistinguishable in both bases. Moreover, the LO Wilson coefficients also coincide in both bases. The effective coefficients are nevertheless different \cite{buras,gorbahn}:
\begin{equation}
C_i^{\rm eff}(\mu) = \left\{ \begin{array}{ll}
C_i(\mu), & \mbox{ for $i = 1, ..., 6$ ,} \\[1mm] 
C_7(\mu) + {\displaystyle\sum_{j=1}^6} y_j C_j(\mu), & \mbox{ for $i = 7$ ,} \\[1mm]
C_8(\mu) + {\displaystyle\sum_{j=1}^6} z_j C_j(\mu), & \mbox{ for $i = 8$ ,}
\end{array} \right.
\end{equation}
where $\vec{y} = (0, 0, 0, 0, -\frac{1}{3}, -1)$ and $\vec{z} = (0, 0, 0, 0, 1, 0)$. In the following, the Wilson coefficients at scale $\mu_W$ are the non-effective coefficients given in the previous section.\\
\\
The Wilson coefficients at LO relevant for the isospin asymmetry calculation are:
\begin{eqnarray}
\label{coeffs}
C_j^{(0)}(\mu_b) & = & C_2^{(0)}(\mu_W) \sum_{i=1}^8 k_{ji} \eta^{a_i} \qquad (j=1,\ldots,6) \;,\\
C_{7}^{(0)eff}(\mu_b) & = & \eta^\frac{16}{23} C_{7}^{(0)}(\mu_W) + \frac{8}{3} \left(\eta^\frac{14}{23} - \eta^\frac{16}{23}\right) C_{8}^{(0)}(\mu_W) + C_2^{(0)}(\mu_W)\sum_{i=1}^8 h_i \eta^{a_i}\;,\\
C_{8}^{(0)eff}(\mu_b) & = & \eta^\frac{14}{23} C_{8}^{(0)}(\mu_W) + C_2^{(0)}(\mu_W) \sum_{i=1}^8 \bar h_i \eta^{a_i}\;,
\end{eqnarray}
with
\begin{equation}
\eta = \frac{\alpha_s(\mu_W)}{\alpha_s(\mu_b)}\;,
\end{equation}
and where the numbers $k_{ji}$, $h_i$ and $\bar h_i$ are given in Table~\ref{traditionalcoef} and the $a_i$ in Table~\ref{tabai}.
\\
\begin{table}[p]
\begin{center}
\begin{tabular}{|c|c|c|c|c|c|c|c|c|}
\hline
$i$ & 1 & 2 & 3 & 4 & 5 & 6 & 7 & 8 \\[1mm]
\hline
$h_i $&$ 2.2996 $&$ - 1.0880 $&$ -3/7 $&$ -1/14 $&$ -0.6494 $&$ -0.0380 $&$ -0.0185 $&$ -0.0057 $\\[1mm]
$\bar h_i $&$ 0.8623 $&$ 0 $&$ 0 $&$ 0 $&$ -0.9135 $&$ 0.0873 $&$ -0.0571 $&$ 0.0209 $\\[1mm]
\hline
$k_{1i} $& $0$ & $0$&$ 1/2 $&$ -1/2 $&$0 $&$ 0 $&$ 0 $&$ 0 $\\[1mm]
$e_{1i} $& $0$ & $0$&$ 0 $&$ 0 $&$0 $&$ 0 $&$ 0 $&$ 0 $\\[1mm]
$f_{1i} $& $0$ & $0$&$ 0.8136 $&$ 0.7142 $&$0 $&$ 0 $&$ 0 $&$ 0 $\\[1mm]
$g_{1i} $& $0$ & $0$&$ 1.0197 $&$ 2.9524 $&$0 $&$ 0 $&$ 0 $&$ 0 $\\[1mm]
\hline
$k_{2i} $& $0$ & $0$& $ 1/2 $&$ 1/2 $&$0 $&$ 0 $&$ 0 $&$ 0 $\\[1mm]
$e_{2i} $& $0$ & $0$&$ 0 $&$ 0 $&$0 $&$ 0 $&$ 0 $&$ 0 $\\[1mm]
$f_{2i} $& $0$ & $0$&$ 0.8136 $&$ - 0.7142 $&$0 $&$ 0 $&$ 0 $&$ 0 $\\[1mm]
$g_{2i} $& $0$ & $0$&$ 1.0197 $&$ - 2.9524 $&$0 $&$ 0 $&$ 0 $&$ 0 $\\[1mm]
\hline
$k_{3i} $& $0$ & $0$& $ -1/14 $&$ 1/6 $&$0.0510 $&$ - 0.1403 $&$ - 0.0113 $&$ 0.0054 $\\[1mm]
$e_{3i} $& $0$ & $0$&$ 0 $&$ 0 $&$0.1494 $&$ -0.3726 $&$ 0.0738 $&$ -0.0173 $\\[1mm]
$f_{3i} $& $0$ & $0$&$ -0.0766 $&$ - 0.1455 $&$-0.8848 $&$ 0.4137 $&$ -0.0114 $&$ 0.1722 $\\[1mm]
$g_{3i} $& $0$ & $0$&$ -0.1457 $&$ - 0.9841 $&$0.2303 $&$ 1.4672 $&$ 0.0971 $&$ -0.0213 $\\[1mm]
\hline
$k_{4i} $& $0$ & $0$& $ - 1/14 $&$ -1/6 $&$0.0984 $&$ 0.1214 $&$ 0.0156 $&$ 0.0026 $\\[1mm]
$e_{4i} $& $0$ & $0$&$ 0 $&$ 0 $&$0.2885 $&$ 0.3224 $&$ -0.1025 $&$ -0.0084 $\\[1mm]
$f_{4i} $& $0$ & $0$&$ -0.2353 $&$ - 0.0397 $&$0.4920 $&$ -0.2758 $&$ 0.0019 $&$-0.1449 $\\[1mm]
$g_{4i} $& $0$ & $0$&$ -0.1457 $&$ 0.9841 $&$0.4447 $&$ -1.2696 $&$ -0.1349 $&$ -0.0104 $\\[1mm]
\hline
$k_{5i} $& $0$ & $0$& $ 0 $&$  0 $&$- 0.0397 $&$ 0.0117 $&$ - 0.0025 $&$ 0.0304 $\\[1mm]
$e_{5i} $& $0$ & $0$&$ 0 $&$ 0 $&$-0.1163 $&$ 0.0310 $&$ 0.0162 $&$ -0.0975 $\\[1mm]
$f_{5i} $& $0$ & $0$&$ 0.0397 $&$  0.0926 $&$0.7342 $&$ -0.1262 $&$ -0.1209 $&$ -0.1085 $\\[1mm]
$g_{5i} $& $0$ & $0$&$ 0 $&$ 0 $&$-0.1792 $&$ -0.1221 $&$ 0.0213 $&$ -0.1197 $\\[1mm]
\hline
$k_{6i} $& $0$ & $0$&$ 0 $&$  0 $&$0.0335 $&$ 0.0239 $&$ - 0.0462 $&$ -0.0112 $\\[1mm]
$e_{6i} $& $0$ & $0$&$ 0 $&$ 0 $&$0.0982 $&$ 0.0634 $&$ 0.3026 $&$ 0.0358 $\\[1mm]
$f_{6i} $& $0$ & $0$&$ -0.1191 $&$ - 0.2778 $&$-0.5544 $&$ 0.1915 $&$ -0.2744 $&$ 0.3568 $\\[1mm]
$g_{6i} $& $0$ & $0$&$ 0 $&$ 0 $&$0.1513 $&$ -0.2497 $&$ 0.3983 $&$ 0.0440 $\\[1mm]
\hline
&&&&&&&&\\[-4mm]
$e_{7i}$ &$\frac{4661194}{816831}$&$ -\frac{8516}{2217}$ &$  0$ &$  0$ & $ -1.9043$  & $  -0.1008$ & $ 0.1216$  &$ 0.0183$\\[1mm]
$f_{7i}$ & $-17.3023$ & $8.5027 $ & $ 4.5508$  & $ 0.7519$& $  2.0040 $ & $  0.7476$  &$ -0.5385$  & $ 0.0914$\\[1mm]
$g_{7i}$ & $14.8088$ &  $ -10.8090$  &$ -0.8740$  & $ 0.4218$ & $  -2.9347$   & $ 0.3971$  & $ 0.1600$  & $ 0.0225$ \\[1mm]
$l_i$ & $0.5784$ &  $ -0.3921$  &$ -0.1429$  & $ 0.0476$ & $  -0.1275$   & $ 0.0317$  & $ 0.0078$  & $ -0.0031$ \\[1mm]
\hline
\end{tabular}
\end{center}
\caption[Useful numbers]{Useful numbers \cite{buras}.\label{traditionalcoef}}
\end{table}%
\\
At NLO, the relevant coefficients read, for $j=1,\cdots,6$
\begin{equation}
C_j^{(1)}(\mu_b)=\sum_{i=3}^8 \left[ e_{ji} \frac{C_{1}^{(1)}(\mu_W)}{15} \eta + C_{2}^{(0)}(\mu_W) f_{ji} + \frac{C_{1}^{(1)}(\mu_W)}{15} g_{ji}\eta\right] \eta^{a_i} \;,
\end{equation}
and
\begin{eqnarray}
C^{(1)eff}_7(\mu_b) &=& \eta^{\frac{39}{23}} C^{(1)eff}_7(\mu_W) + \frac{8}{3} \left( \eta^{\frac{37}{23}} - \eta^{\frac{39}{23}} \right) C^{(1)eff}_8(\mu_W) \\ 
&&+ \sum_{i=1}^8 \left(e_{7i} C_4^{(1)}(\mu_W) \eta + f_{7i} C_2^{(0)}(\mu_W) + g_{7i} \frac{C_1^{(1)}(\mu_W)}{15} \eta\right) \eta^{a_i} \nonumber\\
&&+ \left( \frac{297664}{14283} \eta^{\frac{16}{23}}-\frac{7164416}{357075} \eta^{\frac{14}{23}} 
+\frac{256868}{14283} \eta^{\frac{37}{23}} -\frac{6698884}{357075} \eta^{\frac{39}{23}} \right) C_8^{(0)}(\mu_W) \nonumber \\ 
&&+\frac{37208}{4761} \left( \eta^{\frac{39}{23}} - \eta^{\frac{16}{23}} \right) C_7^{(0)}(\mu_W) +\Delta C^{(1)}_7(\mu_b)\;, \nonumber
\end{eqnarray}
where
\begin{equation}
\Delta C^{(1)}_7(\mu_b)=\sum_{i=1}^8 \left(\frac{2}{3}e_{7i}  + 6 l_i \right) \eta^{a_i+1}\ln\frac{\mu_W^2}{M_W^2}\;.
\end{equation}
The numbers $e_{ji}$, $f_{ji}$, $g_{ji}$ and $l_i$ appearing in the above equations are given in Table \ref{traditionalcoef}.
\append{Calculation of flavor observables}%
We provide here the detailed calculation of all the implemented observables for reference. The meson masses, lifetimes and form factors, as well as the CKM matrix elements which appear in the following are given in Appendix~\ref{parameters}. Finally, in Appendix~\ref{constraints}, we provide also some suggested limits for the observables which can be used in order to constrain the implemented models.

\subsection{Branching ratio of $B \to X_s \gamma$}
\label{bsgamma}
The decay $B \to X_s \gamma$ proceeds through electromagnetic penguin loops, involving $W$ boson in the Standard Model, in addition to charged Higgs boson, chargino, neutralino and gluino loops in supersymmetric models. The contribution of neutralino and gluino loops is negligible in minimal flavor violating scenarios.\\
The branching ratio of $B \to X_s \gamma$ can be written as \cite{misiak2}
\begin{equation}
\rm{BR}(\bar{B} \to X_s \gamma)= \mathrm{BR}(\bar{B} \to X_c e \bar{\nu})_{\rm exp} \left| \frac{ V^*_{ts} V_{tb}}{V_{cb}} \right|^2 \frac{6 \alpha}{\pi C} \left[ P(E_0) + N(E_0) + \epsilon_{em} \right] \;,
\end{equation}
with
\begin{equation}
C = \left| \frac{V_{ub}}{V_{cb}} \right|^2 \frac{\Gamma[\bar{B} \to X_c e \bar{\nu}]}{\Gamma[\bar{B} \to X_u e \bar{\nu}]} \;.
\end{equation}
$P(E_0)$ and $N(E_0)$ denote respectively the perturbative and non perturbative contributions, where $E_0$ is a cut on the photon energy. The numerical values of $C$, ${\rm BR}(\bar{B} \to X_c e \bar{\nu})_{\rm exp}$ and $E_0$ are given in Appendix~\ref{parameters}.\\
\\
Following \cite{misiak2}, we can expand $P(E_0)$ as
\begin{eqnarray}
P(E_0) &=& P^{(0)}(\mu_b) + \left(\frac{\alpha_s(\mu_b)}{4\pi}\right) \left[ P_1^{(1)}(\mu_b) + P_2^{(1)}(E_0,\mu_b) \right]\\ 
&+& \left(\frac{\alpha_s(\mu_b)}{4\pi}\right)^2 \left[ P_1^{(2)}(\mu_b) + P_2^{(2)}(E_0,\mu_b) + P_3^{(2)}(E_0,\mu_b) \right]
+ {\cal O}\left(\alpha_s^3(\mu_b)\right)\;,\nonumber
\end{eqnarray}
where
\begin{eqnarray}
P^{(0)}(\mu_b)   &=&   \left( C_7^{(0)\rm eff}(\mu_b)\right)^2\;, \nonumber\\
P_1^{(1)}(\mu_b) &=& 2 C_7^{(0)\rm eff}(\mu_b) C_7^{(1)\rm eff}(\mu_b)\;, \nonumber\\
P_1^{(2)}(\mu_b) &=&   \left( C_7^{(1)\rm eff}(\mu_b)\right)^2 + 2 C_7^{(0)\rm eff}(\mu_b) C_7^{(2)\rm eff}(\mu_b)\;,
\end{eqnarray}
and
\begin{eqnarray}
P_2^{(1)}(E_0,\mu_b) &=& \sum_{i,j=1}^{8} C_i^{(0)\rm eff}(\mu_b)\; C_j^{(0)\rm eff}(\mu_b) \; K_{ij}^{(1)}(E_0,\mu_b)\;,\\[2mm] 
P_3^{(2)}(E_0,\mu_b) &=& 2 \sum_{i,j=1}^{8} C_i^{(0)\rm eff}(\mu_b)\; C_j^{(1)\rm eff}(\mu_b) \; K_{ij}^{(1)}(E_0,\mu_b)\;,
\end{eqnarray}
where the $K_{ij}^{(1)}$ can be written in the form:
\begin{eqnarray}
K_{i7}^{(1)} &=&  {\rm Re}\,r_i^{(1)} - \frac{1}{2} \gamma^{(0)\rm eff}_{i7} \ln \left( \frac{\mu_b}{m_b^{1S}} \right)^2 + 2 \phi_{i7}^{(1)}(\delta)\;,
\hspace{1cm} {\rm for~} i \leq 6\;, \label{K1i7}\\
K_{77}^{(1)} &=& -\frac{182}{9} + \frac{8}{9}\pi^2 - \gamma^{(0)\rm eff}_{77} \ln \left( \frac{\mu_b}{m_b^{1S}} \right)^2 + 4\,\phi_{77}^{(1)}(\delta)\;, \\
K_{78}^{(1)} &=&  \frac{44}{9} - \frac{8}{27}\pi^2 -\frac{1}{2} \gamma^{(0)\rm eff}_{87} \ln \left( \frac{\mu_b}{m_b^{1S}} \right)^2  + 2\,\phi_{78}^{(1)}(\delta)\;,\\
K_{ij}^{(1)} &=&  2 (1 + \delta_{ij}) \phi_{ij}^{(1)}(\delta)\;,\hspace{34mm} {\rm for~} i,j \neq 7\;.
\end{eqnarray}
\\
The matrix $\hat{\gamma}^{(0)\rm eff}$ and the quantities $r_i^{(1)}$ are given by \cite{buras2}:\\
\begin{equation}
\begin{array}{rcl}
r_1^{(1)} &=& \dfrac{833}{729} - \dfrac{1}{3} [a(z)+b(z)] + \dfrac{40}{243} i \pi\;,\\[6mm]
r_2^{(1)} &=& -\dfrac{1666}{243} + 2[a(z)+b(z)] - \dfrac{80}{81} i \pi\;,\\[6mm]
r_3^{(1)} &=& \dfrac{2392}{243} + \dfrac{8\pi}{3\sqrt{3}} + \dfrac{32}{9} X_b - a(1) + 2 b(1) + \dfrac{56}{81} i \pi\;,\\[6mm]
r_4^{(1)} &=& -\dfrac{761}{729} - \dfrac{4\pi}{9\sqrt{3}} - \dfrac{16}{27} X_b + \dfrac{1}{6} a(1) + \dfrac{5}{3} b(1) + 2 b(z) - \dfrac{148}{243} i \pi\;,\\[6mm]
r_5^{(1)} &=& \dfrac{56680}{243} + \dfrac{32\pi}{3\sqrt{3}} + \dfrac{128}{9} X_b - 16 a(1) + 32 b(1) 
         + \dfrac{896}{81} i \pi\;,\\[6mm]
r_6^{(1)} &=& \dfrac{5710}{729} - \dfrac{16\pi}{9\sqrt{3}} - \dfrac{64}{27} X_b - \dfrac{10}{3} a(1) + \dfrac{44}{3} b(1) + 12 a(z) + 20 b(z) - \dfrac{2296}{243} i \pi\;,\\[6mm]
r_8^{(1)} &=& \dfrac{44}{9} - \dfrac{8}{27} \pi^2 + \dfrac{8}{9} i \pi\;,
\end{array}
\end{equation}\\
where
\begin{equation}
z= \left(\frac{m_c(\mu_c)}{m_b^{1S}}\right)^2 \;,
\end{equation}
and the constant $X_b$ can be written as\\
\begin{equation}
X_b = \int_0^1 dx \int_0^1 dy \int_0^1 dv \; x y \ln\bigl( v + x (1-x) (1-v) (1-v + v y)\bigr) \simeq -0.1684 \;,
\end{equation}
and
\begin{equation}
\hat{\gamma}^{(0)\mathrm{eff}} = \left[
\begin{array}{cccccccc}
\vspace{0.2cm}
-4 & \dfrac{8}{3} &       0     &   -\dfrac{2}{9} &      0    &     0     & -\dfrac{208}{243} &  \dfrac{173}{162} \\ 
\vspace{0.2cm}
12 &     0    &       0     &    \dfrac{4}{3} &      0    &     0     &   \dfrac{416}{81} &    \dfrac{70}{27} \\ 
\vspace{0.2cm}
 0 &     0    &       0     &  -\dfrac{52}{3} &      0    &     2     &  -\dfrac{176}{81} &    \dfrac{14}{27} \\ 
\vspace{0.2cm}
 0 &     0    &  -\dfrac{40}{9} & -\dfrac{100}{9} &  \dfrac{4}{9} &  \dfrac{5}{6} & -\dfrac{152}{243} & -\dfrac{587}{162} \\ 
\vspace{0.2cm}
 0 &     0    &       0     & -\dfrac{256}{3} &      0    &    20     & -\dfrac{6272}{81} &  \dfrac{6596}{27} \\ 
\vspace{0.2cm}
 0 &     0    & -\dfrac{256}{9} &   \dfrac{56}{9} & \dfrac{40}{9} & -\dfrac{2}{3} & \dfrac{4624}{243} &  \dfrac{4772}{81} \\ 
\vspace{0.2cm}
 0 &     0    &       0     &       0     &      0    &     0     &     \dfrac{32}{3} &        0      \\ 
\vspace{0.2cm}
 0 &     0    &       0     &       0     &      0    &     0     &    -\dfrac{32}{9} &     \dfrac{28}{3} \\
\end{array} \right]\;.
\end{equation}\\
The small-$m_c$ expansions of $a(z)$ and $b(z)$ up to ${\cal O}(z^4)$ read\\
\begin{eqnarray}
a(z)  &=& \frac{16}{9} \left\{ 
\left[ \frac{5}{2} -\frac{\pi^2}{3} -3 \zeta(3) + \left( \frac{5}{2} - \frac{3\pi^2}{4} \right) \ln z\,
+ \frac{1}{4} (\ln z)^2 + \frac{1}{12} (\ln z)^3 \right] z \right.\\
&& + \left(\frac{7}{4} +\frac{2\pi^2}{3} -\frac{\pi^2}{2} \ln z\,  -\frac{1}{4} (\ln z)^2 +\frac{1}{12} (\ln z)^3 \right) z^2 
+ \left[ -\frac{7}{6} -\frac{\pi^2}{4} + 2\ln z\, - \frac{3}{4}(\ln z)^2 \right] z^3\nonumber\\
&& + \left( \frac{457}{216} - \frac{5\pi^2}{18} -\frac{1}{72} \ln z\, -\frac{5}{6} (\ln z)^2 \right) z^4 + i \pi \left[ \left( 4 -\frac{\pi^2}{3} + \ln z\, + (\ln z)^2 \right) \frac{z}{2} \right. \nonumber\\
&&\left.\left. + \left( \frac{1}{2} -\frac{\pi^2}{6} - \ln z\,\! + \frac{1}{2} (\ln z)^2 \right) z^2 + z^3 
+ \frac{5}{9} z^4  \right]\right\} + {\cal O}(z^5 (\ln z)^2)\;,\nonumber\\[2mm]
b(z) &=& -\frac{8}{9} \left\{ \left( -3 +\frac{\pi^2}{6} - \ln z\, \right) z - \frac{2\pi^2}{3} z^{3/2}
+ \left( \frac{1}{2} + \pi^2 - 2 \ln z\, - \frac{1}{2} (\ln z)^2 \right) z^2  \right.\\
&&+ \left. \left(-\frac{25}{12} -\frac{1}{9} \pi^2 - \frac{19}{18} \ln z\, + 2 (\ln z)^2 \right) z^3 
+ \left( -\frac{1376}{225} + \frac{137}{30} \ln z\, + 2 (\ln z)^2 + \frac{2\pi^2}{3} \right) z^4
\right.\nonumber\\
&&+ \left. i \pi \left[ -z + (1-2 \ln z\,) z^2  + \left(-\frac{10}{9} + \frac{4}{3} \ln z\, \right) z^3 + z^4\right]\right\} + {\cal O}(z^5 (\ln z)^2)\;,\nonumber
\end{eqnarray}
where $\zeta(3)$ is the Riemann zeta function given in Eq. (\ref{zeta3}). Defining 
\begin{equation}
\delta \equiv 1 - 2E_0/m_b^{1S}\;,
\end{equation}
we can write
\begin{eqnarray}
\phi_{47}^{(1)}(\delta) &=& -\frac{1}{54} \delta \left( 1 - \delta + \frac{1}{3} \delta^2 \right)~
+~ \frac{1}{12}~ \lim_{m_c \to m_b} \phi_{27}^{(1)}(\delta)\;,\\
\phi_{48}^{(1)}(\delta) &=& -\frac{1}{3} \phi_{47}^{(1)}(\delta)\;.
\end{eqnarray}
\\
The remaining NNLO correction to $P(E_0)$ is composed of two parts:
\begin{equation}
P_2^{(2)}(E_0,\mu_b) = P_2^{(2)\beta_0}(E_0,\mu_b) +P_2^{(2)\rm rem}(E_0,\mu_b)\;,
\end{equation}
with\\
\begin{equation}
P_2^{(2)\beta_0}(E_0,\mu_b) \simeq \sum_{i,j=1,2,7,8} C_i^{(0)\rm eff}(\mu_b)\; C_j^{(0)\rm eff}(\mu_b)\; K_{ij}^{(2)\beta_0}(E_0,\mu_b)\;,
\end{equation}\\
where the contribution for $i,j=3,4,5,6$ are neglected, and\\
\begin{eqnarray}
K_{27}^{(2)\beta_0} &=& \beta_0\, 
{\rm Re}\left\{ -\frac{3}{2} r_2^{(2)}(z) +2\left[a(z)+b(z)-\frac{290}{81}\right] L_b 
- \frac{100}{81} L_b^2 \right\}\\
&& \hspace*{0.5cm} + 2 \phi^{(2)\beta_0}_{27}(\delta)\;,\nonumber\\[3mm]
K_{17}^{(2)\beta_0} &=& -\frac{1}{6} K_{27}^{(2)\beta_0},\\[5mm]
K_{77}^{(2)\beta_0} &=& \beta_0 \left\{ -\frac{3803}{54} -\frac{46}{27} \pi^2 +\frac{80}{3} \zeta(3) 
+ \left( \frac{8}{9} \pi^2 - \frac{98}{3} \right) L_b -\frac{16}{3} L_b^2 \right\}\\
&& \hspace*{0.5cm} + 4\phi_{77}^{(2)\beta_0}(\delta)\;,\nonumber\\[3mm]
K_{78}^{(2)\beta_0} &=& \beta_0 \left\{ \frac{1256}{81} -\frac{64}{81} \pi^2 - \frac{32}{9} \zeta(3)
+ \left( \frac{188}{27} -\frac{8}{27} \pi^2 \right) L_b + \frac{8}{9} L_b^2  \right\}\\
&& \hspace*{0.5cm} + 2\phi_{78}^{(2)\beta_0}(\delta)\;,\nonumber\\[3mm]
K_{ij}^{(2)\beta_0} &=&
2(1+\delta_{ij})\phi^{(2)\beta_0}_{ij}(\delta)\;,\hspace{0.5cm} {\rm for}~i,j \neq 7\;.
\end{eqnarray}
The small-$m_c$ expansion of ${\rm Re}\,r_2^{(2)}(z)$ up to ${\cal O}(z^4)$ leads to
\begin{eqnarray}
{\rm Re}\,r_2^{(2)}(z) &=& \frac{67454}{6561} - \frac{124\pi^2}{729} -\frac{4}{1215} \left(11280 - 1520\pi^2 - 171\pi^4 - 5760 \zeta(3) \right.\nonumber\\
&&+ 6840\ln z - 1440\pi^2\ln z\, - 2520\zeta(3)\ln z\, + 120(\ln z)^2 + 100 (\ln z)^3  \nonumber\\
&& \left.- 30 (\ln z)^4\right) z -\frac{64\pi^2}{243} \left( 43 - 12 \ln 2 -3\ln z\,\right) z^{3/2} 
- \frac{2}{1215} \left(11475 - 380\pi^2 \right. \nonumber\\
&&  + 96\pi^4 +7200 \zeta(3) - 1110\ln z\, - 1560\pi^2\ln z\, + 1440 \zeta(3)\ln z\, +990(\ln z)^2 
\nonumber\\
&& \left. +260(\ln z)^3 -60(\ln z)^4\right)z^2 + \frac{2240\pi^2}{243} z^{5/2} - \frac{2}{2187} \left(62471 -2424\pi^2 \right. \nonumber\\
&& \left. - 33264\zeta(3) - 19494\ln z\, - 504\pi^2\ln z\, - 5184(\ln z)^2 +2160(\ln z)^3\right) z^3
\nonumber\\
&& - \frac{2464}{6075} \pi^2 z^{7/2} + \left( - \frac{15103841}{546750} + \frac{7912}{3645} \pi^2 + \frac{2368}{81} \zeta(3) + \frac{147038}{6075} \ln z \right.\nonumber\\
&&\left. + \frac{352}{243} \pi^2 \ln z\, + \frac{88}{243} (\ln z)^2 - \frac{512}{243} (\ln z)^3 \right) z^4 + {\cal O}(z^{9/2} (\ln z)^4)\;.
\end{eqnarray}
where
\begin{equation}
L_b = \ln \left( \frac{\mu_b}{m_b^{1S}} \right)^2\;. 
\end{equation}
The function $\phi_{77}^{(2)\beta_0}(\delta)$ takes the form\\
\begin{equation}
\phi_{77}^{(2)\beta_0}(\delta) = \beta_0 \left[ \phi_{77}^{(1)}(\delta) L_b + 4 \int_0^{1-\delta} dx\; F^{(2,nf)} \right]\;,
\end{equation}
where $F^{(2,nf)}$ is given by \cite{melnikov}
\begin{eqnarray}
&& F^{(2,{\rm nf})} = S_{\rm nf}\delta(1-z)
-\frac{1}{2}\left[\frac{\ln^2(1-z)}{1-z}\right]_+-\frac{13}{36}\left[\frac{\ln(1-z)}{1-z}\right]_++\left(-\frac{\pi^2}{18} +\frac{85}{72}\right)\left[\frac{1}{1-z}\right]_+ \nonumber\\
&&+\frac{z^2-3}{6(z-1)}{\rm Li}_2(1-z)+ \frac{z^2-3}{6(z-1)}\ln(1-z)\ln(z)-\frac{1+z}{4}\ln^2(1-z) \nonumber\\
&& - \frac{6z^3-25z^2-z-18}{36z}\ln(1-z) -(1+z)\frac{\pi^2}{36} +\frac{-49+38z^2-55z}{72}\;,
\end{eqnarray}\\
where the $[\ln^n(1-x)/(1-x)]_+$ are the plus-distributions defined in the standard way, and $S_{\rm nf}= 49/24+\pi^2/8-2\zeta(3)/3 \approx 2.474$. The remaining $\phi_{ij}^{(2)\beta_0}(\delta)$ functions are neglected.\\
\\
The $P_2^{(2)\rm rem}$ term is more difficult to calculate, as its analytic expression is only known in the limit $m_c \gg m_b/2$. In this limit, we have\\
\begin{equation}
P_2^{(2)\rm rem}(E_0,\mu_b) \simeq \sum_{i,j=1,2,7,8} C_i^{(0)\rm eff}(\mu_b)\; C_j^{(0)\rm eff}(\mu_b)\; K_{ij}^{(2)\rm rem}(E_0,\mu_b)\;, \label{P2rem}
\end{equation}
where\\
\begin{eqnarray}
K_{22}^{(2)\rm rem} &=& 36\,K_{11}^{(2)\rm rem} + {\cal O}\left(\frac{1}{z}\right) ~=~
-6\,K_{12}^{(2)\rm rem} + {\cal O}\left(\frac{1}{z}\right)  \\
&=& \left(K_{27}^{(1)}\right)^2 + {\cal O}\left(\frac{1}{z}\right) ~=~ \left[ \frac{218}{243} - \frac{208}{81} L_D\right]^2 + {\cal O}\left(\frac{1}{z}\right)\;,\nonumber\\[2mm]
K_{27}^{(2)\rm rem} &=& K_{27}^{(1)} K_{77}^{(1)} + \left( \frac{127}{324} - \frac{35}{27} L_D \right) K_{78}^{(1)} + \frac{2}{3} (1 - L_D ) K_{47}^{(1)\rm rem} \\ 
&& - \frac{4736}{729} L_D^2 + \frac{1150}{729} L_D - \frac{1617980}{19683} + \frac{20060}{243} \zeta(3) 
+ \frac{1664}{81} L_c + {\cal O}\left(\frac{1}{z}\right)\;,\nonumber\\[2mm]
K_{28}^{(2)\rm rem} &=& K_{27}^{(1)} K_{78}^{(1)} + \left( \frac{127}{324} - \frac{35}{27} L_D \right) K_{88}^{(1)} + \frac{2}{3} (1 - L_D ) K_{48}^{(1)}  + {\cal O}\left(\frac{1}{z}\right)\;,
\end{eqnarray}
\ \\[-1cm]
\begin{eqnarray}
K_{17}^{(2)\rm rem} &=&\!\! -\frac{1}{6} K_{27}^{(2)\rm rem} + \left(\frac{5}{16}-\frac{3}{4}L_D\right)K_{78}^{(1)} -\frac{1237}{729} +\frac{232}{27}\zeta(3)+\frac{70}{27}L_D^2\\
&& -\frac{20}{27} L_D + {\cal O}\left(\frac{1}{z}\right)\;,\nonumber\\
K_{18}^{(2)\rm rem} &=&\!\! -\frac{1}{6} K_{28}^{(2)\rm rem} + \left(\frac{5}{16}-\frac{3}{4}L_D\right)K_{88}^{(1)} + {\cal O}\left(\frac{1}{z}\right)\;,\\
K_{77}^{(2)\rm rem} &=& \left( K_{77}^{(1)} - 4 \phi_{77}^{(1)}(\delta)+\frac{2}{3} \ln z\, \right) K_{77}^{(1)} -\frac{32}{9} L_D^2 + \frac{224}{27} L_D -\frac{628487}{729}\\
&&  - \frac{628}{405} \pi^4 + \frac{31823}{729} \pi^2 + \frac{428}{27} \pi^2 \ln 2 + \frac{26590}{81} \zeta(3) - \frac{160}{3} L_b^2  \nonumber\\
&& - \frac{2720}{9} L_b + \frac{256}{27} \pi^2 L_b + \frac{512}{27}\pi\alpha_\Upsilon\; +\; 4 \phi_{77}^{(2)\rm rem}(\delta)\;
+ {\cal O}\left(\frac{1}{z}\right)\;, \nonumber\\[2mm]
K_{78}^{(2)\rm rem} &=&  \left(-\frac{50}{3} + \frac{8}{3} \pi^2 - \frac{2}{3} L_D \right) K_{78}^{(1)} 
+ \frac{16}{27} L_D^2 -\frac{112}{81} L_D + \frac{364}{243} + {\cal O}\left(\frac{1}{z}\right)\;, \\[2mm]
K_{88}^{(2)\rm rem} &=& \left(-\frac{50}{3} + \frac{8}{3} \pi^2 - \frac{2}{3} L_D \right) K_{88}^{(1)} + {\cal O}\left(\frac{1}{z}\right)\;,
\end{eqnarray}
with
\begin{eqnarray}
K_{47}^{(1)\rm rem} &=& K_{47}^{(1)} - \beta_0 \left( \frac{26}{81} - \frac{4}{27} L_b \right),\\[2mm]
L_c &=& \ln \left( \frac{\mu_c}{m_c(\mu_c)}\right)^2\;, 
\end{eqnarray}
and
\begin{equation}
L_D \equiv L_b - \ln z\, = \ln \left( \frac{\mu_b}{m_c(\mu_c)}\right)^2\;.
\end{equation}
The function $\phi_{77}^{(2)\rm rem}(\delta)$ reads
\begin{eqnarray}
\phi_{77}^{(2)\rm rem}(\delta) &=& -4\int_0^{1-\delta} dx\; \left[ \frac{16}{9} F^{(2,a)} + 4 F^{(2,na)} +\frac{29}{3} F^{(2,nf)} \right] \label{phi77rem}\\
&& -\frac{8\pi\,\alpha_\Upsilon }{27\,\delta} \left[ 2 \delta \ln^2 \delta 
+ \left(4 +\! 7 \delta -\! 2 \delta^2 +\! \delta^3\right)\ln\delta+ 7 - \frac{8}{3}\delta - 7\delta^2 + 4\delta^3 - \frac{4}{3}\delta^4 \right]\nonumber \;,
\end{eqnarray}
where $F^{(2,a)}$ and $F^{(2,na)}$ are given by \cite{melnikov}:\\
\begin{eqnarray}
&& F^{(2,{\rm a})} =
S_{\rm a}\delta(1-z)+\frac{1}{2}\left[\frac{\ln^3(1-z)}{1-z}\right]_++\frac{21}{8}\left[\frac{\ln^2(1-z)}{1-z}\right]_++\left(-\frac{\pi^2}{6}+\frac{271}
{48}\right)\left[\frac{\ln(1-z)}{1-z}\right]_+
\nonumber\\
&&+\left(\frac{425}{96}-\frac{\pi^2}{6}-\frac{\zeta(3)}{2}
\right)\left[\frac{1}{1-z}\right]_++\frac{4z-4z^2+1+z^3}{2(z-1)} \left[
{\rm Li}_3\left(\frac{z}{2-z}\right) - {\rm Li}_3\left(-\frac{z}{2-z}\right) \right.
\nonumber\\
&& \left.  -
2{\rm Li}_3\left(\frac{1}{2-z}\right) +\frac{\zeta(3)}{4} \right] -2(z-1)^2{\rm Li}_3(z-1) +\left[\frac{z^3-2z^2+2z-3}{2(z-1)}\ln(1-z)\right.\nonumber\\
&& \left. -\frac{-140z^4+219z^3-124z^2+28z+27z^5+9z^6+z^8-6z^7-6}{12z(z-1)^3}\right]{\rm Li}_2(z-1)\nonumber\\
&&+ \left[ \frac{2z^3-9z^2-2z+11}{4(z-1)}\ln(1-z)-\frac{-27z^2+8z^6-9+21z-3z^3+64z^4-
46z^5}{12z(z-1)^3}\right]{\rm Li}_2(1-z)\nonumber\\
&&-\frac{-17z^2+4z+4z^3+11}{4(z-1)}{\rm Li}_3(1
-z)-\frac{2z^3+13-9z^2}{4(z-1)}{\rm Li}_3(z)+\frac{4z-4z^2+1+z^3}{6(z-1)}\ln^3(2-z
)\nonumber\\
&&+\left[-\frac{-140z^4+219z^3-124z^2+
28z+27z^5+9z^6+z^8-6z^7-6}{12z(z-1)^3}\ln(1-z)\right.\nonumber\\
&&\left. -\frac{4z-4z^2+1+z^3}{2(z-1)}\ln^2(1-z)-\frac{4z-4z^2+1+z^3}{z-1}\frac{\pi^2}{12}\right]\ln(2-z)\nonumber\\
&& +\frac{z^3-2z^2+2z+1}{4z}\ln^3(1-z)+
\frac{z^5-3z^4+5z^3+7z^2+5z-9}{24z}\ln^2(1-z)\nonumber\\
&&+\left[-\frac{z^2+8z-11}{8(z-1)}\ln^2(1-z)-\frac{-27z^2+8z^6-9+21z-3z^3
+64z^4-46z^5}{12z(z-1)^3}\ln(1-z)\right]\ln(z)\nonumber\\
&&+\left[(-z^2+z-3)\frac{\pi^2}{12} -
\frac{4z^5+151z+2z^4-48z^2-41z^3-36}{48z(z-1)}\right]\ln(1-z)\nonumber\\
&&+\frac{z^3-11z^2-2z+18}{4(z-1)}\zeta(3)-\frac{8z^4-244z^3+ 175z^2+598z-569}{96(z-1)}\nonumber\\
&&-\frac{(z-2)(z^4-z^3-11z^2+13z+3)}{z}\frac{\pi^2}{72} \;,
\end{eqnarray}
and
\begin{eqnarray}
&& F^{(2,{\rm na})} =
S_{\rm na}\delta(1-z)+\frac{11}{8}\left[\frac{\ln^2(1-z)}{1-z}\right]_++\left(\frac{\pi^2}{12}+
\frac{95}{144}\right)\left[\frac{\ln(1-z)}{1-z}\right]_+\nonumber\\
&& +\left(\frac{\zeta(3)}{4}-\frac{905}{288}+\frac{17\pi^2}{72}\right)\left[\frac{1}{1-z}\right]_+
+(z-1)^2{\rm Li}_3(z-1) \nonumber\\
&&-\frac{4z-4z^2+1+z^3}{4(z-1)}\left[{\rm Li}_3\left(\frac{z}{2-z}\right) -{\rm Li}_3\left(- \frac{z}{2-z}\right)-2{\rm Li}_3\left(\frac{1}{2-z}\right) +\frac{\zeta(3)}{4}\right]  \nonumber\\
&&+\left[\frac{-140z^4+219z^3-124z^2+28z+27z^5+9z^6+z^8-6z^7-6}{24z(z-1)^3}\right.\nonumber\\
&&\left.-\frac{z^3-2z^2+2z-3}{4(z-1)}\ln(1-z)\right]{\rm Li}_2(z-1) + \left[\frac{(1+z)(2z^4-29z^3+73z^2-57z+15)}{24(z-1)^3} \right. \nonumber\\
&&\left.+\frac{z(3-z)}{4}\ln(1-z)\right]{\rm Li}_2(1-z)
+\frac{4z-4z^2+1+z^3}{4(z-1)}{\rm Li}_3(z)+\frac{(z-3)z}{2}{\rm Li}_3(1-z)\nonumber\\
&&-\frac{4z-4z^2+1+z^3}{12(z-1)}\ln^3(2-z) +\left[\frac{4z-4z^2+1+z^3}{4(z-1)}\ln^2(1-z) +\frac{4z-4z^2+1+z^3}{z-1}\frac{\pi^2}{24}\right.\nonumber\\
&& \left. +\frac{-140z^4+219z^3-124z^2+28z+27z^5+9z^6+z^8-6z^7-6}{24z(z-1)^3} \ln(1-z) \right]\ln(2-z) \nonumber\\
&&+\frac{(1+z)(2z^4-29z^3+73z^2-57z+15)}{24(z-1)^3}\ln(1-z)\ln(z)-\frac{(z-1)^2}{8}\ln^3(1-z)\nonumber\\
&&-\frac{(z+2)(z^3-5z^2+9z-35)}{48}\ln^2(1-z) \frac{z^5-3z^4-3z^3+34z^2-24z+3}{z}\frac{\pi^2}{144}
\nonumber\\
&& +\left[(z^2-z+3)\frac{\pi^2}{24} +\frac{6z^5+72-392z^3+51z^4+219z^2+92z}{144z( z-1)} \right]\ln(1-z)+
\nonumber\\
&&-\frac{z^3-10z^2+6z+7}{8(z-1)}\zeta(3)+
\frac{12z^4-754z^3+1191z^2+264z-761}{288(z-1)}\;,
\end{eqnarray}
where $S_{{\rm a}}= 1.216$, $S_{\rm na }= -4.795$ and ${\rm Li}_3(x) = \int \limits_{0}^{x} dy {\rm Li}_2(y)/y$.\\
\\
$\alpha_{\Upsilon}$ in Eq. (\ref{phi77rem}) is defined as
\begin{equation}
\alpha_{\Upsilon} \equiv \alpha_s^{(4)}(\mu=m_b^{1S}) \;\;.
\end{equation}
The missing $K_{ij}$ and $\phi_{ij}$ are neglected.\\
\\
To estimate the value of $P_2^{(2)\rm rem}$ for $m_c < m_b/2$, it is necessary to use an interpolation method. We consider the linear combination:
\begin{eqnarray}
P_2^{(2)\rm rem}(z) &=& x_1\, [|r_2^{(1)}(z)|^2-|r_2^{(1)}(0)|^2] + x_2\, {\rm Re}\,[r_2^{(2)}(z)-r_2^{(2)}(0)]\nonumber\\
&& + x_3\, {\rm Re}\,[r_2^{(1)}(z)-r_2^{(2)}(0)] + x_4\, z\frac{d}{dz}{\rm Re}\,r_2^{(1)}(z) + x_5 \;. \label{interpol}
\end{eqnarray}
In this equation, $x_1, \cdots, x_5$ are constants to be determined. Noting that $P_2^{(2)\rm rem}(0)=x_5$, we make two different assumptions:\\
(a) $x_5 = 0$,\\
(b) $x_5 = - P_1^{(2)}(z=0) - P_3^{(2)}(z=0)$.\\
\\
To determine the four other $x_i$, we impose that for $z \gg 1$, Eqs. (\ref{P2rem}) and (\ref{interpol}) coincide, and by matching the different terms in both equations, the $x_i$ can be worked out. In particular, we can show that
\begin{equation}
x_1=\bigl(C_2^{(0)\rm eff}\bigr)^2 + \frac{1}{36} \bigl(C_1^{(0)\rm eff}\bigr)^2 - \frac{1}{3} \bigl(C_1^{(0)\rm eff}\bigr)\bigl(C_2^{(0)\rm eff}\bigr)\;,
\end{equation}
and
\begin{equation}
x_2=C_7^{(0)\rm eff}\left(\frac{4019}{486}C_1^{(0)\rm eff} -\frac{1184}{81} C_2^{(0)\rm eff} - 4 C_7^{(0)\rm eff}+\frac{4}{3} C_8^{(0)\rm eff}\right)\;.
\end{equation}
Finally, $x_3$ and $x_4$ can be determined by choosing two large values for $z$ and requiring matching between Eqs. (\ref{P2rem}) and (\ref{interpol}) for both values.
\\
The two possible determinations of $x_5$ lead to different results, and we therefore compute the branching ratio in both cases (a) and (b), and then we give the average value as output, as advised in \cite{misiak2}.\\
\\
The non-perturbative correction $N(E_0)$ reads \cite{gambino}
\begin{equation}
N(E_0) = -\frac{1}{18} \left( K_c^{(0)} + r K_t^{(0)} \right)
\left( \eta^{\frac{6}{23}} + \eta^{-\frac{12}{23}} \right) \; \frac{\lambda_2}{m_c^2} 
+ ...\;,
\end{equation}
where $r=\overline{m}_b(\mu_W)/m_b^{1S}$, $\lambda_2 \approx (m_{B^*}^2-m_B^2)/4$ is given in Appendix~\ref{parameters}, and
\begin{equation}
K_c^{(0)} = \sum_{i=1}^8 d_i \eta^{a_i} \label{Kc0}\;,
\end{equation}
with $d_i$ given in Table~\ref{tabdi}, and $a_i$ in Table~\ref{tabai}, 
\begin{table}[t]
\begin{tabular}{|c|c|c|c|c|c|c|c|c|}
\hline
$i$ & 1 & 2 & 3 & 4 & 5 & 6 & 7 & 8 \\ 
\hline
\ &&&&&&&& \\[-4mm]
$d_i$ &     1.4107 & $-$0.8380 & $-$0.4286 & $-$0.0714 
      &  $-$0.6494 & $-$0.0380 & $-$0.0185 & $-$0.0057 \\[1.5mm]
\hline
\end{tabular}
\caption[Useful numbers for $K_c^{(0)}$]{Useful numbers for $K_c^{(0)}$ \cite{gambino}.\label{tabdi}}
\end{table}%
and
\begin{equation}
K_t^{(0)} = \left(C_7(\mu_W) + \frac{23}{36}\right) \eta^{\frac{4}{23}} - \frac{8}{3} \left( C_8(\mu_W) + \frac{1}{3} \right) (\eta^{\frac{4}{23}} - \eta^{\frac{2}{23}}) \;. \label{Kt0}
\end{equation}
In Eqs. (\ref{Kc0}) and (\ref{Kt0}), $\eta=\alpha_s(\mu_W)/\alpha_s(\mu_b)$.\\
\\
The electromagnetic correction $\epsilon_{em}$ can be written as \cite{micromegas}:
\begin{equation}
\epsilon_{em}= \frac{\alpha}{\alpha_s(\mu_b)} \bigg( 2\,\left[ C_7^{({\rm em})}(\mu_b)\,C_7^{(0)}(\mu_b) \right] - k_{\rm SL}^{({\rm em})}(\mu_b)\,|C_7^{(0)}(\mu_b)|^2\bigg) \;,
\end{equation}
where
\begin{equation}
k_{\rm SL}^{({\rm em})}(\mu_b) = \frac{12}{23} \left( \eta^{-1} - 1 \right) = \frac{2\alpha_s(\mu_b)}{\pi} \ln\frac{\mu_W}{\mu_b} \;,
\end{equation}
and
\begin{eqnarray}
C_7^{({\rm em})}(\mu_b) &=& \left( \frac{32}{75}\,\eta^{-\frac{9}{23}} - \frac{40}{69}\,\eta^{-\frac{7}{23}} + \frac{88}{575}\,\eta^{\frac{16}{23}} \right) C_7^{(0)}(\mu_W)\\
&& + C_8^{({\rm em})}(\mu_b) C_8^{(0)}(\mu_W) + C_2^{({\rm em})}(\mu_b) \nonumber \;,
\end{eqnarray}
with
\begin{eqnarray}
C_8^{({\rm em})}(\mu_b) &=& -\frac{32}{575}\,\eta^{-\frac{9}{23}} + \frac{32}{1449}\,\eta^{-\frac{7}{23}}
+ \frac{640}{1449}\,\eta^{\frac{14}{23}}  - \frac{704}{1725}\,\eta^{\frac{16}{23}}\;,\\
C_2^{({\rm em})}(\mu_b) &=& -\frac{190}{8073}\,\eta^{-\frac{35}{23}} - \frac{359}{3105}\,\eta^{-\frac{17}{23}}+ \frac{4276}{121095}\,\eta^{-\frac{12}{23}} + \frac{350531}{1009125}\,\eta^{-\frac{9}{23}} \nonumber\\
&&+ \frac{2}{4347}\,\eta^{-\frac{7}{23}} - \frac{5956}{15525}\,\eta^{\frac{6}{23}}
+ \frac{38380}{169533}\,\eta^{\frac{14}{23}} - \frac{748}{8625}\,\eta^{\frac{16}{23}}\;.
\end{eqnarray}
\\
Using all the above equations, the inclusive branching ratio of $B \to X_s \gamma$ can be obtained.

\subsection{Isospin asymmetry of $B \to K^* \gamma$}%
\label{isospin}
The isospin asymmetry $\Delta_{0}$ in $B \to K^* \gamma$ decays arises when the photon is emitted from the spectator quark. The contribution to the decay width depends therefore on the charge of the spectator quark and is different for charged and neutral $B$ meson decays:\\
\begin{equation}
\Delta_{0\pm}=\dfrac{\Gamma(\bar B^0\to\bar K^{*0}\gamma) - \Gamma(B^\pm \to K^{*\pm}\gamma)}{\Gamma(\bar B^0\to\bar K^{*0}\gamma) + \Gamma(B^\pm\to K^{*\pm}\gamma)}\;,
\end{equation}\\
which can be written as \cite{kagan}:
\begin{equation}
\Delta_{0} =\mbox{Re}(b_d-b_u) \;,
\end{equation} 
where the spectator dependent coefficients $b_q$ take the form:
\begin{equation}
b_q = \frac{12\pi^2 f_B\,Q_q}{\overline{m}_b\,T_1^{B\to K^*} a_7^c}\left(\frac{f_{K^*}^\perp}{\overline{m}_b}\,K_1+ \frac{f_{K^*} m_{K^*}}{6\lambda_B m_B}\,K_{2q} \right)\;.
\end{equation}
In the same way as for $b \to s \gamma$ branching ratio, the SUSY contributions induced by charged Higgs and chargino loops must be taken into account for the calculation of isospin symmetry breaking. \\
\\
The functions $K_1$ and $K_{2q}$ can be written in terms of the Wilson coefficients $C_i$ in the traditional basis (see Appendix \ref{wilsonbasis2}) at scale $\mu_b$ \cite{kagan}:
\begin{eqnarray}
K_1 &=& -\left( C_6(\mu_b) + \frac{C_5(\mu_b)}{N} \right) F_\perp+ \frac{C_F}{N}\,\frac{\alpha_s(\mu_b)}{4\pi}\,\left\{\left( \frac{m_b}{m_B} \right)^2 C_8(\mu_b)\,X_\perp \right.\\
&&\left. -C_2(\mu_b) \left[ \left(\frac43\ln\frac{m_b}{\mu_b} + \frac23 \right) F_\perp - G_\perp(x_{cb})\right] + r_1 \right\}\nonumber\;,\\
\nonumber\\
K_{2q} &=& \frac{V_{us}^* V_{ub}}{V_{cs}^* V_{cb}}\left( C_2(\mu_b) + \frac{C_1(\mu_b)}{N} \right) \delta_{qu} + \left( C_4(\mu_b) + \frac{C_3(\mu_b)}{N} \right) \\
&&+ \frac{C_F}{N}\,\frac{\alpha_s(\mu_b)}{4\pi} \left[ C_2(\mu_b)\left( \frac43\ln\frac{m_b}{\mu_b} + \frac23 - H_\perp(x_{cb}) \right) + r_2 \right] \nonumber\;,
\end{eqnarray}
where $N=3$ and $C_F=4/3$ are color factors, and:
\begin{eqnarray}
r_1 &=& \left[\, \frac83\,C_3(\mu_b) + \frac43\,n_f \bigl(C_4(\mu_b)+C_6(\mu_b)\bigr) - 8\bigl(N C_6(\mu_b)+C_5(\mu_b)\bigr) \right] F_\perp \ln\frac{\mu_b}{\mu_0} + ... \;, \nonumber\\
r_2 &=& \left[ -\frac{44}{3}\,C_3(\mu_b) - \frac43\,n_f\bigl(C_4(\mu_b)+C_6(\mu_b)\bigr) \right]\ln\frac{\mu_b}{\mu_0} + ... \;.
\end{eqnarray}
Here the number of flavors $n_f=5$, and $\mu_0=O(m_b)$ is an arbitrary normalization scale.\\
\\
The coefficient $a_7^c$ reads \cite{bosch}:
\begin{eqnarray}
a^c_7(K^*\gamma) &=& C_7(\mu_b) + \frac{\alpha_s(\mu_b) C_F}{4\pi} \bigl( C_2(\mu_b) G_2(x_{cb})+ C_8(\mu_b) G_8\bigr)\label{a7c}\\
&& +\frac{\alpha_s(\mu_h) C_F}{4\pi} \bigl( C_2(\mu_h) H_2(x_{cb})+ C_8(\mu_h) H_8\bigr) \nonumber\;, 
\end{eqnarray}
where $\mu_h=\sqrt{\Lambda_h \mu_b}$ is the spectator scale, and
\begin{equation}
G_2(x_{cb}) = -\frac{104}{27}\ln\frac{\mu_b}{m_b}+ g_2(x_{cb})\;, \qquad \;\;G_8 = \frac{8}{3}\ln\frac{\mu_b}{m_b} + g_8 \;,
\end{equation}
with
\begin{eqnarray}
g_8 &=& \frac{11}{3}-\frac{2\pi^2}{9}+\frac{2i\pi}{3} \;,\\
g_2(x) &=& \frac{2}{9} \bigg[ 48+30i\pi-5\pi^2-2i\pi^3 -36\zeta(3) +\left( 36+6i\pi-9\pi^2\right)\ln x\nonumber\\
&&+\left( 3+6i\pi\right) \ln^2\! x+\ln^3\! x \bigg] x\\
&& {}+\frac{2}{9} \bigg[ 18+2\pi^2 -2i\pi^3 +\left( 12-6\pi^2 \right)\ln x +6i\pi\ln^2\! x+\ln^3\! x\bigg] x^2\nonumber \\
&& {}+\frac{1}{27} \bigg[ -9+112 i\pi-14\pi^2+\left(182-48i\pi\right)\ln x-126\ln^2\! x\bigg] x^3 \nonumber\\
&&-\frac{833}{162}-\frac{20i\pi}{27} +\frac{8\pi^2}{9} x^{3/2}\;,  \nonumber
\end{eqnarray}
where $\zeta(3)$ is given in Eq. (\ref{zeta3}) and $x_{cb}=\displaystyle\frac{m^2_c}{m^2_b}\;$. The function $H_2(x)$ in Eq. (\ref{a7c}) is defined as:
\begin{equation}
H_2(x)=-\frac{2\pi^2}{3 N}\frac{f_B f^\perp_{K^*}}{T_1^{B\to K^*} m^2_B}\int^1_0 d\xi\frac{\Phi_{B1}(\xi)}{\xi}\int^1_0 dv\, h(\bar v,x)\Phi_\perp(v) \;,
\end{equation}
where $h(u,x)$ is the hard-scattering function:
\begin{equation}
h(u,x)=\displaystyle\frac{4x}{u^2}\left[\mbox{Li}_2\!\left(\displaystyle\frac{2}{1-\sqrt{\displaystyle\frac{u-4x+i\varepsilon}{u}}}\right)+\mbox{Li}_2\!\left(\displaystyle\frac{2}{1+\sqrt{\displaystyle\frac{u-4x+i\varepsilon}{u}}}\right)\right]-\frac{2}{u} \;,
\end{equation}
and $\mbox{Li}_2$ is the usual dilogarithm function given in Eq. (\ref{Li2}).\\
$\Phi_\perp$ is the light-cone wave function with transverse polarization, which can be written in the form \cite{ball}:
\begin{equation}
\Phi_\perp (u) = 6 u \bar{u} \left[ 1 + 3 a_1^\perp\, \xi + a_2^\perp\, \frac{3}{2} ( 5\xi^2-1) \right]\;,
\end{equation}
where $\bar u=1-u$ and $\xi=2u-1$, and $\Phi_{B1}$ is the distribution amplitude of the $B$ meson involved in the leading-twist projection. Finally:
\begin{equation}
H_8=\frac{4\pi^2}{3 N}\frac{f_B f^\perp_{K^*}}{T_1^{B\to K^*} m^2_B}\int^1_0 d\xi\frac{\Phi_{B1}(\xi)}{\xi}\int^1_0 dv\frac{\Phi_\perp(v)}{v} \;.
\end{equation}
The first negative moment of $\Phi_{B1}$ can be parametrized by the quantity $\lambda_B$ such as 
\begin{equation}
\int^1_0 d\xi\frac{\Phi_{B1}(\xi)}{\xi}=\frac{m_B}{\lambda_B}\;. 
\end{equation}
The convolution integrals of the hard-scattering kernels with the meson distribution amplitudes are as follows:
\begin{eqnarray}
F_\perp &=& \int_0^1 dx\,\frac{\phi_\perp(x)}{3\bar x} \;,\nonumber\\
G_\perp(s_c) &=& \int_0^1 dx\,\frac{\phi_\perp(x)}{3\bar x} \, G(s_c,\bar x) \;, \\
H_\perp(s_c) &=& \int_0^1 dx\,\bigg( g_\perp^{(v)}(x) - \frac{g_\perp^{\prime\,(a)}(x)}{4} \bigg)\,G(s_c,\bar x) \;,\nonumber\\
X_\perp &=& \int_0^1 dx\,\phi_\perp(x)\,\frac{1+\bar x}{3\bar x^2}\;,\nonumber
\end{eqnarray}
with $s_c=(m_c/m_b)^2$, and
\begin{equation}
G(s,\bar x) = -4 \int_0^1 du\,u\bar u\,\ln(s-u\bar u\bar x-i\epsilon)\;,
\end{equation}
and the Gegenbauer momenta \cite{ball}:
\begin{eqnarray}
g_\perp^{(a)}(u) & = & 6 u \bar u \left[ 1 + a_1^\parallel \xi + \left\{\frac{1}{4}a_2^\parallel +
\frac{5}{3}\, \zeta^A_{3} \left(1-\frac{3}{16}\,\omega^A_{1,0}\right) + \frac{35}{4} \zeta^V_{3}\right\}(5\xi^2-1)\right]\nonumber\\
& & + 6\, \tilde{\delta}_+ \,  (3u \bar{u} + \bar{u} \ln \bar{u} + u \ln u ) + 6\, \tilde{\delta}_- \,  (\bar{u} \ln \bar{u} - u \ln u)\;,\\
g_\perp^{(v)}(u) & = & \frac{3}{4}(1+\xi^2)+ a_1^\parallel\,\frac{3}{2}\, \xi^3  + \left(\frac{3}{7} \, a_2^\parallel + 5 \zeta_{3}^A \right) \left(3\xi^2-1\right) \nonumber\\
&& + \left( \frac{9}{112}\, a_2^\parallel + \frac{105}{16}\, \zeta_{3}^V - \frac{15}{64}\, \zeta_{3}^A \omega_{1,0}^A \right) \left( 3 - 30 \xi^2 + 35\xi^4\right)\nonumber\\
&& +\frac{3}{2}\,\tilde{\delta}_+\,(2+\ln u + \ln\bar{u}) +\frac{3}{2}\,\tilde{\delta}_-\, ( 2 \xi + \ln\bar{u} - \ln u) \;.
\end{eqnarray}
To compute $X_\perp$, the parameter $\displaystyle X=\ln(m_B/\Lambda_h)\,(1+\varrho\,e^{i\varphi})$ is introduced to parametrize the logarithmically divergent integral $\int_0^1 dx/(1-x)$. $\varrho\le 1$ and the phase $\varphi$ are arbitrary, and $\Lambda_h \approx 0.5$ GeV is a typical hadronic scale. The remaining parameters are given in Appendix~\ref{parameters}.\\
\\
\verb?SuperIso? first computes numerically all the integrals and the Wilson coefficients, and then calculates the isospin asymmetry of $B \rightarrow K^* \gamma$ using all the above equations.

\subsection{Branching ratio of $B_u \to \tau \nu_\tau$}%
\label{Btaunu}
The purely leptonic decay $B_u \to \tau \nu_\tau$ occurs via $W^+$ and $H^+$ mediated annihilation processes. This decay is helicity suppressed in the SM, but there is no such suppression for the charged Higgs exchange at high $\tan\beta$, and the two contributions can therefore be of similar magnitudes. This decay is thus very sensitive to charged Higgs boson and provide important constraints.\\
\\
The branching ratio of $B_u \to \tau \nu_\tau$ in supersymmetry is given by \cite{isidori}
\begin{equation}
\mathrm{BR}(B_u\to\tau\nu_\tau)=\frac{G_F^2f_B^2|V_{ub}|^2}{8\pi}\tau_B m_B m_\tau^2\left(1-\frac{m_\tau^2}{m_B^2}\right)^2 \left[1-\left(\frac{m_B^2}{m_{H^+}^2}\right)\frac{\tan^2\beta}{1+\epsilon_0\tan\beta}\right]^2 \;,\label{btaunu1}
\end{equation}
where $\epsilon_0$ is given in Eq. (\ref{eps0}), and $\tau_B$ is the $B^\pm$ meson lifetime which is given in Appendix~\ref{parameters} together with the other constants in this equation.\\
\\
The following ratio is usually considered to express the new physics contributions:
\begin{equation}
R^{\mathrm{MSSM}}_{\tau\nu_\tau}=\frac{\mathrm{BR}(B_u\to\tau\nu_\tau)_{\mathrm{MSSM}} }{\mathrm{BR}(B_u\to\tau\nu_\tau)_{\mathrm{SM}}}=\left[1-\left(\frac{m_B^2}{m_{H^+}^2}\right)\frac{\tan^2\beta}{1+\epsilon_0\tan\beta}\right]^2 \;,\label{btaunu2}
\end{equation}
which is also implemented in \verb?SuperIso?.\\
\\
In the 2HDM, Eq. (\ref{btaunu1}) takes the form
\begin{equation}
\mathrm{BR}(B_u\to\tau\nu_\tau)=\frac{G_F^2f_B^2|V_{ub}|^2}{8\pi}\tau_B m_B m_\tau^2\left(1-\frac{m_\tau^2}{m_B^2}\right)^2 \left[1-\left(\frac{m_B^2}{m_{H^+}^2}\right)\lambda_{bb}\lambda_{\tau\tau}\right]^2 \;,
\end{equation}
where the Yukawa couplings $\lambda_{bb},\lambda_{\tau\tau}$ can be found in Table \ref{2hdmyuk} for the four types of 2HDM Yukawa sectors.
%
\subsection{Branching ratio of $B \to D \tau \nu_\tau$}%
\label{BDtaunu}
The semileptonic decay $B \to D \tau \nu_\tau$ is similar to $B_u \to \tau \nu_\tau$. The SM helicity suppression here occurs only near the kinematic endpoint. The branching ratio of $B \to D \tau \nu_\tau$ on the other hand is about 50 times larger that the branching ratio of $B_u \to \tau \nu_\tau$ in the SM.\\
\\
In supersymmetry, the partial rate of the transition $B\to D\ell \nu_\ell$ (where $\ell=e,\mu$ or $\tau$) can be written in function of $w$ as \cite{kamenik}
\begin{eqnarray}
\frac{d\Gamma(B\to D\ell \nu_\ell)}{dw} &=& \frac{G_F^2|V_{cb}|^2 m_B^5}{192\pi^3}\rho_V(w) \label{partialbdtaunu}\\
&& \times\left[1 - \frac{m_{\ell}^2}{m_B^2}\, \left\vert 1- t(w)\, \frac{m_b}{(m_b-m_c)m^2_{H^{+}}}\,\frac{\tan^2\beta}{1+\epsilon_0\tan\beta} \right\vert^2 \rho_S(w) \right]\;,\nonumber 
\end{eqnarray}
where $w$ is a kinematic variable defined as:
\begin{equation}
w = \frac{1 + (m_D/m_B)^2-(p_B - p_D)^2/m_B^2}{2 m_D/m_B}\;,
\end{equation}
with $p_D$ and $p_B$ the meson four-momenta, and $t(w) = m_B^2 + m_D^2 - 2 w \,m_D \,m_B$. Again, $\epsilon_0$ is given in Eq. (\ref{eps0}), and
\begin{equation}
t(w) = m_B^2+ m_D^2 - 2 w m_D m_B \;\;.
\end{equation}
In general 2HDM, Eq. (\ref{partialbdtaunu}) is replaced by
\begin{eqnarray}
\frac{d\Gamma(B\to D\ell \nu_\ell)}{dw} &=& \frac{G_F^2|V_{cb}|^2 m_B^5}{192\pi^3}\rho_V(w)\\
&& \times\left[1 - \frac{m_{\ell}^2}{m_B^2}\, \left\vert 1-t(w)\, \frac{m_b\lambda_{bb}-m_c\lambda_{cc}}{(m_b-m_c)m^2_{H^{+}}}\lambda_{\ell\ell} \right\vert^2 \rho_S(w) \right]\;,\nonumber 
\end{eqnarray}
where the Yukawa couplings $\lambda_{bb},\lambda_{\ell\ell}$ can be found in Table \ref{2hdmyuk} for the four types of 2HDM Yukawa sectors.\\
\\
The vector and scalar Dalitz density contributions read \cite{kamenik}
\begin{eqnarray}
\rho_V(w)  &=& 4\,\left(1+\frac{m_D}{m_B}\right)^2\left(\frac{m_D}{m_B}\right)^3\left(w^2-1\right)^{\frac{3}{2}}
\left(1-\frac{m_\ell^2}{t(w)}\right)^2 \left(1+\frac{m^2_\ell}{2 t(w)}\right)\,G(w)^2\;,~~~\\
\rho_S(w) & =& \frac{3}{2} \frac{m_B^2}{t(w)} \,\left(1+\frac{m^2_\ell}{2 t(w)}\right)^{-1} \frac{1+w}{1-w}\,\Delta(w)^2\;,
\end{eqnarray}
where $G(w)$ and $\Delta(w)$ are hadronic form factors. $G(w)$ can be parametrized as
\begin{equation}
G(w) = G(1)\times [1 - 8 \rho^2 z(w) + (51 \rho^2 -10)z(w)^2 - (252\rho^2 - 84)z(w)^3]\;,
\end{equation} 
with 
\begin{equation}
z(w)=\frac{\sqrt{w+1}-\sqrt 2}{\sqrt{w+1}+\sqrt2} \;,
\end{equation} 
and $\Delta(w)$ is \cite{divitiis}
\begin{equation}
\Delta(w) = 0.46 \pm 0.02\;.
\end{equation}
The parameters $G(1)$ and $\rho^2$ are given in Appendix~\ref{parameters}. Integrating Eq. (\ref{partialbdtaunu}) over $w$ leads as a result to the value of the branching ratio.\\
\\
The following ratio
\begin{equation}
\xi_{D\ell\nu} = \frac{\mathrm{BR}(B\to D^0\tau\nu_\tau)}{\mathrm{BR}(B\to D^0 e \nu_e)}
\end{equation}
is also considered in order to reduce some of the theoretical uncertainties. It can be calculated using Eq. (\ref{partialbdtaunu}).
%
\subsection{Branching ratio of $B_s \to \mu^+ \mu^-$}%
\label{Bsmumu}
The rare decay $B_s \to \mu^+ \mu^-$ proceeds via $Z^0$ penguin and box diagrams in the SM, and the branching ratio is therefore highly suppressed. In supersymmetry, for large values of $\tan\beta$ this decay can receive large contributions from neutral Higgs bosons in chargino, charged Higgs and $W$-mediated penguins.\\
\\
The relevant operators for this decay are
\begin{eqnarray}
O_{10}&=&\frac{e^2}{16 \pi^2} (\bar{s}_\alpha \gamma^{\mu} P_L b_\alpha) (\bar{l} \gamma_{\mu}\gamma_5 l)\;,\nonumber\\
O_S&=&\frac{e^2}{16 \pi^2} m_b (\bar{s}_\alpha P_R b_\alpha) (\bar{l}l)\;,\\
O_P&=&\frac{e^2}{16 \pi^2} m_b (\bar{s}_\alpha P_R b_\alpha)(\bar{l} \gamma_5 l)\;,\nonumber
\end{eqnarray}
where $\alpha$ is a color index and $P_{L,R}=(1\mp \gamma_5)/2$.\\
The branching fraction for $\mathrm{BR}(B_s \to \mu^+ \mu^-)$ is given by \cite{bobeth2,ellis}
\begin{eqnarray}
\mathrm{BR}(B_s \to \mu^+ \mu^-) &=& \frac{G_F^2 \alpha^2}{64 \pi^3} f_{B_s}^2 \tau_{B_s} M_{B_s}^3 |V_{tb}V_{ts}^*|^2 \sqrt{1-\frac{4 m_\mu^2}{M_{B_s}^2}} \nonumber \\
&\times& \left\{\left(1-\frac{4 m_\mu^2}{M_{B_s}^2}\right) M_{B_s}^2 | C_S |^2 + \left |C_P M_{B_s} -2 \, C_A \frac{m_\mu}{M_{B_s}} \right |^2\right\} \;,
\end{eqnarray} 
where $f_{B_s}$ is the $B_s$ decay constant, $M_{B_s}$ is the $B_s$ meson mass and $\tau_{B_s}$ is the $B_s$ mean life, all given in Appendix~\ref{parameters}.\\
\\
$C_A$ encloses the SM contributions \cite{ellis}:
\begin{equation}
C_A=\frac{1.033}{\sin^2 \theta_W}\left( \frac{\overline{m}_t(\overline{m}_t)}{170 \, \rm{GeV}} \right)^{1.55}\;,
\end{equation} 
which includes both LO and NLO QCD corrections. 
%
\subsubsection{Supersymmetry}
$C_S$ and $C_P$ include the SUSY loop effects due to Higgs boson contributions, box and penguin SUSY diagrams and counter-terms.
The charged Higgs contribution is given by \cite{bobeth2}
\begin{equation}
C_S^{H^\pm} = -C_P^{H^\pm} = \frac{m_\mu m_t^2 \tan^2\beta}{4 m^2_{H^\pm} M_W^2\sin^2\theta_W} D_3\left(\frac{m_{H^\pm}^2}{m_t^2}\right)\;.
\label{cs_H}
\end{equation}
The SUSY contributions are expressed as \cite{bobeth2}
\begin{eqnarray}
C_{S,P}^{\rm{Box}} = &\mp& \frac{m_\mu\tan^2\beta} {2M_W}\sum_{i,j=1}^{2}\sum_{a=1}^{6}\sum_{k,m,n=1}^{3}
\frac{1}{m_{\tilde{\chi}_i^{\pm}}^2}\Bigg[ (R^\dagger_{\tilde{\nu}})_{lk}(R_{\tilde{\nu}})_{kl} (\Gamma^{U_L})_{am}U_{j2}\Gamma^a_{imn}\nonumber\\
&&\times\Bigg(y_{ai} U_{j2}^{\ast}V_{i1}^{\ast}\pm\frac{m_{\tilde{\chi}_j^{\pm}}}{m_{\tilde{\chi}_i^{\pm}}}
  U_{i2}V_{j1}\Bigg)D_1(x_{ki},y_{ai},z_{ji})\Bigg]\;,
\end{eqnarray} 
\begin{eqnarray}
C_{S,P}^{\rm{Peng}} = &\pm&\frac{m_\mu\tan^2\beta}{M_W^2 m_A^2} \sum_{i,j=1}^{2}\sum_{a,b=1}^{6}\sum_{k,m,n=1}^{3}\Gamma^a_{imn}(\Gamma^{U_L})_{bm}U_{j2} \\
&\times&\Bigg[M_W\Bigg(y_{aj} U_{j2}^{\ast}V_{i1}^{\ast}\pm \frac{m_{\tilde{\chi}_i^{\pm}}}{m_{\tilde{\chi}_j^{\pm}}}U_{i2}V_{j1}\Bigg)D_2(y_{aj},z_{ij})\delta_{ab}\delta_{km}\nonumber\\
&&-\frac{ m_{q_k}}{\sqrt{2}m_{\tilde{\chi}_i^{\pm}}}[\mu^*(\Gamma^{U_R})_{ak}(\Gamma^{U_L\dagger})_{kb}
\pm\mu(\Gamma^{U_L})_{ak}(\Gamma^{U_R\dagger})_{kb}] D_2(y_{ai},y_{bi})\delta_{ij}\Bigg]\;,\nonumber
\end{eqnarray}
and the counter term contribution is
\begin{eqnarray}
C_{S,P}^{\rm{Count}} &=&\mp \frac{m_\mu\tan^3\beta}{\sqrt{2}M_W^2m_A^2}\sum_{i=1}^{2}\sum_{a=1}^{6}\sum_{m,n=1}^{3} \Bigl[ m_{\tilde{\chi}_i^{\pm}} D_3(y_{ai})U_{i2}(\Gamma^{U_L})_{am}\Gamma^a_{imn} \Bigr]\;,
\end{eqnarray} 
where $m_A$ is the mass of the CP-odd Higgs in the MSSM or the mass of the heavy CP-odd Higgs in the NMSSM, $m_{q_k} = (m_u, m_c, m_t)$ and
\begin{equation}
\Gamma^a_{imn}= \frac{1}{2\sqrt{2}\sin^2\theta_W} \, \frac{V_{mb}^{}V_{ns}^*}{V_{tb}^{}V_{ts}^*} \,\bigl[\sqrt{2}M_W V_{i1}(\Gamma^{U_L\dagger})_{na}-m_{q_n}\,V_{i2}(\Gamma^{U_R\dagger})_{na}\bigr]  \;,
\end{equation} 
with
\begin{equation}
x_{ki}=\frac{m_{\tilde{\nu}_k}^2}{m_{\tilde{\chi}_i^{\pm}}^2}\;,\quad
y_{ai}=\frac{m_{{\tilde{u}_a}}^2}{m_{\tilde{\chi}_i^{\pm}}^2}\;,\quad
z_{ij}=\frac{m_{\tilde{\chi}_i^{\pm}}^2}{m_{\tilde{\chi}_j^{\pm}}^2}\;.
\end{equation}
$U_{ij}$ and $V_{ij}$ refer to the chargino mixing matrix elements, and $R_{\tilde{\nu}}$ to the sneutrino mixing matrix. The functions $D_1(x,y,z)$, $D_2(x,y)$ and $D_3(x)$ are defined as
\begin{equation}
D_1(x,y,z)= \frac{x\ln x}{(1-x)(x-y)(x-z)} + (x\leftrightarrow y) + (x\leftrightarrow z)\;,
\end{equation} 
\begin{equation}
D_2(x,y)= \frac{x\ln x}{(1-x)(x-y)}+(x\leftrightarrow y)\;,
\end{equation} 
\begin{equation}
D_3(x)= \frac{x\ln x}{1-x}\;.
\end{equation}  
Note that the function $D_2(x,y)$ coincides with $H_2(x,y)$ in Eq. (\ref{Hxy}). $\Gamma^{U_L}$ and $\Gamma^{U_R}$ are given by
\begin{eqnarray}
(\Gamma^{U_L})^{\rm{T}}&=&\left(\begin{array}{cccccc}
1&0&0&0&0&0\\
0&1&0&0&0&0\\
0&0& \cos\theta_{\tilde{t}}&0&0&-\sin\theta_{\tilde{t}}
\end{array}\right)\;, \\
(\Gamma^{U_R})^{\rm{T}}&=&\left(\begin{array}{cccccc}
0&0&0&1&0&0\\
0&0&0&0&1&0\\
0&0& \sin\theta_{\tilde{t}}&0&0&\cos\theta_{\tilde{t}}
\end{array}\right)\;,  
\end{eqnarray} 
with the stop mixing angle $(-\pi/2\leqslant \theta_{\tilde{t}} \leqslant \pi/2$).\\
\\
In the NMSSM, additional contributions arise from the light CP-odd Higgs $a_1^0$ \cite{hiller2004}:
\begin{eqnarray}
C_P^{a_1^0}& =& m_\ell \tan\beta \frac{1}{\sqrt{2} m_W^2}
\left( \frac{v \delta_- }{x} \right) \frac{1}{m_{B_s}^2-m_{a_1^0}^2} 
\nonumber \\
&\times  &\left\{ -\frac{v \delta_-}{x} 
\sum_{i=1}^2\sum_{a=1}^6\sum_{m,n=1}^3 \Bigl[ m_{\tilde{\chi}_i^{\pm}} D_3(y_{ai})U_{i2}(\Gamma^{U_L})_{am}\Gamma^a_{imn} \Bigr]\right.\\
& +& \left. \frac{\lambda v}{\sqrt{2} \sin^2\theta_W} \sum_{i,j=1}^2\sum_{a=1}^6\sum_{k=1}^3
m_{q_k} (\Gamma^{U_R})_{ak} (\Gamma^{U_L\dagger})_{ka} U_{j2} V_{i2} \Bigl[ y_{aj} U_{j2}^* V_{i1}^* - \frac{m_{\tilde{\chi}_i^{\pm}}^2}{m_{\tilde{\chi}_j^{\pm}}} U_{i2} V_{j1} \Bigr] D_2(y_{aj,z_{ij}})\right\}\nonumber \;,
\end{eqnarray}
where $\lambda v$ is related to the particle masses by
\begin{equation}
m^2_{H^\pm}=m_A^2+m_W^2- \lambda^2 v^2  \;,
\end{equation}
and $v \delta_-/x$ is related to the CP-odd Higgs mixing angle (see Eq. (\ref{Amix})):
\begin{equation}
\theta_A =  \frac{\pi}{2}+ \frac{v}{x \tan \beta} \delta_- \;.
\end{equation}\\
To take into account high $\tan\beta$ corrections, we multiply $C_P$ and $C_S$ by \cite{ellis,dedes}
\begin{equation}
\frac{1}{(1+\epsilon_b \tan\beta)^2}\;,
\end{equation}
where $\epsilon_b$ is given in Eq. (\ref{epsb}).\\
\\
Adding all the contributions to the Wilson coefficients, we can then calculate the branching ratio.
%
\subsubsection{2HDM}
In pure 2HDM without supersymmetry, Eq.~(\ref{cs_H}) does not hold in general and we have to calculate separately different contributions to this equation. For this purpose we extend the results of \cite{logan} to general Yukawa couplings. The most important contributions (for large Yukawa couplings) are given in the following.\\
\\
The first contribution to $C_S$ and $C_P$ is from box diagrams involving $H^+$ and $W^+$:
\begin{equation}
C_S^a = -C_P^a = -\frac{m_{\mu}}{4 M_W^2 \sin^2\theta_W} \left( \lambda_{bb} + \frac{m_s}{m_b} \lambda_{ss} - 2 \frac{m_t}{m_b} \lambda_{tt} \right) \lambda_{\mu\mu} \, B_+(x_{H^+W},x_{tW})\;,
\end{equation}\\
where $x_{H^+W} = M^2_{H^+}/M^2_W$, $x_{tW} = \overline{m}_t^2(m_t)/M_W^2$, and
\begin{equation}
B_+(x,y) = \frac{y}{x-y}\left( \frac{\log y}{y-1} - \frac{\log x}{x-1} \right)\;.
\end{equation}
The second contribution comes from penguin diagrams mediated by neutral Higgs bosons with $H^+$ and $W^+$ in the loop:
\begin{eqnarray}
C_S^b &=& - \frac{m_{\mu}}{4 \sin^2\theta_W} \left( \frac{\sin^2 \!\alpha}{M_{h^0}^2} + \frac{\cos^2 \!\alpha}{M_{H^0}^2} \right) \left( \lambda_{bb} + \frac{m_s}{m_b} \lambda_{ss} - 2 \frac{m_t}{m_b} \lambda_{tt} \right) \lambda_{\mu\mu} \, P_+ \! \left( x_{H^+W}, x_{tW} \right) \;, \nonumber\\
C_P^b &=& \frac{m_{\mu}}{4 M_{A_0}^2\sin^2\theta_W} \left( \lambda_{bb} + \frac{m_s}{m_b} \lambda_{ss} - 2 \frac{m_t}{m_b} \lambda_{tt} \right) \lambda_{\mu\mu} \, P_+ \! \left( x_{H^+W}, x_{tW} \right) \;,
\end{eqnarray}\\
where $\alpha$ is the Higgs mixing angle. The third contribution originates from penguin diagrams mediated by neutral Higgs bosons, involving $H^+$ and $G^+$ in the loops:
\begin{eqnarray}
C_S^c &=& \frac{m_{\mu}}{4 \sin^2\theta_W} \left[ \, \frac{\sin^2\!\alpha}{M^2_{h^0}} 
	\frac{(M_{H^+}^2 - M_{h_0}^2)}{M_W^2} + \frac{\cos^2\!\alpha}{M^2_{H^0}}
	\frac{(M_{H^+}^2 - M_{H_0}^2)}{M_W^2} \right]\nonumber\\
&& \times \left( \lambda_{bb} + \frac{m_s}{m_b} \lambda_{ss} - 2 \frac{m_t}{m_b} \lambda_{tt} \right) \lambda_{\mu\mu} P_+(x_{H^+W},x_{tW}) \;, \\
C_P^c &=& -\frac{m_{\mu}}{4 M^2_{A^0}\sin^2\theta_W} \left( \, \frac{M_{H^+}^2 - M_{A^0}^2}{M_W^2} \right) \left( \lambda_{bb} + \frac{m_s}{m_b} \lambda_{ss} - 2 \frac{m_t}{m_b} \lambda_{tt} \right) \lambda_{\mu\mu} P_+(x_{H^+W},x_{tW}) \;. \nonumber 
\end{eqnarray}\\
These results involve the loop function
\begin{equation}
P_+(x,y) = \frac{y}{x-y} \left( \frac{x \, \log x}{x-1} - \frac{y\, \log y}{y-1} \right)\;.
\end{equation}
The last contribution is from self-energy diagrams:
\begin{eqnarray}
C_S^d &=& -\frac{m_{\mu}}{4 \sin^2\theta_W} \left(\frac{\sin^2\! \alpha}{M_{h^0}^2} + 
         \frac{\cos^2\! \alpha}{M_{H^0}^2} \right) \left( \lambda_{bb} + \frac{m_s}{m_b} \lambda_{ss}\right) \lambda_{\mu\mu}  \nonumber\\
&& \times \left[ x_{H^+W} + \left( \lambda_{bb} + \frac{m_s}{m_b} \lambda_{ss}\right) \lambda_{tt} \right] \, P_+ (x_{H^+W}, x_{tW}) \;, \end{eqnarray}
\begin{equation}
	C_P^d = 
    	\frac{m_{\mu}}{4 M_{A_0}^2\sin^2\theta_W} \, \left( \lambda_{bb} + \frac{m_s}{m_b} \lambda_{ss}\right) \lambda_{\mu\mu} \, 
	\left[ x_{H^+W} + \left( \lambda_{bb} - \frac{m_s}{m_b} \lambda_{ss}\right) \lambda_{tt} \right] \, P_+ (x_{H^+W}, x_{tW})\;. \nonumber
\end{equation}\\
The Yukawa couplings $\lambda_{ii}$ are given in Table \ref{2hdmyuk} for the four types of 2HDM Yukawa sectors. Adding the four contributions, the charged Higgs contribution to the Wilson coefficients can be obtained.

\subsection{Branching ratio of $K \to \mu \nu_{\mu}$}%
\label{Kmunu}
The leptonic kaon decay $K \to \mu \nu_{\mu}$ is also very similar to $B_u \to \tau \nu_\tau$, and is mediated via $W^+$ and $H^+$ annihilation processes. The charged Higgs contribution is however reduced as $H^+$ couples to lighter quarks in this case.\\
\\
We consider the following ratio in \verb?SuperIso? in order to reduce the theoretical uncertainties from $f_K$ \cite{flavianet}, which reads in supersymmetry
\begin{eqnarray}
\dfrac{\rm{BR}(K \to \mu \nu_\mu)}{\rm{BR}(\pi \to \mu \nu_\mu)}&=& 
\frac{\tau_K}{\tau_\pi}\left|\frac{V_{us}}{V_{ud}} \right|^2 \frac{f^2_K}{f^2_\pi} \frac{m_K}{m_\pi}\left(\frac{1-m^2_\ell/m_K^2}{1-m^2_\ell/m_\pi^2}\right)^2 \nonumber \\
&& \times \left[1-\frac{m^2_{K^+}}{M^2_{H^+}}\left(1 - \frac{m_d}{m_s}\right)\frac{\tan^2\beta}{1+\epsilon_0\tan\beta}\right]^2 \left(1+\delta_{\rm em}\right)\;,\label{kmunu1}
\end{eqnarray}
where $\delta_{\rm em} = 0.0070 \pm 0.0035$ is a long distance electromagnetic correction factor, the ratio $f_K/f_\pi$ is given in Appendix~\ref{parameters}, and $\epsilon_0$ for the second generation of quarks reads:
\begin{equation}
\epsilon_0 =-\frac{2\,\alpha_s \,\mu}{3\,\pi \, m_{\tilde{g}}} H_2\left(\frac{m^2_{\tilde{q}_L}}{m^2_{\tilde{g}}},\frac{m^2_{\tilde{d}_R}}{m^2_{\tilde{g}}}\right)\;,
\end{equation}  
where $H_2(x,y)$ is given in Eq. (\ref{Hxy}).\\
\\
The additional quantity $R_{\ell 23}$ \cite{flavianet} is also implemented in \verb?SuperIso?
\begin{equation}
R_{\ell 23}=\left| \frac{V_{us}(K_{\ell 2})}{V_{us}(K_{\ell 3})} \times \frac{V_{ud}(0^+ \to 0^+)}{V_{ud}(\pi_{\ell 2})} \right|=\left|1-\frac{m^2_{K^+}}{M^2_{H^+}}\left(1 - \frac{m_d}{m_s}\right)\frac{\tan^2\beta}{1+\epsilon_0\tan\beta}\right|\;,\label{kmunu2}
\end{equation}
where $\ell i$ refers to leptonic decays with $i$ particles in the final state, and $0^+ \to 0^+$ corresponds to nuclear beta decay.\\
\\
In general 2HDM, Eq. (\ref{kmunu2}) reads
\begin{equation}
R_{\ell 23}=\left|1-\frac{m^2_{K^+}}{M^2_{H^+}}\left(1 - \frac{m_d}{m_s}\right)\lambda_{ss}\lambda_{\mu\mu}\right|\;,
\end{equation}
where the Yukawa couplings $\lambda_{ss},\lambda_{\mu\mu}$ can be found in Table \ref{2hdmyuk} for the four types of 2HDM Yukawa sectors.
%
\subsection{Branching ratio of $D_s \to \ell \nu_\ell$}%
\label{Dslnu}
The purely leptonic decays  $D_s\to \ell\nu_\ell$ are very similar to $K \to \mu \nu_{\mu}$ and proceed via annihilation of the heavy meson into $W^+$ and $H^+$. The charged Higgs boson contribution can only suppress the branching ratio and is therefore slightly disfavored. In supersymmetry the branching fraction is given by (where $\ell=e,\mu$ or $\tau$) \cite{akeroyd,akeroyd2}:
\begin{eqnarray}
\rm{BR}(D_s\to \ell\nu_\ell) &=& \frac{G_F^2}{8\pi} \left|V_{cs}\right|^2 f_{D_s}^2 m_{\ell}^2 M_{D_s} \tau_{D_s} 
\left(1-\frac{m_{\ell}^2}{M_{D_s}^2}\right)^2 \nonumber \\
&& \times \left[1+\left(\frac{1}{m_c+m_s}\right)\left(\frac{M_{D_s}}{m_{H^+}}\right)^2\left(m_c-
\frac{m_s\tan^2\beta}{1+\epsilon_0 \tan\beta}\right) \right]^2  \;.
\label{Ds}
\end{eqnarray}
where $\tau_{D_s}$ and $f_{D_s}$ are the $D_s^\pm$ meson lifetime and decay constant respectively, which are given in Appendix~\ref{parameters} together with the other constants in this equation, and
\begin{equation}
\epsilon_0 =-\frac{2\,\alpha_s \,\mu}{3\,\pi \, m_{\tilde{g}}} H_2\left(\frac{m^2_{\tilde{q}_L}}{m^2_{\tilde{g}}},\frac{m^2_{\tilde{u}_R}}{m^2_{\tilde{g}}}\right)\;,
\end{equation}  
with $H_2(x,y)$ given in Eq. (\ref{Hxy}).\\
\\
In general 2HDM, Eqs. (\ref{Ds}) becomes
\begin{eqnarray}
\rm{BR}(D_s\to \ell\nu_\ell) &=& \frac{G_F^2}{8\pi} \left|V_{cs}\right|^2 f_{D_s}^2 m_{\ell}^2 M_{D_s} \tau_{D_s} 
\left(1-\frac{m_{\ell}^2}{M_{D_s}^2}\right)^2 \nonumber \\
&& \times \left[1-\left(\frac{M_{D_s}}{m_{H^+}}\right)^2\frac{m_s\lambda_{ss}-m_c\lambda_{cc}}{(m_c+m_s)}\lambda_{\ell\ell} \right]^2  \;.
\end{eqnarray}where the Yukawa couplings $\lambda_{cc},\lambda_{\ell\ell}$ can be found in Table \ref{2hdmyuk} for the four types of 2HDM Yukawa sectors.

%
\append{Muon anomalous magnetic moment}%
\label{muon}
The magnetic moment of the muon can be written as
\begin{equation}
M = (1 + a_\mu) \frac{e\hbar}{2m_\mu}\;,
\end{equation}
where $e\hbar/2m_\mu$ is the Dirac moment and the small higher order correction to the tree level is called the anomalous magnetic moment $a_\mu = (g-2)/2$.
%
\subsection{Supersymmetry}
Supersymmetry can contribute to the anomaly through chargino-sneutrino and neutralino-smuon loops. The one-loop SUSY contributions to $a_\mu$ are \cite{martin}
\begin{equation}
\delta a_\mu^{\chi^0} = \frac{m_\mu}{16\pi^2} \sum_{i=1}^{n_{\chi^0}} \sum_{m=1}^2\left\{ -\frac{m_\mu}{ 12 m^2_{\tilde\mu_m}} (|n^L_{im}|^2+ |n^R_{im}|^2)F^N_1(x_{im})  +\frac{m_{\chi^0_i}}{3 m^2_{\tilde \mu_m}} {\rm Re}[n^L_{im}n^R_{im}] F^N_2(x_{im})\right\}\;,
\end{equation}
and
\begin{equation}
\delta a_{\mu}^{\chi^\pm} = \frac{m_\mu}{16\pi^2}\sum_{k=1}^2 \left\{ \frac{m_\mu}{ 12 m^2_{\tilde\nu_\mu}} \Bigl(|c^L_k|^2+ |c^R_k|^2\Bigr)F^C_1(x_k) +\frac{2m_{\chi^\pm_k}}{3m^2_{\tilde\nu_\mu}}\,{\rm Re}[ c^L_kc^R_k]\, F^C_2(x_k)\right\}\;,
\end{equation}
where $n_{\chi^0}$ is 4 in the MSSM and 5 in the NMSSM. $i$, $m$ and $k$ are neutralino, smuon and chargino mass eigenstate labels respectively, and
\begin{eqnarray}
n^R_{im} & = &  \sqrt{2} g' N_{i1} X_{m2} + y_\mu N_{i3} X_{m1} \;,\\
n^L_{im} & = &  \frac{1}{\sqrt{2}} \left (g N_{i2} + g' N_{i1} \right ) X_{m1}^* - y_\mu N_{i3} X^*_{m2}\;,\\
c^R_k & = & y_\mu U_{k2}\;,\\
c^L_k & = & -g V_{k1}\;,
\end{eqnarray}
where $y_\mu = g m_\mu/\sqrt{2} m_W \cos\beta$ is the muon Yukawa coupling, and the $X$ is the smuon mixing matrix. The functions $F_i^N$ and $F_i^C$ depend respectively on $x_{im}=m^2_{\chi^0_i}/m^2_{\tilde\mu_m}$ and $x_k=m^2_{\chi^\pm_k}/m^2_{\tilde\nu_\mu}$ as
\begin{eqnarray}
F^N_1(x) & = &\frac{2}{(1-x)^4}\left[ 1-6x+3x^2+2x^3-6x^2\ln x\right]\;, \\
F^N_2(x) & = &\frac{3}{(1-x)^3}\left[ 1-x^2+2x\ln x\right]\;, \\
F^C_1(x) & = &\frac{2}{(1-x)^4}\left[ 2+ 3x - 6 x^2 + x^3 +6x\ln x\right]\;, \\
F^C_2(x) & = & -\frac{3}{2(1-x)^3}\left[ 3-4x+x^2+2\ln x\right]\;.
\end{eqnarray}
The anomalous magnetic moment can also receive at one loop contributions from the Higgs bosons, which can be large in the NMSSM \cite{ellwanger2008}:
\begin{eqnarray}
\delta a_{\mu}^{H^0}&=&\frac{G_\mu m_{\mu}^2}{4\sqrt{2}\pi^2} \sum_{i=1}^3 
\frac{(U^H_{i2})^2}{ \cos^2\beta}\int_0^1{\frac{x^2(2-x)\,dx}{x^2+ \left(\frac{m_{h_i}}{m_{\mu}}\right)^2(1-x)}} \;,\\
\delta a_{\mu}^{A^0}&=&-\frac{G_\mu m_{\mu}^2}{4\sqrt{2}\pi^2}
\sum_{i=1}^2 (U^A_{i1})^2\tan^2\beta\int_0^1{\frac{x^3\,dx}{x^2+
\left(\frac{m_{a_i}}{m_{\mu}}\right)^2(1-x)}} \;,\\
\delta a_{\mu}^{H^+}&=& \frac{G_\mu m_{\mu}^2}{4\sqrt{2}\pi^2}\tan^2\beta 
\int_0^1{\frac{x(x-1)\,dx}{x-1+ \left(\frac{m_{H^{\pm}}}{m_{\mu}}\right)^2}}\;,
\end{eqnarray}
where $G_\mu=g/4\sqrt2 M_W^2$, and $U^H$ and $U^A$ are respectively the CP-even and CP-odd Higgs mixing matrices given in Eqs. (\ref{Hmix}) and (\ref{Amix}), and $m_{h_i}$ and $m_{a_i}$ refer respectively to the masses of the three CP-even and the two CP-odd Higgs bosons.\\
In addition to these contributions, we also consider the leading logarithm QED correction from two-loop evaluation \cite{degrassi}:
\begin{equation}
a^{\rm SUSY}_{\mu,\, 2\,{\rm loop}}=a^{\rm SUSY}_{\mu,\, 1\,{\rm loop}}\,\left(1-\frac{4\alpha}{\pi}\ln\frac{M_{\rm SUSY}}{m_\mu} \right)\;,
\end{equation}
where $M_{\rm SUSY}$ is a typical superpartner mass scale.\\
\\
The dominant two loop contributions from the photonic Barr-Zee diagrams with physical Higgs bosons are as follows \cite{stockinger}\\
\begin{eqnarray}
a_\mu^{(\chi\gamma H)} &=& \frac{\alpha^2 m_\mu^2}{8\pi^2 M_W^2 \sin^2\theta_W}\
\sum_{k=1,2}\Bigg\{\sum_{a_i}{\rm Re}[ \lambda_\mu^{a_i} \lambda_{\chi^+_k}^{a_i}]
\ f_{PS}(m_{\chi^+_k}^2/M_{a_i}^2) \nonumber\\
&& + \sum_{h_i} {\rm Re}[\lambda_\mu^{h_i} \lambda_{\chi^+_k}^{h_i} ] \ f_S(m_{\chi^+_k}^2/M_{h_i}^2) \Bigg\} \;,\\
a_\mu^{(\tilde{f}\gamma H)} &=& \frac{\alpha^2 m_\mu^2}{8\pi^2 M_W^2 \sin^2\theta_W}\
\sum_{\tilde{f}=\tilde{t},\tilde{b},\tilde{\tau}}(N_c Q^2)_{\tilde{f}} \sum_{j=1,2} \sum_{{h_i}} 
{\rm Re}[ \lambda_\mu^{h_i} \lambda_{\tilde{f}_j}^{h_i}] \ f_{\tilde{f}}(m_{\tilde{f}_j}^2/M_{h_i}^2) \;,
\end{eqnarray}
where $h_i$ and $a_i$ stand respectively for $(h^0,H^0)$ and $A^0$ in the MSSM, and $(h^0,H^0,H_3^0)$ and $(a_1^0,A_2^0)$ in the NMSSM. $N_c$ is the color number and $Q$ the electric charge.\\
\\
The couplings of the Higgs to muon, charginos and sfermions in the MSSM are given by:\\
\begin{eqnarray}
\lambda_\mu^{[h^0,H^0,A^0]} &=& \left[ -\frac{\sin\alpha}{\cos\beta},\frac{\cos\alpha}{\cos\beta},\tan\beta\right]\;,\\
\lambda_{\chi^+_k}^{[h^0,H^0,A^0]} &=& \frac{\sqrt2 M_W}{m_{\chi^+_k}} \Big\{U_{k1}V_{k2}\big[ \cos\alpha, \sin\alpha, -\cos\beta\big] \nonumber \\
&& +U_{k2}V_{k1}\big[ -\sin\alpha,\cos\alpha,-\sin\beta \big]\Big\}\;, \\[0.2cm]
\lambda_{\tilde{t}_i}^{[h^0,H^0]} &=& \frac{2m_t}{m_{\tilde{t}_i}^2 \sin\beta} \Big\{+\mu^*\big[\sin\alpha,-\cos\alpha\big]+A_t\big[\cos\alpha,\sin\alpha\big]\Big\} \ (D^{\tilde{t}}_{i1})^* D^{\tilde{t}}_{i2}\; ,\\
\lambda_{\tilde{b}_i}^{[h^0,H^0]} &=& \frac{2m_b} {m_{\tilde{b}_i}^2 \cos\beta} \Big\{-\mu^* \big[\cos\alpha,\sin\alpha\big] +A_b\big[-\sin\alpha,\cos\alpha\big]\Big\} (D^{\tilde{b}}_{i1})^* D^{\tilde{b}}_{i2}\; ,\\
\lambda_{\tilde{\tau}_i}^{[h^0,H^0]} &=& \frac{2m_\tau} {m_{\tilde{\tau}_i}^2 \cos\beta} \Big\{-\mu^* \big[\cos\alpha,\sin\alpha\big] +A_\tau\big[-\sin\alpha,\cos\alpha\big]\Big\}
\  (D^{\tilde{\tau}}_{i1})^* D^{\tilde{\tau}}_{i2} \;,
\end{eqnarray}
where $U$ and $V$ are the chargino mixing matrices, and $D_{\tilde{f}}$ is the sfermion $\tilde{f}$ mixing matrix.\\
\\
In the NMSSM, these couplings can be generalized as \cite{ellwanger2008}:
\begin{eqnarray}
\lambda_\mu^{h_i}&=&\frac{U^H_{i2}}{\cos\beta}\\
\lambda_\mu^{a_i}&=&U^A_{i2}\tan\beta\\
\lambda_{\chi^{\pm}_k}^{h_i}&=&\frac{\sqrt{2}M_W}{g\, m_{\chi^+_k}} \Big[\lambda U_{k2}V_{k2}U^H_{i3}+g
\left(U_{k1}V_{k2}U^H_{i1}+U_{k2}V_{k1}U^H_{i2}\right)\Big] \;,\\
\lambda_{\chi^{\pm}_k}^{a_i}&=&\frac{\sqrt{2}M_W}{g\, m_{\chi^+_k}} \Big[\lambda U_{k2}V_{k2}U^A_{i2}-g
\left(U_{k1}V_{k2}\cos\beta+U_{k2}V_{k1}\sin\beta\right)U^A_{i1}\Big] \;,\\
\nonumber\\
\lambda_{\tilde{t}_k}^{h_i} &=& \frac{2\sqrt{2}M_W}{g m_{\tilde{t}_i}^2} \Biggl\{h_t\big[A_t U^H_{i1}-\lambda (x U^H_{i2}+v_d U^H_{i3} ) \big] \mbox{Re}[(D^{\tilde{t}}_{k1})^* D^{\tilde{t}}_{k2}] \nonumber\\ 
&& +\Big[h_t^2v_u U^H_{i1}-\frac{g'^2}{3} \left(v_u U^H_{i1}-v_d U^H_{i2}\right)\Big]\left|D^{\tilde{t}}_{k2}
 \right|^2\nonumber\\
&& +\Big[h_t^2v_u U^H_{i1} -\frac{3g^2-g'^2}{12} (v_u U^H_{i1}-v_d U^H_{i2})\Big] \left|D^{\tilde{t}}_{k1}\right|^2 \Biggr\} \;,\\
\nonumber\\
\lambda_{\tilde{b}_k}^{h_i}&=&\frac{2\sqrt{2}M_W}{g m_{\tilde{b}_i}^2}
\Biggl\{h_b\big[A_b U^H_{i2}-\lambda (x U^H_{i1}+v_u U^H_{i3} )\big]
\mbox{Re}[(D^{\tilde{b}}_{k1})^* D^{\tilde{b}}_{k2}] \nonumber\\
&&  +\Big[h_b^2v_d U^H_{i2}+\frac{g'^2}{6} \left(v_u U^H_{i1}-v_d U^H_{i2}\right)\Big]
 \left|D^{\tilde{b}}_{k2}\right|^2\nonumber\\
&& +\Big[h_b^2v_d U^H_{i2} +\frac{3g^2+g'^2}{12}\left(v_u U^H_{i1}-v_d U^H_{i2}\right)\Big]
 \left|D^{\tilde{b}}_{k1}\right|^2\Biggr\} \;,\\
\nonumber\\
\lambda_{\tilde{\tau}_k}^{h_i}&=& \frac{2\sqrt{2}M_W}{g m_{\tilde{\tau}_i}^2} \Biggl\{h_{\tau}\left[A_{\tau} U^H_{i2}-\lambda\left(x U^H_{i1}+v_u U^H_{i3}
\right)\right]\mbox{Re}[(D^{\tilde{\tau}}_{k1})^* D^{\tilde{\tau}}_{k2}] \nonumber\\
&& +\left[h_{\tau}^2v_d U^H_{i2} +\frac{g'^2}{2}\left(v_u U^H_{i1}-v_d U^H_{i2}\right)\right]
 \left|D^{\tilde{\tau}}_{k2}\right|^2\nonumber\\
&& +\left[h_{\tau}^2v_d U^H_{i2} +\frac{g^2-g'^2}{4}\left(v_u U^H_{i1}-v_d U^H_{i2}\right)\right]
 \left|D^{\tilde{\tau}}_{k1}\right|^2\Biggr\} \;,
\end{eqnarray}
where $v_u$, $v_d$ and $x$ are the VEV of $H_u$, $H_d$ and $S$ such as
\begin{equation}
 v_u^2 = \frac{\sin^2\beta}{\sqrt{2} G_F} \qquad , \qquad \tan\beta=\frac{v_u}{v_d} \;.
\end{equation}
\\
The loop integral function $f_{PS}$ is given by:
\begin{equation}
f_{PS}(x) = x \int_0^1 {\rm d}z \frac{1}{z(1-z)-x} \ln\frac{z(1-z)}{x} =
\frac{2x}{y}\Big[{\rm Li}_2\Big(1-\frac{1-y}{2x}\Big) -{\rm Li}_2\Big(1-\frac{1+y}{2x}\Big) \Big]\;.
\end{equation}\\
with $y=\sqrt{1-4x}$. The other loop functions are related to $f_{PS}$ as
\begin{eqnarray}
f_S(x) &=& (2x-1)f_{PS}(x) - 2x(2+\log x)\;,\\
f_{\tilde{f}}(x) &=& \frac{x}{2}\Big[2+\log x-f_{PS}(x)\Big]\;.
\end{eqnarray}
\\
The contribution from the bosonic electroweak two loop diagrams can be written as \cite{heinemeyer,ellwanger2008}:
\begin{equation}
\delta a_{\mu}^{bos}=
\frac{5\, G_F\,m_{\mu}^2\,\alpha}{24\sqrt{2}\,\pi^3}
\left(c_L\ln \frac{m_\mu^2}{M_W^2}+c_0\right)\; ,
\end{equation}
where
\begin{equation}
c_L=\frac{1}{30} \left[98+9c_L^h+23\left(1-4s_W^2\right)^2\right]\;.
\end{equation}
$c_L^h$ in the MSSM is given by
\begin{equation}
c_L^h = \frac{\cos{2\beta}\, M_Z^2}{\cos\beta} \left[ \frac{\cos\alpha
\cos(\alpha+\beta)}{m_H^2} + \frac{\sin\alpha
\sin(\alpha+\beta)}{m_h^2} \right]\; ,
\end{equation}
and in the NMSSM is extended to:
\begin{equation}
c_L^h = \cos{2\beta}\, M_Z^2 \left[ \sum_{i=1}^3 \frac{U^H_{i2} ( U^H_{i2} - \tan\beta\, U^H_{i1} )}{m_{h_i}^2} \right] \;,
\end{equation}
from which the SM bosonic electroweak two loop contributions have to be deduced:
\begin{equation}
\delta a_{\mu}^{SM}=
\frac{5\, G_F\,m_{\mu}^2\,\alpha}{24\sqrt{2}\,\pi^3}
\left(c_L^{SM}\ln \frac{m_\mu^2}{M_W^2}+c_0^{SM}\right)\; ,
\end{equation}
with
\begin{equation}
c_L^{SM}=\frac{1}{30} \left[107+23\left(1-4s_W^2\right)^2\right]\;,
\end{equation}
and $c_0$ and $c_0^{SM}$ are neglected.
%
\subsection{2HDM}
In the 2HDM the dominant contribution can be obtained by generalizing the results of \cite{cheung}:
\begin{eqnarray}
\delta a_\mu^{H} &=& \sum_{f}\frac{\alpha \,m_\mu m_f}{8\pi^3}(N_c Q^2)_f \Biggl\{ -\frac{2\, T_3^f }{m_{A_0}^2}  \rho^f\rho^\mu g\left(x_{fA_0}\right)\\
&& -\frac{1}{m_{h_0}^2} \Bigl[\kappa^f\sin(\beta-\alpha)+\rho^f\cos(\beta-\alpha)\Bigr]\Bigl[\kappa^\mu\sin(\beta-\alpha)+\rho^\mu\cos(\beta-\alpha)\Bigr] f\left(x_{fh_0}\right) \nonumber\\
&& -\frac{1}{m_{H_0}^2} \Bigl[\kappa^f\cos(\beta-\alpha)-\rho^f\sin(\beta-\alpha)\Bigr]\Bigl[\kappa^\mu\cos(\beta-\alpha)-\rho^\mu\sin(\beta-\alpha)\Bigr] f\left(x_{fH_0}\right)\Biggr\} \nonumber
\end{eqnarray}
where $\kappa^f=\sqrt2m_f/v$, $\rho^f=\lambda_{ff} \kappa^f$, $x_{fX}=m_f^2/m_{X}^2$, and $T_3^f$ is the third component of the weak isospin, $-1/2$ for down-type fermions and $1/2$ for up-type fermions. Finally, the $f$ and $g$ functions are given by
\begin{equation}
f(x)=\int_0^1\mathrm{d}y\frac{1-2y(1-y)}{y(1-y)-x}\ln\frac{y(1-y)}{x}\; ,
\end{equation}
and
\begin{equation}
g(x)=\int_0^1\mathrm{d}y\frac{1}{y(1-y)-x}\ln\frac{y(1-y)}{x}\; .
\end{equation}
%
\append{LHA file format for 2HDM}%
\label{lha_2hdm}
SuperIso needs a Les Houches Accord (LHA) inspired input file for the calculations in 2HDM. This file, in addition to the usual MODSEL, SMINPUTS, GAUGE, MASS and ALPHA blocks of SLHA format, needs three additional blocks to specify the Yukawa coupling matrices for up-type and down-type quarks and for leptons. Also, the 2HDM model must be specified in the MODSEL block with the entry 0 followed by a positive integer. Such a file can be generated by \verb?2HDMC?. An example is given in the following:\\
\begin{verbatim}

      Block MODSEL # Select Model
          0   10     #  10 = THDM
      Block SMINPUTS  # Standard Model inputs
          1        1.27910000e+02   # 1/alpha_em(MZ) SM MSbar
          2        1.16637000e-05   # G Fermi
          3        1.17600000e-01   # alpha_s(MZ) SM MSbar
          4        9.11876000e+01   # MZ
          5        4.24680224e+00   # mb(mb)
          6        1.71200000e+02   # mt (pole)
          7        1.77700000e+00   # mtau(pole)
      Block GAUGE  # SM Gauge couplings
          1        3.58051564e-01   # g'
          2        6.48408288e-01   # g
          3        1.47780518e+00   # g_3
      Block MASS      #  Mass spectrum (kinematic masses)
      #  PDG      Mass
          25        9.11123204e+01   # Mh1, lightest CP-even Higgs
          35        5.00013710e+02   # Mh2, heaviest CP-even Higgs
          36        4.99999960e+02   # Mh3, CP-odd Higgs
          37        5.06332023e+02   # Mhc
      Block ALPHA     # Effective Higgs mixing parameter
                    3.12022092e+00   # alpha


      Block UCOUPL
          1     1     0.00000000e+00   # LU_{1,1}
          1     2     0.00000000e+00   # LU_{1,2}
          1     3     0.00000000e+00   # LU_{1,3}
          2     1     0.00000000e+00   # LU_{2,1}
          2     2     2.00000000e-02   # LU_{2,2}
          2     3     0.00000000e+00   # LU_{2,3}
          3     1     0.00000000e+00   # LU_{3,1}
          3     2     0.00000000e+00   # LU_{3,2}
          3     3     2.00000000e-02   # LU_{3,3}
      Block DCOUPL
          1     1     0.00000000e+00   # LD_{1,1}
          1     2     0.00000000e+00   # LD_{1,2}
          1     3     0.00000000e+00   # LD_{1,3}
          2     1     0.00000000e+00   # LD_{2,1}
          2     2    -5.00000000e+01   # LD_{2,2}
          2     3     0.00000000e+00   # LD_{2,3}
          3     1     0.00000000e+00   # LD_{3,1}
          3     2     0.00000000e+00   # LD_{3,2}
          3     3    -5.00000000e+01   # LD_{3,3}
      Block LCOUPL
          1     1    -5.00000000e+01   # LL_{1,1}
          1     2     0.00000000e+00   # LL_{1,2}
          1     3     0.00000000e+00   # LL_{1,3}
          2     1     0.00000000e+00   # LL_{2,1}
          2     2    -5.00000000e+01   # LL_{2,2}
          2     3     0.00000000e+00   # LL_{2,3}
          3     1     0.00000000e+00   # LL_{3,1}
          3     2     0.00000000e+00   # LL_{3,2}
          3     3    -5.00000000e+01   # LL_{3,3}


\end{verbatim}
The UCOUPL, DCOUPL and LCOUPL blocks contain respectively the Yukawa couplings $\lambda_{UU}$, $\lambda_{DD}$ and $\lambda_{LL}$. If some entries in these blocks are missing, they will be set to 0. The Yukawa couplings for the types I--IV are specified in Table~\ref{2hdmyuk}.
\newpage
\append{Suggested limits}%
\label{constraints}
The limits on the masses of Higgs and SUSY particles from direct searches at colliders are given in Table~\ref{masslimits}. Some of the limits are subject to auxiliary conditions (see \cite{PDG}) which are also taken into account in the program. These values are encoded in \verb?src/excluded_masses.c? and can be updated by the user if necessary.\\
\\
Finally, in Table \ref{limits}, we present our suggested limits for each observable, which can be used to constrain SUSY parameters.\\
We would like to stress however that some of the inputs in this table suffer from large uncertainties from the determination of CKM matrix elements and/or hadronic parameters. The constraints obtained using these observables should therefore not be over-interpreted (see \cite{eriksson} for more details).
\begin{table}[!h]
\begin{center}
\hspace*{1.5cm}\begin{tabularx}{15.cm}{|X|X||X|X|X} 
\cline{1-4}
\multicolumn{4}{|c|}{}\\[-1mm]
\multicolumn{4}{|c|}{\textbf{Lower bounds on the particle masses in GeV}}\\[3mm]
\cline{1-4}
\multicolumn{4}{c}{\vspace*{-4mm}}\\
\cline{1-4}
$h^0$  & 111 & $\tilde{e}_R$ & 73 &\\[3mm]
\cline{1-4}
$H^+$  & 79.3 & $\tilde{\mu}_R$ & 94 &\\[3mm]
\cline{1-4}
$A^0$  & 93.4 & $\tilde{\tau}_1$ & 81.9 &\\[3mm]
\cline{1-4}
$\chi^0_1$  & 46 & $\tilde{\nu}$  & 94 &\\[3mm]
\cline{1-4}
$\chi^0_2$  & 62.4 & $\tilde{t}_1$ & 95.7 &\\[3mm]
\cline{1-4}
$\chi^0_3$  & 99.9 & $\tilde{b}_1$  & 89 &\\[3mm]
\cline{1-4}
$\chi^0_4$ & 116 & $\tilde{q}$ & 379 &\\[3mm]
\cline{1-4}
$\chi^\pm_1$ & 94 & $\tilde{g}$ & 308 &\\[3mm]
\cline{1-4}
\end{tabularx}
\end{center}
\caption[Higgs and MSSM particles mass limits]{ Limit on the masses of Higgs and MSSM particles from direct searches \cite{PDG}. The limit on the $h^0$ mass includes a 3 GeV intrinsic uncertainty. \label{masslimits}}
\end{table}
\begin{table}[p]
\begin{center}
\hspace*{-2.1cm}\begin{tabularx}{15cm}{|c|c|c|X}
\cline{1-3}
\textbf{Observable} & \textbf{Combined experimental value} & \textbf{95\% C.L. Bound}& \\[3mm]
\cline{1-3}&&\\
${\rm BR}(B \to X_s \gamma) $& $ (3.52\pm0.23\pm0.09)\times10^{-4}$ \cite{HFAG} &  $2.15\times 10^{-4}\leq \mathrm{BR(}b\to s\gamma\mathrm{)}\leq 4.89\times 10^{-4}  $\\[3mm]
\cline{1-3}&&\\
$\Delta_0(B \to K^* \gamma)$ & $(3.1 \pm 2.3)\times 10^{-2}$ (a) &$-1.7\times 10^{-2} < \Delta_0 < 8.9\times 10^{-2}$\\[3mm]
\cline{1-3}&&\\
${\rm BR}(B_u \to \tau \nu_\tau)$ &$ (1.41 \pm 0.43) \times 10^{-4}$ \cite{HFAG} & $ 0.39 \times 10^{-4} < {\rm BR}(B_u \to \tau \nu_\tau) < 2.42 \times 10^{-4}  $\\[3mm]
$R_{\tau\nu_\tau}$ & $ 1.28\pm 0.38 $ (b) &$ 0.52 < R_{\tau\nu_\tau} < 2.04$ \\[3mm]
\cline{1-3}&&\\
${\rm BR}(B \to D^0 \tau \nu_\tau)$ &$ (8.6 \pm 2.4 \pm 1.1 \pm 0.6) \times 10^{-3} $ \cite{babar2} &$ 2.9\times 10^{-3} < {\rm BR}(B \to D^0 \tau \nu_\tau) < 14.2\times 10^{-3} $\\[3mm]
$\xi_{D \ell \nu}$ &$ 0.416 \pm 0.117 \pm 0.052$ \cite{babar2} &$ 0.151 < \xi_{D \ell \nu} < 0.681 $\\[3mm]
\cline{1-3}&&\\
${\rm BR}(B_s \to \mu^+ \mu^-)$ &$ < 5.8 \times 10^{-8}$ \cite{CDF}&$ {\rm BR}(B_s \to \mu^+ \mu^-) < 6.6 \times 10^{-8}$\\[3mm]
\cline{1-3}&&\\
$\dfrac{{\rm BR}(K \to \mu \nu)}{{\rm BR}(\pi \to \mu \nu)}$ & $0.6358\pm0.0011$ (c) & $0.6257 < \dfrac{{\rm BR}(K \to \mu \nu)}{{\rm BR}(\pi \to \mu \nu)} < 0.6459$\\[5mm]
$R_{\ell 23}$ & $1.004 \pm 0.007$ \cite{flavianet} &$ 0.990 < R_{\ell 23} < 1.018$ (d)\\[3mm]
\cline{1-3}&&\\
${\rm BR}(D_s \to \tau \nu_\tau)$ &$ (5.7 \pm 0.4) \times 10^{-2}$ \cite{akeroyd2} & $ 4.8 \times 10^{-2} < {\rm BR}(D_s \to \tau \nu_\tau) < 6.6 \times 10^{-2}$\\[3mm]
${\rm BR}(D_s \to \mu \nu_\mu)$ & $ 5.8\pm 0.4  \times 10^{-3}$ \cite{akeroyd2} &$ 4.9 \times 10^{-3} < {\rm BR}(D_s \to \mu \nu_\mu) < 6.7 \times 10^{-3}$ \\[3mm]
\cline{1-3}&&\\
$\delta a_\mu$&  $(2.95\pm 0.88)\times 10^{-9}$ \cite{reviewmuon} &$ 1.15\times 10^{-9} <\delta a_\mu < 4.75\times 10^{-9} $\\[3mm]
\cline{1-3}
\end{tabularx}
\end{center}
\caption[Limits for implemented observables]{ Suggested limits for the observables implemented in SuperIso v3.0. The 95\% C.L. bounds presented in this table include both the experimental and the theoretical uncertainties.\label{limits}}\vspace*{0.5cm}
(a) Value obtained combining the most recent measurement of Babar \cite{babar4} with the results of \cite{belle,PDG}.\\
(b) Value deduced from \cite{HFAG}.\\
(c) Value obtained combining the results of \cite{flavianet,PDG}.\\
(d) See \cite{eriksson} for a discussion on the uncertainties.
\end{table}%
\append{Useful parameters}%
\label{parameters}
The masses of quarks and mesons, as well as some other useful parameters such as lifetimes, CKM matrix elements or decay constants, are given in Table \ref{tabparam}.
\begin{table}[p]
\begin{center}
\begin{tabularx}{162mm}{|X|X|X|X|X|X|X|}
\hline
\multicolumn{7}{|c|}{\textbf{Meson masses in MeV} \cite{PDG}}\\
\hline
$m_{\pi}$ & $m_{K}$ & $m_{K^*}$ & $m_{D}$ & $m_{D_s}$ & $m_{B}$ & $m_{B_s}$\\[1mm]
\hline
139.6 & 493.7 & 891.7 & 1864.8 & 1968.5 & 5279.1 & 5366.3\\
\hline
\end{tabularx}\\[5mm]
\begin{tabularx}{162mm}{|X|X|X|X|X|}
\hline
\multicolumn{5}{|c|}{\textbf{Meson lifetimes in s} \cite{PDG}}\\
\hline
$\tau_{\pi}$ & $\tau_{K}$ & $\tau_{B}$ & $\tau_{B_s}$& $\tau_{D_s}$\\[1mm]
\hline
$2.6033 \times 10^{-8}$ & $1.2380 \times 10^{-8}$ & $1.638 \times 10^{-12}$ & $1.425 \times 10^{-12}$ & $5.00 \times 10^{-13}$\\
\hline
\end{tabularx}\\[5mm]
\begin{tabularx}{162mm}{|X|X|X|X|X|}
\hline
\multicolumn{5}{|c|}{\textbf{Quark masses in GeV and $\mathbf{\alpha_s}$} \cite{PDG}}\\
\hline
$\overline{m}_b(\overline{m}_b)$ & $\overline{m}_c(\overline{m}_c)$ & $m_s$ & $m_t^{\rm pole}$ \cite{mtop} & $\alpha_s(M_Z)$\\
\hline
$4.20^{+0.17}_{-0.07}$ & $1.27^{+0.07}_{-0.11}$ & $0.104^{+0.026}_{-0.034}$ & $172.4 \pm 1.2$ & $0.1176\pm0.002$\\[1mm]
\hline
\end{tabularx}\\[5mm]
\begin{tabularx}{162mm}{|c|X|X|X|X|X|X|}
\hline
\multicolumn{7}{|c|}{\textbf{Hadronic parameters in MeV} ($\mu = 1$ GeV)}\\
\hline
$f_K/f_{\pi}$ \cite{HPQCD} & $f_{K^*}$ \cite{ball2} & $f^\perp_{K^*}$ \cite{ball2} & $f_B$ \cite{lubicz} & $\lambda_B$ \cite{braun} & $f_{B_s}$ \cite{lubicz} & $f_{D_s}$ \cite{follana}\\[1mm]
\hline
$1.189\pm0.007$ & $220 \pm 5$ & $185 \pm 10$ & $200\pm20$ & $460\pm110$ & $245\pm25$ & $241\pm3$\\
\hline
\end{tabularx}\\[5mm]
\begin{tabularx}{162mm}{|X|X|X|X|X|X|}
\hline
\multicolumn{6}{|c|}{\textbf{Meson mass and coupling parameters} ($\mu = 1$ GeV) \cite{ball}}\\
\hline
$\zeta^A_{3}$ & $\zeta^V_{3}$ & $\zeta^T_{3}$ & $\omega_{1,0}^A$ & $\tilde{\delta}_+$ & $\tilde{\delta}_-$\\[1mm]
\hline
0.032 & 0.013 & 0.024 & -2.1 & 0.16 & -0.16\\
\hline
\end{tabularx}\\[5mm]
\begin{tabularx}{162mm}{|X|X|X|X|}
\hline
\multicolumn{4}{|c|}{\textbf{Meson related parameters} ($\mu = 1$ GeV)}\\
\hline
$a_1^\perp(K^*)$ \cite{ball3} & $a_2^\perp(K^*)$ \cite{ball2} & $a_1^\parallel(K^*)$ \cite{ball3} & $a_2^\parallel(K^*)$ \cite{ball3}\\[1mm]
\hline
$0.04 \pm 0.03$ & $0.15 \pm 0.10$ & $0.03 \pm 0.02$ & $0.11 \pm 0.09$\\
\hline\hline
$G(1)$ \cite{divitiis} & $\rho^2$ \cite{kamenik} & $\Delta(w)$ \cite{divitiis} & $T_1^{B\to K^*}$ \cite{ball2}\\[1mm]
\hline
$1.026\pm0.017$ & $1.17\pm0.18$ & $0.46\pm0.02$ & $0.31 \pm 0.04$\\
\hline
\end{tabularx}\\[5mm]
\begin{tabularx}{162mm}{|X|X|X|X|}
\hline
\multicolumn{4}{|c|}{\textbf{CKM parameters}}\\
\hline
$|V_{us}|$ \cite{flavianet} & $|V_{ub}|$ \cite{PDG} & $|V_{cb}|$ \cite{PDG} & $|V_{cs}|$ \cite{PDG}\\[1mm]
\hline
$0.2253\pm0.0009$ & $(3.95\pm0.35)\times 10^{-3}$ & $(4.12\pm0.11) \times 10^{-2}$ & $ 0.97334\pm0.00023$\\
\hline
\end{tabularx}
\begin{tabularx}{162mm}{|X|X|X|}
\hline
$|V_{us}/V_{ud}|$ \cite{flavianet} & $|V_{ts} V_{tb}/V_{cb}|^2$ \cite{misiak2} & $\mbox{Re}(V_{us}^* V_{ub}/V_{cs}^*V_{cb})$ \cite{kagan}\\[1mm]
\hline
$0.2321\pm0.0015$ & $0.9676 \pm 0.0033$ & $0.011\pm 0.005$\\
\hline
\end{tabularx}\\[5mm]
\begin{tabularx}{162mm}{|X|X|X|X|}
\hline
\multicolumn{4}{|c|}{\textbf{$\mathbf{b \to s \gamma}$ related parameters}}\\
\hline
$C$ \cite{bauer} & $E_0$ (GeV) \cite{misiak2} & $\lambda_2$ (GeV$^2$) \cite{misiak2,PDG} & \mbox{$\rm{BR}(B\to X_c e \bar\nu)_{\rm{exp}}$ \cite{babar3}}\\[1mm]
\hline
$0.58 \pm 0.01$ & 1.6 & 0.12 & $0.1061 \pm 0.0017$\\
\hline
\end{tabularx}
\end{center}
\caption[Input parameters]{Input parameters.\label{tabparam}}
\end{table}
%
%
\newpage
\section*{Acknowledgments}
\noindent The author would like to thank Alexandre Arbey for many useful discussions, and for his technical helps throughout the development of the program. Thanks also to Ben Allanach and Peter Skands for their help regarding the SLHA2 format, Pietro Slavich for useful discussions, and the members of THEP group at Uppsala University for interesting discussions and in particular Oscar St{\aa}l for useful discussions on 2HDM. The author is also grateful to Cyril Hugonie for his help with NMSSMTools and his comments.\\
\\
The literature on flavor physics observables and indirect searches for new physics is very rich, therefore it is impossible to refer to the complete list here. Since the present manuscript is meant to serve as a manual for the program, only the articles used directly in the program are acknowledged for clarity. We refer to \cite{allanach_tools} for a fairly complete review of available tools in SUSY.

\end{document}